\documentclass[12pt]{article}

\usepackage{multirow, bm, euscript, url, verbatim}
\usepackage{amsmath,amssymb,graphicx, graphics, color} 
\usepackage{bbm}
\usepackage{cite}
\usepackage{latexsym}
\usepackage{array}
\usepackage{cancel}

\usepackage{hyperref}

\usepackage{enumerate}

\usepackage[normalem]{ulem} 

\usepackage{dsfont}

\newcommand{\centeron}[2]{{\setbox0=\hbox{#1}\setbox1=\hbox{#2}\ifdim
\wd1>\wd0\kern.5\wd1\kern-.5\wd0\fi \copy0
\kern-.5\wd0\kern-.5\wd1\copy1\ifdim\wd0>\wd1
                                   \kern.5\wd0\kern-.5\wd1\fi}}
\newcommand{\ltap}{\>\centeron{\raise.35ex\hbox{$<$}}
                           {\lower.65ex\hbox{$\sim$}}\>}
\newcommand{\gtap}{\>\centeron{\raise.35ex\hbox{$>$}}
                           {\lower.65ex\hbox{$\sim$}}\>}

\newcommand\ZZ{\hbox{\zfont Z\kern-.4emZ}}
\font\zfont = cmss10 

\newcommand{\fref}[1]{Fig.\ \ref{f.#1}}
\newcommand{\eref}[1]{Eq.\ (\ref{e.#1})}

\newcommand{\sref}[1]{Section \ref{s.#1}}
\newcommand{\ssref}[1]{Section \ref{ss.#1}}

\newcommand{\cref}[1]{Chapter \ref{c.#1}}
\newcommand{\tref}[1]{Table \ref{t.#1}}

\newcommand{\tempstringone}{}
\newcommand{\tempstringtwo}{}

\newcommand{\ba}{\begin{array}}
\newcommand{\ea}{\end{array}}

\newcommand{\beq}{\begin{eqnarray}}
\newcommand{\eeq}{\end{eqnarray}}
\newcommand{\beqs}{\begin{eqnarray*}}
\newcommand{\eeqs}{\end{eqnarray*}}

\newcommand{\bal}{\begin{align}} 
\newcommand{\eal}{\end{align}}

\def\bi{\begin{itemize}}
\def\ei{\end{itemize}}
\def\ben{\begin{enumerate}}
\def\een{\end{enumerate}}
\def\bc{\begin{center}}
\def\ec{\end{center}}
\def\bt{\begin{table}}
\def\et{\end{table}}
\def\btb{\begin{tabular}}
\def\etb{\end{tabular}}

\def\gev{\, {\rm GeV}}

\def\tev{\, {\rm TeV}}

\def\mass2{mass${}^2$}

\begin{document}
\bibliographystyle{unsrt}
\begin{titlepage}

\vskip2.5cm
\begin{center}
\vspace*{5mm}
{\huge \bf Mixing It Up With $M_{T2}$:} \vspace{5mm}

{\huge \bf Unbiased Mass Measurements} \vspace{5mm}

{\huge \bf at Hadron Colliders}
\end{center}
\vskip0.5cm

\begin{center}
{\bf David Curtin}
\end{center}
\vskip 8pt

\begin{center}
{\it Department of Physics, YITP,  Stony Brook University, Stony Brook, NY 11794.}\\
\vspace*{0.5cm}
{\it Department of Physics, LEPP, Cornell University, Ithaca, NY 14853.} 
\\

\vspace*{0.9cm}

{\tt curtin@insti.physics.sunysb.edu}

\end{center}

\vglue 0.9truecm
\begin{abstract}
Recently, much progress has been made on techniques to measure the masses of new particles with partially-invisible decays at a hadron collider. 
We examine for the first time the realistic application of $M_{T2}$-based measurement methods to a fully hadronic final state from a symmetric two-step decay chain with maximal combinatorial uncertainty. Several problems arise in such an analysis: the $M_{T2}$ variables are powerful but fragile, with shallow edges that are easily washed out or faked by ubiquitous combinatorics background. Traditional methods of both cleaning up the distribution and determining edge position can fail badly.
To perform successful mass measurements we introduce several new techniques: the Edge-to-Bump method of extracting an edge from a distribution by analyzing a distribution of fits rather than a single fit; a very simple yet high-yield method for determining decay-chain assignments event-by-event; and a systematic procedure to obtain $M_{T2}$ edge measurements in the presence of heavy combinatorics background, they key element being the parallel use of at least two independent methods of reducing combinatorics background to avoid fake measurements.
All of these techniques are developed in a Monte Carlo study of the decay $\tilde g \tilde g \rightarrow 2 \tilde b + 2 b  \rightarrow 4 b + 2 \chi^0_1$ and verified in a second blind study with a different spectrum. In both cases, the gluino and sbottom masses are measured to a precision of $\sim 10\%$ with $\mathcal{O}(100 \mathrm{fb}^{-1})$ at the LHC14 (assuming pessimistic $b$-tag efficiencies).
\end{abstract}
\end{titlepage}

\tableofcontents

\section{Introduction}
\label{s.intro} \setcounter{equation}{0} \setcounter{footnote}{0}

There are very good reasons to believe that the Standard Model (SM) of particle physics is an incomplete description of nature. We expect the Large Hadron Collider (LHC) to soon find evidence of beyond-Standard-Model (BSM) physics, and after a discovery is made the next order of business is measuring the properties of the new particles. 

Supersymmetry is one of the most promising extensions of the Standard Model. It solves the hierarchy problem and allows for perturbative gauge coupling unification. Its simplest incarnation, the Minimal Supersymmetric Standard Model (MSSM), has a discrete symmetry under which all superpartners are charged. This makes the lightest supersymmetric particle (LSP) stable, and if it is neutral the LSP can be a viable dark matter (DM) candidate. Furthermore, this implies that any produced superpartners must decay into pairs of LSPs. (Many other BSM theories also feature a discrete symmetry that stabilizes a DM candidate and forces it to be pair-produced, so while we use the language of supersymmetry for familiarity our discussion applies to those cases as well.)

The noisy environment of a hadron collider makes any measurement challenging. 
If the final state of a particle collision can be fully reconstructed, the masses of intermediate particles can often be determined by looking for resonances in the invariant mass spectrum. But in SUSY and other theories which produce final states with missing transverse energy (MET), mass determination requires the use of more sophisticated methods of analyzing the decay chain. One way is to look for kinematic edges in the distributions of different invariant mass combinations of the daughter particles \cite{kinematicedges}. The locations of these edges reveal information about the unknown particle masses, and if enough of these are measured in a long decay chain, complete mass determination is possible. Another approach is the polynomial method \cite{polynomialmethod}, which involves solving the four-momentum equations of all the measured signal events simultaneously to determine all the masses. The third option is to use the family of $M_{T2}$-based kinematic variables \cite{MT2original, MT2subsystem, MT2analytical, MT2perp, MT2diffmothers, MT2diffdaughters, MT2other,MT2hemispheremethod,ZurekMT2perp}, which are generalizations of the simple transverse mass to the case of two massive invisible particles in the decay chain. Complete mass determination is possible in a chain as short as two decays by measuring the endpoints/edges in the distributions of the various $M_{T2}$-subsystem variables \cite{MT2subsystem} one can construct. (Exploiting the dependence of these variables on the total $p_T$ carried away by initial state radiation (ISR) can even make it possible to determine all the masses in a single-step decay chain \cite{MT2perp, ZurekMT2perp}.)

There is still much work to be done in translating all of these ideas into realistic applications. 
In this paper we concentrated on the invariant-mass-edge and $M_{T2}$ based approaches and the problems that arise in their application to a fully hadronic final state with maximal combinatorial uncertainty. $M_{T2}$ endpoints are much harder to measure than invariant mass edges. They are more vulnerable to combinatorics background, since for these variables it is both very ubiquitous as well as possessing of internal structure. This makes fake edge measurements very hard to avoid. Even if this issue is addressed, traditional methods of extracting endpoints from distributions fail for realistic distributions of $M_{T2}$ subsystem variables, since their edges are very shallow.

We addressed these issues in a Monte Carlo study of the decay  $\tilde g \tilde g \rightarrow 2 \tilde b + 2 b \rightarrow 4 b + 2 \chi^0_1$ with the aim of extracting all the unknown masses. This led to the development of three new measurement techniques:\vspace{-2mm}
\begin{enumerate} 
\itemsep=0mm
\item Extracting an endpoint from a distribution is traditionally done by fitting a kink-like function to some subset of the data. For shallow $M_{T2}$ edges (with possibly several fake edges in the distribution), this introduces unacceptable levels of systematic error and human bias into the process.  Our approach is to analyze a \emph{distribution of many simple fits}, rather than a single sophisticated fit. We implement this idea in the ``Edge-to-Bump'' method which turns the problem of edge-measurement into bump-hunting and can be used to extract multiple edge measurements with meaningful error bars from any kind of distribution. We also make a Mathematica implementation the algorithm publicly available. 

\item We outline an extremely simple and high-yield procedure to deduce correct decay chain assignments for $\mathcal{O}(10\%)$ of events, given a known ${M_{jj}}$ edge. While there are other methods of dealing with unknown decay chain assignments \cite{MixedEventTechnique, MT2analytical, MT2hemispheremethod, combinatoricsmatrixelement, combinatoricsRY, combinatoricsBKMN, combinatoricsCGP}, to the best of our knowledge this is the only event-by-event method with 100\% purity at parton-level (without measurement errors).

\item We introduce two simple methods of cleaning up $M_{T2}$ distributions with combinatorics background: one uses the above decay chain assignment, the second simply drops the largest few $M_{T2}$ possibilities per event. While these methods work well some of the time, we argue that in principle no single method can be trusted to reliably reveal an $M_{T2}$ edge and avoid fake measurements. The only way to avoid such false positives is the simultaneous use of (at least) \emph{two separate methods} of reducing combinatorics background. The edges obtained from each method are used to cross-check the other, and the measurement is only kept if they agree. 
\end{enumerate}

We first encountered these issues in \cite{susysumrule}, where we conducted a parton-level Monte Carlo study of the same decay to measure the light stop and sbottom masses and show that the SUSY-Yukawa sum rule could provide meaningful constraints on the stop and sbottom mixing angles. Our method of determining decay-chain assignments was presented in that earlier work, as well as the basic idea of using two methods of reducing combinatorics background to cross-check $M_{T2}$ measurements, but a fully consistent application required the development of the Edge-to-Bump method. 

The purpose of this article is to flesh out all these basic ideas and develop them into realistic measurement techniques, which is done in Sections \ref{s.edgemeasurement}, \ref{s.kinedge} and \ref{s.MT2}. The Monte Carlo study used to develop these techniques, which includes showering/hadronization and detector effects, is discussed in \sref{casestudy1}. To ensure that our analysis was not inadvertently `fine-tuned' for one particular spectrum, we performed a second \emph{blind} Monte Carlo study in \sref{casestudy2}, which was successful and demonstrates the general applicability of our measurement techniques.  We conclude with \sref{conclusion}, and provide additional plots from the collider studies in the Appendix.

\section{The Edge-to-Bump  Measurement Method}
\label{s.edgemeasurement} \setcounter{equation}{0} \setcounter{footnote}{0}

The simplest example of a kinematic edge  arises when considering the decay chain $A \rightarrow j_1 B$, $B \rightarrow j_2 X$, where $X$ is invisible and $j_1, j_2$ are some SM particles. Neglecting the mass of the SM daughters and assuming the decay is on-shell, it is easy to show that their invariant mass cannot exceed $M_{jj}^{max} = \sqrt{(m_A^2 - m_B^2)(m_B^2 - m_X^2)/m_B^2}$. The $M_{jj}$ distribution will feature an endpoint or edge at $M_{jj} = M_{jj}^{max}$, and measuring the location of this feature reveals information about the masses of $A, B$ and $X$. In practice this is complicated by combinatorics background and various smearing effects, but since the kinematic edge tends to be reasonably steep and the combinatorics background fairly flat, the extraction of such kinematic edges is well understood \cite{kinematicedges}. 

As we have already mentioned, the family of $M_{T2}$-type kinematic variables  \cite{MT2original, MT2subsystem, MT2analytical, MT2perp, MT2diffmothers, MT2diffdaughters, MT2other,MT2hemispheremethod,ZurekMT2perp} is potentially much more powerful than the simple invariant mass, allowing for complete mass measurement in a two- or maybe even a one-step decay chain. These variables also feature endpoints in their distribution which reveal the mass information, but by nature of their construction they are much less robust, yielding shallower edges that are more difficult to measure and more vulnerable to combinatorics background, which itself can have unwanted features that introduce artifacts into the total distribution. Measuring these edges reliably in a realistic setting for a fully hadronic final state was one of the main challenges of this paper. In working our way towards a working solution we had to reconsider the basic procedure for extracting edges from a distribution, leading to the development of the Edge-to-Bump method.

\subsection{The Basic Idea}
Edges are by their very nature problematic features to detect. Unlike for bumps, the important part of the edge is defined by only very few events, with most of the data carrying little information. Since we usually do not know the full shape of the distribution a global fit is out of the question, so the problem is usually approached by fitting a function to a small subset of the data. This function is usually some kind of kink function (the most primitive example being a linear kink, two joined lines with different gradients), and the hope is that this fit function is a good approximation of the actual event distribution in the vicinity of the edge. 

The choice of any particular approximate fit function introduces systematic error into the edge measurement that is hard to quantify. Since the usual procedure involves visually identifying a feature and choosing some range of data to fit the function to, this introduces human bias into the process. For most if not all fit functions, the chosen domain of the fit also influences the measurement, again a hard-to-quantify systematic error, and merely fitting the function over some range of domains leaves the choice of range to the human, again a source of bias. The statistical error returned by the fit does not reflect any of these contributions and hence represents a gross overestimation of confidence in the edge position, which can lead to plain false measurements. Needless to say this approach is far from ideal, and while the above mentioned problems might seem peripheral and of limited physical interest they are in fact  prohibitive to conducting realistic $M_{T2}$-based mass measurements in the presence of combinatorics background. This motivates our search for a solution.

The main problem stems from the unknown shape of the distribution and the use of one or a few fits. One might try to ameliorate these problems with ever more sophisticated choices of fit function, but that does not address the basic issue. We instead propose the opposite approach: to use a very basic fit function, but fit it thousands of times to one distribution, over domains of random length and position. This allows us to analyze the \emph{distribution of fits} rather than a single fit itself. The simplest way to proceed (we comment on some possible elaborations below) is to consider the distribution of found edges, which will be peaked around actual physical edges. The problem of \emph{edge detection} has been transformed into the much more tractable problem of \emph{bump hunting}, and the various sources of error -- physical smearing of the edge due to detector effects or initial state radiation (ISR), choice of fit domain and fit function -- are all reflected in the width of found peaks, probed by sheer redundancy. 

This approach, which we call  the ``Edge-to-Bump Method'', has the advantage of being in principle fully automated (removing human bias) and probing the entire distribution, allowing it to find several physical edges in the data if they exist -- they will merely be reflected as multiple peaks in the edge distribution.

Let us now move on to describing our particular implementation of this basic idea, which we will later use in our collider studies. We emphasize that our algorithm should be seen as a working proof-of-concept, with probably much room for optimization or improvement.

\subsection{Detailed Procedure}
\label{ss.edgefitterprocedure}
As an example consider a distribution in some variable, call it $M$, which has two edges or endpoints at  $M = M_A$ and $M_B$, represented schematically in \fref{sampleedge}(a). One or both of these edges might be physically interesting, and we want to determine their position. 

\subsubsection*{Step 1: Generate Random Fit Domains}
Generate many random domains, i.e. line intervals $(M_{start}, M_{end})$, such that the distributions of the line intervals' lengths and midpoints are flat. This avoids introducing bias into the kink distribution obtained from fitting linear kink functions over each of these domains. A typical number of domains to generate is about 5,000.

\subsubsection*{Step 2: Fit Kinks to $M$-Distribution}
Using the linear kink PDF shown in \fref{sampleedge}(b), obtain a measured kink position $K$ for each of the generated fit domains, see \fref{samplekinkdistribution}(a). Many of measured kinks will not be physically meaningful if the $M$-distribution does not contain a real kink inside that fitdomain, but the obtained $K$ values should peak around real kinks in the distribution. 

\begin{figure}
\begin{center}
\begin{tabular}{ccc}
\includegraphics[width=6cm]{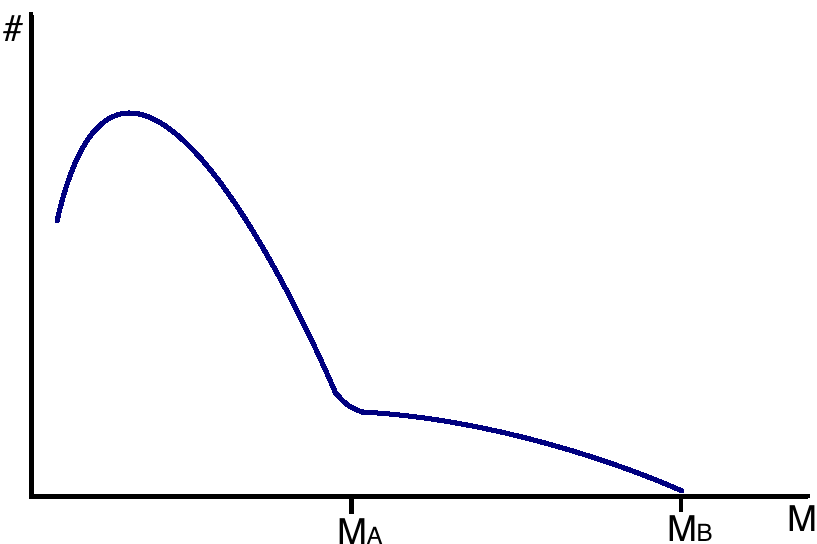}
&
&\includegraphics[width=6cm]{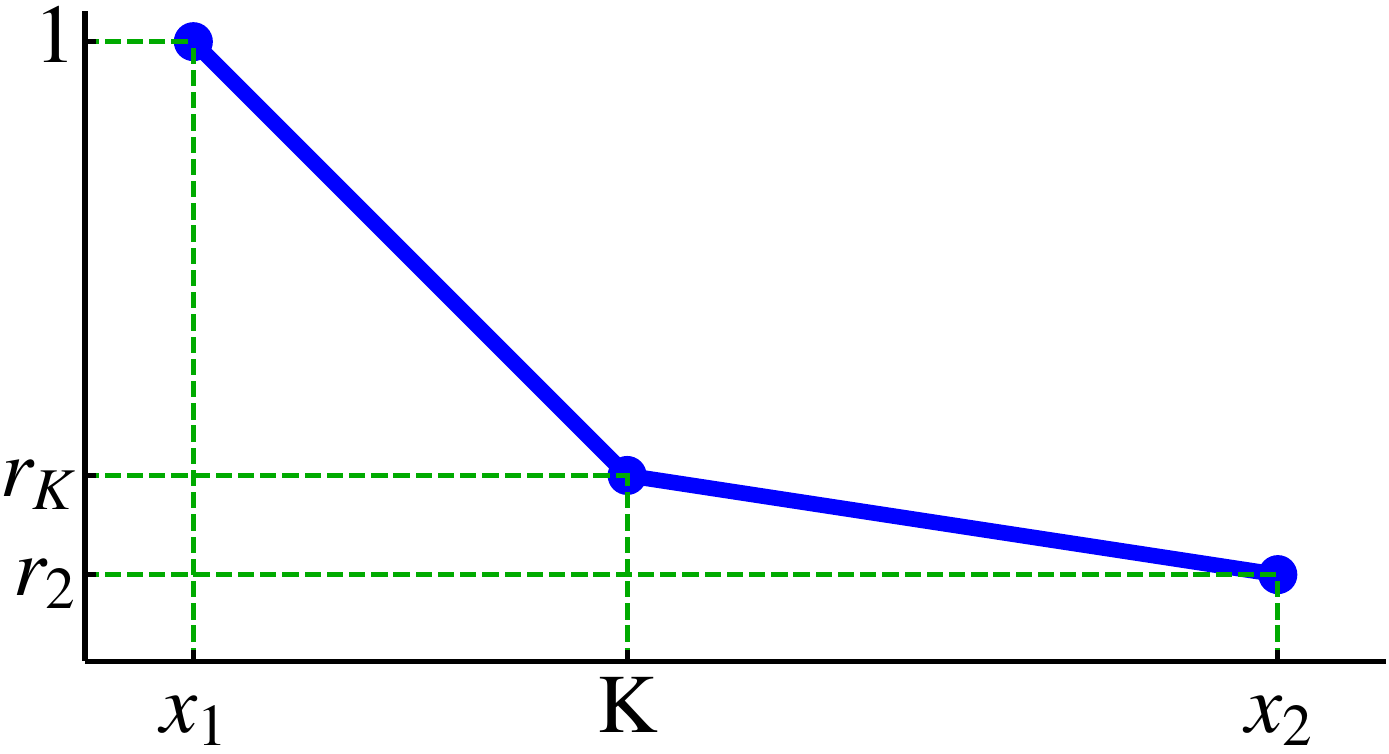}\\ 
\\ (a) & & (b)
\end{tabular}
\end{center}
\caption{(a) A schematic example of a distribution in some kinematic variable $M$ which features two edges/endpoints at $M = M_{A}, M_{B}$. (b) The linear kink fit function used in our procedure. For each fit with chosen fit domain $(x_1, x_2)$, the variables $K, r_K, r_2$ float, with $K$ being the kink position.}
\label{f.sampleedge}
\end{figure}

\begin{figure}
\begin{center}
\begin{tabular}{m{6cm} m{5mm} m{6cm}}
\includegraphics[width=6cm]{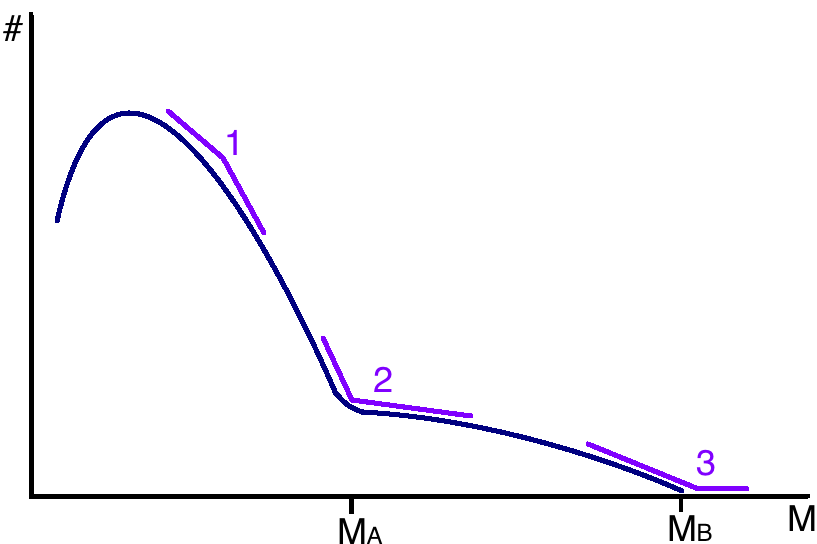}
&$\longrightarrow$ 
&\includegraphics[width=6cm]{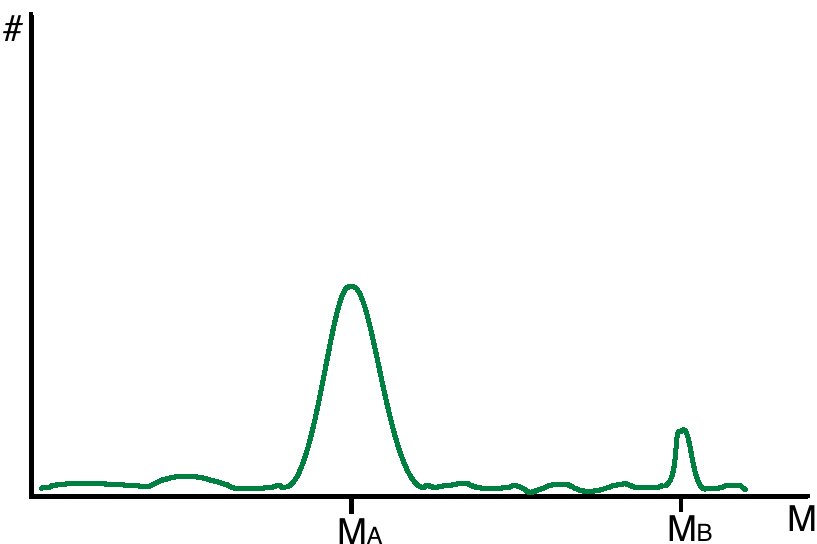}\\ 
\\ \textcolor{white}{fillerfillerfillerfil} (a) & & \textcolor{white}{fillerfillerfillerfil} (b)
\end{tabular}
\end{center}
\caption{(a) The data is fit to the linear kink function over all the generated domains, here shown for three examples. Each domain yields a kink position value $K$. (b) After applying some basic filters we plot the distribution of the obtained values of $K$. Edges in the data show up as peaks in the kink distribution.}
\label{f.samplekinkdistribution}
\end{figure}

\subsubsection*{Step 3: Obtain Kink Distribution}

We now have a collection of $K$-values corresponding to kinks found in each of the fit domains. We want to eliminate kinks that are clearly irrelevant, i.e. obtained from fitting to a small handful of $M$-values at the very end of the distribution, or tiny fluctuations in the data. For this reason we discard kinks that were obtained using less than some number $N_{min}$ of events and kinks that were obtained from a fit domain shorter than some minimum length $L_{min}$. We typically choose $N_{min} = 50$ or so for a distribution of a few thousand values and $L_{min}$ to be a few tens of the minimum possible/sensible bin size. \emph{The exact values of $N_{min}$ and $L_{min}$ will not significantly affect the result.} We also discard kinks where the corresponding fit has a flat likelihood function and kinks that do not correspond to end points (i.e. we require the second gradient in each fit to be smaller than the first). The resulting kink distribution looks something like \fref{samplekinkdistribution}(b), and edges in the $M$-distribution are now visible as peaks in the kink distribution.

If the $M$-distribution were extremely clean and only had one edge we could just use the mean and standard deviation of the entire kink distribution as our measured edge position and uncertainty. In practice, however, there will be a `diffuse background' of irrelevant kinks scattered throughout the kink distribution, and there might be more than one peak (as is the case for our schematic example). We therefore need some way of detecting the separate peaks and analyzing their \emph{shape}.

\subsubsection*{Step 4: Detect Peaks in Kink Distribution}
The remainder of the process deals with detecting peaks in the obtained kink distribution and measuring their position, yielding measurements of the corresponding edges in the $M$-distribution. There are many ways of doing this (one could borrow various `bump-hunting' techniques), and we will only show one method we developed that works well for all the examples we studied.

Consider a general data distribution (in our case, the locations of found kinks). If we look only for very narrow peaks we are likely to miss very wide peaks. It therefore makes sense to define a maximum peak width $w$ that we want to be sensitive to, and scan over $w$ to detect all the peaks of different width in a distribution.  A real peak will show up for all (or many) $w$-values above some $w_{min}$.

Say we want to test whether there is a peak of (at most) width $\sim w$ at position $M_0$ in the data. Define a boundary width $b = 2 w$ and restrict ourselves to the range $(M_0 - \frac{w}{2} - b, M_0 + \frac{w}{2} + b)$. Define $N_L, N_0, N_R$ as the number of data points in the bins $(M_0 - \frac{w}{2} - b, M_0 - \frac{w}{2})$, $(M_0 - \frac{w}{2}, M_0 + \frac{w}{2} )$ and $ (M_0 + \frac{w}{2} , M_0 + \frac{w}{2} + b)$. If the data distribution in our selected range were flat, then we would expect $\langle N_0 \rangle = \frac{w}{w+2b} N_{tot}$, $\langle N_L \rangle = \langle N_R \rangle  = \frac{b}{w+2b}N_{tot}$, where $N_{tot}$ is the total number of points in our selected data range. Assuming $N_{tot} > 0$, we say \emph{there is a peak of (at most) width $w$ in the data range $(M_0 - 0.5 w, M_0 + 0.5 w)$ if the following are all true:}
\begin{eqnarray*}
\langle N_L \rangle - N_L &>&  s \sqrt{\langle N_L \rangle},\\
\langle N_R \rangle - N_R &>&  s \sqrt{\langle N_R \rangle},\\
N_0 -  \langle N_0 \rangle  &>&  s \sqrt{\langle N_0\rangle},\\
\end{eqnarray*}
where we set $s = 3$. In other words, we require there to be 3 $\sigma$ more events in the center bin and 3 sigma fewer events in both side bins than expected for a flat distribution.  If we then scan over the value of $M_0$ we obtain candidate \emph{peak intervals} in which we expect to find peaks. This is shown schematically in \fref{peakintervals}(a).

Since we want to detect kinks of all sizes, we scan over the parameter $w$ and obtain peak intervals for each value. The resulting plot will look something like \fref{peakintervals}(b). The real peaks are reliably detected and distinguished from random noise and show up as up-side down cones growing with $w$.

\begin{figure}
\begin{center}
\hspace*{-1cm}
\begin{tabular}{ccc}
\includegraphics[width=5.6cm]{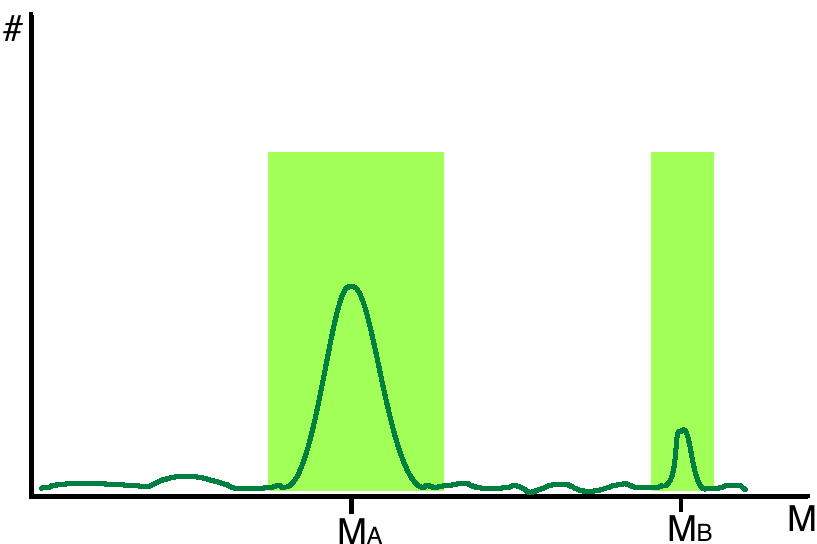}
&
\includegraphics[width=5.6cm]{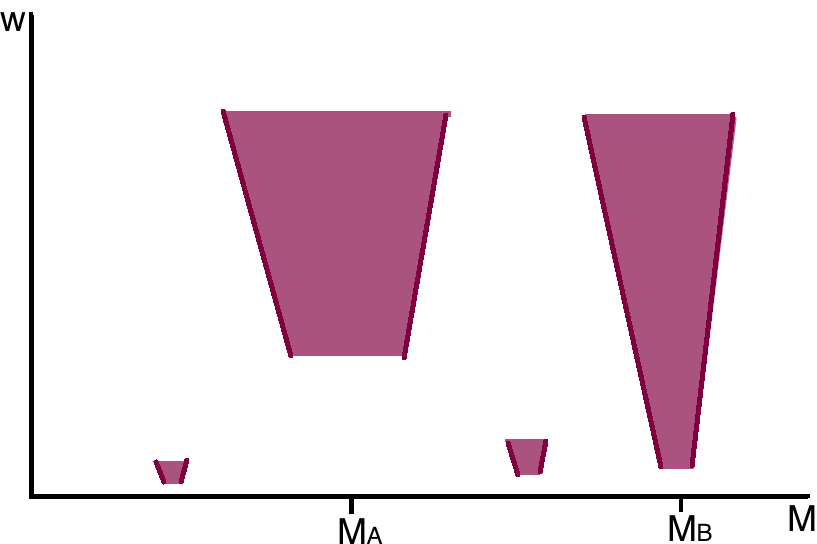}
&
\includegraphics[width=5.6cm]{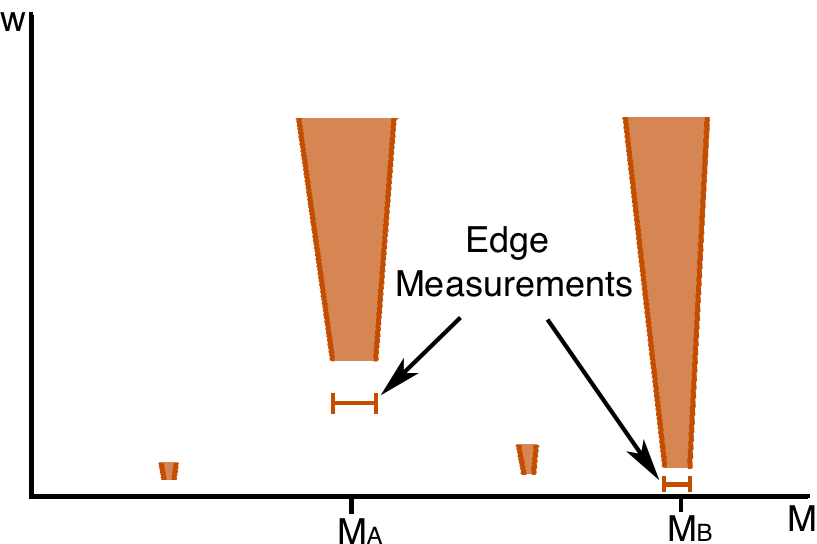}\\
(a)&(b)&(c)
\end{tabular}
\end{center}
\caption{(a) For a given $w$, we detected two peaks in the kink distribution (peak intervals shaded). (b) A plot of the found peak intervals as a function of $w$. Note how the two real peaks are reliably detected above some minimum $w$ (which depends on the size of the peak), while small-scale fluctuations only show up at small $w$. (c) Plot of 1-$\sigma$ confidence level intervals of the peak's position for each $w$. For the two physical peaks, keep the measurement with the smallest error as the final measurement of the peak/edge position.}
\label{f.peakintervals}
\end{figure}

\subsubsection*{Step 5: Obtain Edge Measurements from found Peaks}
We want to turn each peak interval (for each $w$) into a measurement of peak position. To this end, extend it symmetrically in each direction by either $b = 2 w$ or until one boundary hits another peak interval, then take the mean and standard deviation of the data within that extended interval. This will give a measurement of the peak's position with associated 1$\sigma$  error. Plotting the obtained 1-$\sigma$ confidence level intervals (1$\sigma$CLI) of peak position vs $w$ yields \fref{peakintervals}(c). Notice how the real edges show up as `broadening rivers' flowing from small to large $w$, since the consistent detection for $w$ larger than the minimum value is characteristic of a real peak in the edge distribution.

For each of the two physical peaks identified in the previous step, keep the measurement with the smallest error as the measurement of the corresponding kink's position.

\subsubsection*{Comments}

It is very important to point out that our procedure, specifically the kink filtering in step 3, can occasionally produce peaks in the edge distribution that are very clearly \emph{filter artifacts}. This can arise in flat parts near the beginning of the original distribution if it has low statistics: the kink filter that selects for endpoints will keep edges corresponding to downward fluctuations of the flat distribution while discarding edges that correspond to upward fluctuations. This can produce a fake peak in the edge distribution that will show up as an edge in the measurement-vs-peakwidth plot. Such edges are easily identified and should be ignored. 

One could replace steps 4 and 5 by a different procedure for detecting peaks in a distribution, but the method we present works well enough for our studied examples. While our peak-detection method is certainly physically motivated, the choice of the particular values for $s$, $b$ and the amount by which we extend each peak interval to obtain the associated peak measurement were optimized using many artificially generated distributions with edges of varying quality and the first Monte Carlo study presented in this paper. That being said, changing the values generally does not have a large effect on the measurement outcome. 

If there is no clearly `dominant' peak in the peakwidth vs $w$ distribution (see \fref{peakintervals}(b)) this means that no clear edge can be detected in the $M$-distribution. This might seem ambiguous, but in all the examples we have studied the decision is obvious, and certainly much less prone to bias than direct human visual identification of an edge in a messy distribution. We also point out that that extremely sharp, steep edges in the data will likely have their position and associated uncertainty slightly overestimated by our method, since the fit function we use is more suitable for relatively flat edges. Since the detection of extremely well-defined features is less problematic to begin with, our method can be seen as complementary since it focuses on shallower edges that tend to arise in $M_{T2}$ distributions or generally in the presence of smearing and background.

The main idea of the `Edge-to-Bump' method is to find edges (or other features) not by looking at the original distribution but at a distribution of many found fits over random domains.  While we simply plotted the histogram of kink position after some filtering there are many other analyses one could perform on the fit-distribution. For example, one could assign each edge a quality factor and weigh it accordingly, or make use of correlations between kink position and other fit properties, like gradient change. Some very preliminary investigations suggest the latter method especially could simplify and improve the measurement process, and we leave its detailed exploration for future study.

Our method is fairly computationally intensive: (uncompiled) Mathematica on a single 2 GHz CPU core takes several hours to perform the required thousands of fits over a single distribution. 
Implementation in a faster programming language would no doubt improve this by orders of magnitude.

\subsection{Examples}

\newcommand{\temp}{6.7cm}

\begin{figure}
\begin{center}
\begin{tabular}{cc}
\includegraphics[width=\temp]{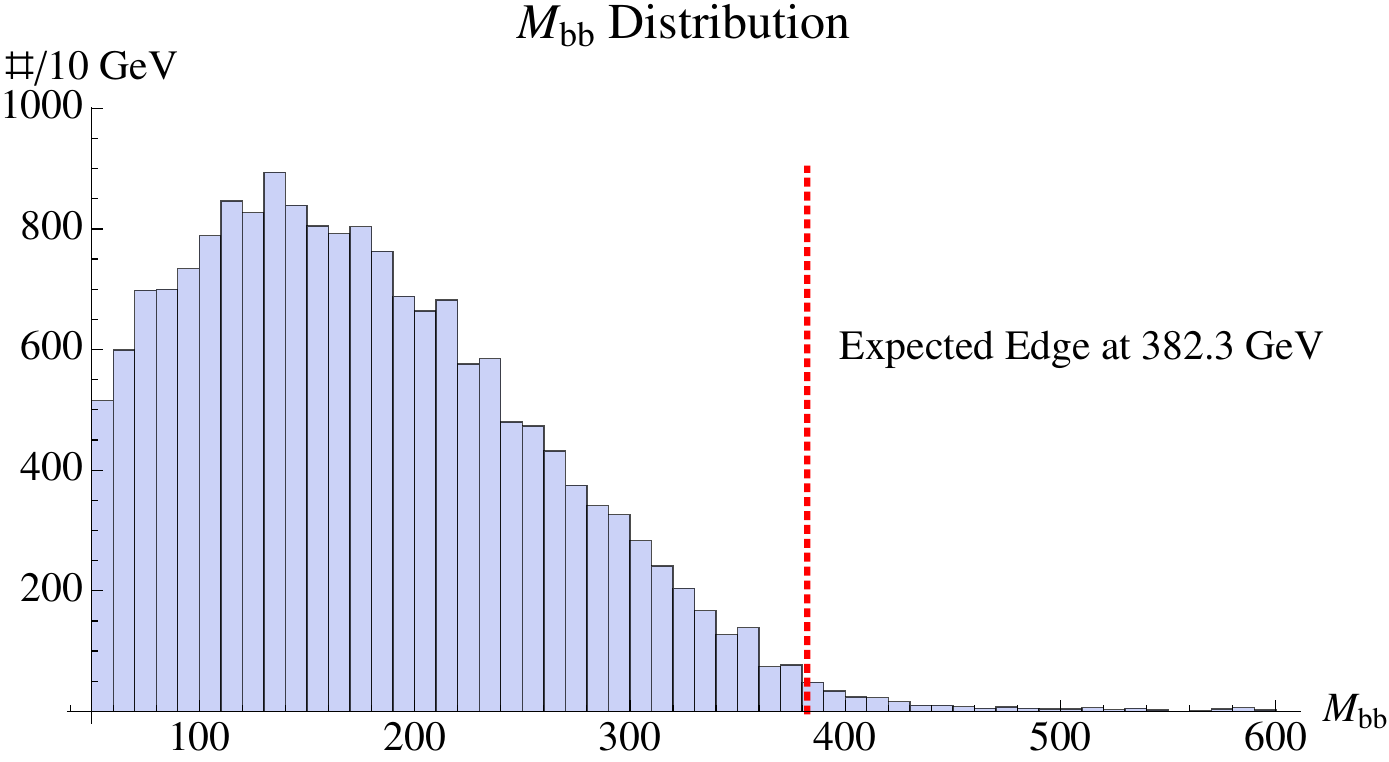}
&
\includegraphics[width=\temp]{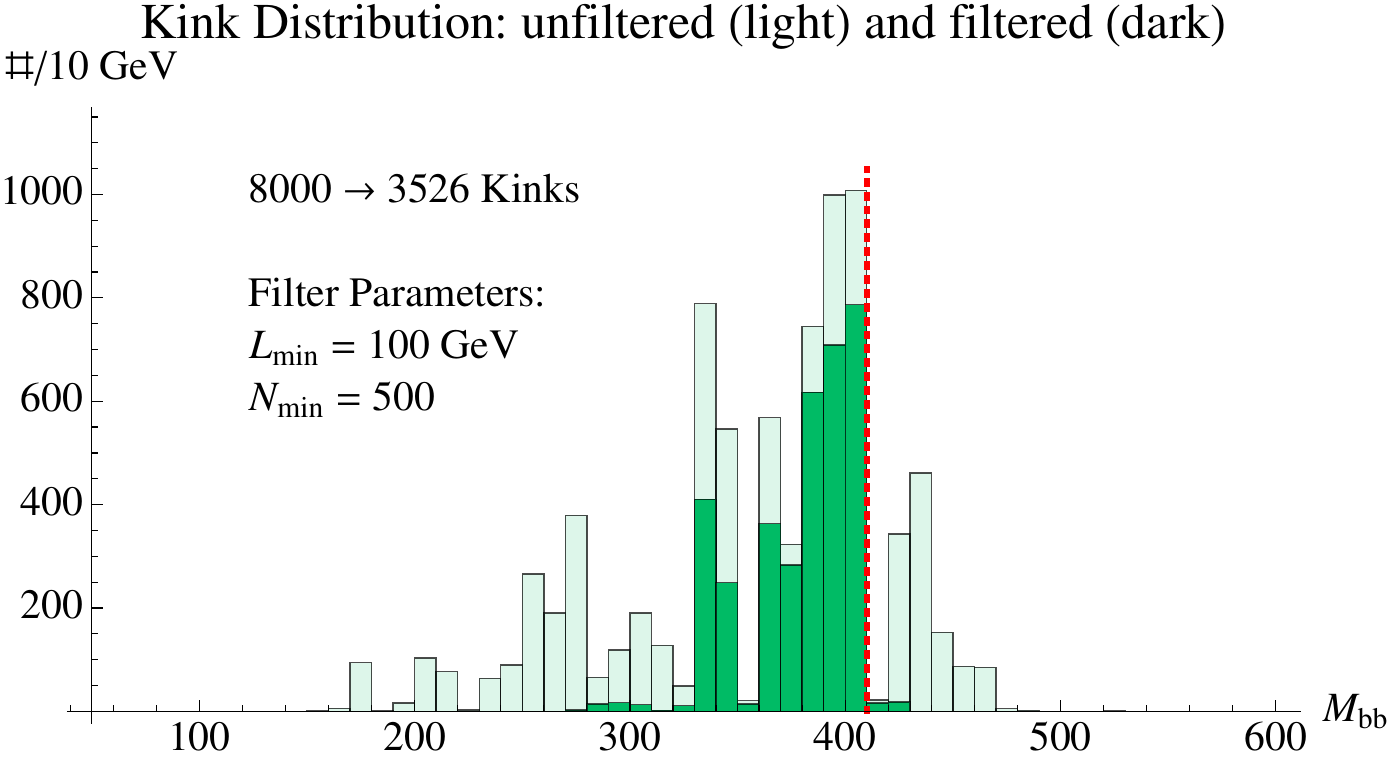}\\
\includegraphics[width=\temp]{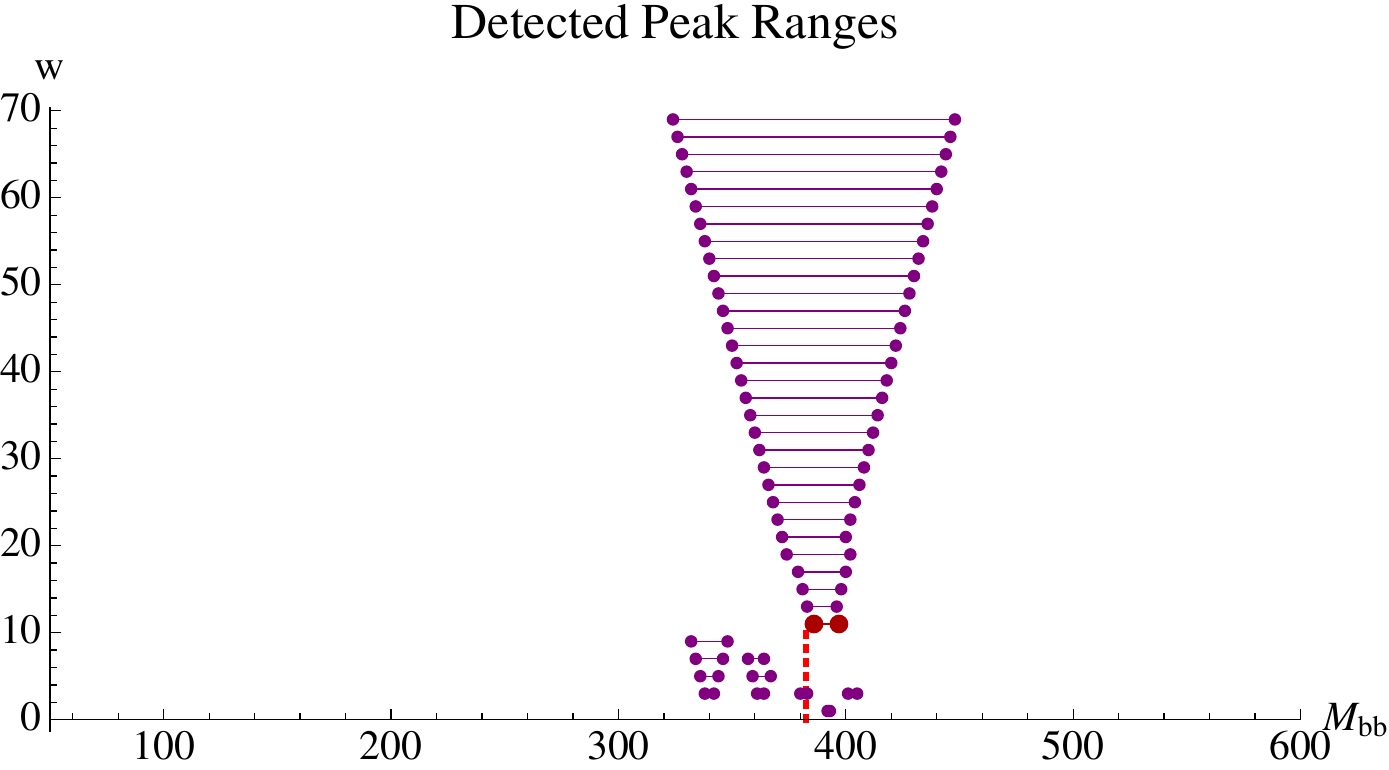}
&
\includegraphics[width=\temp]{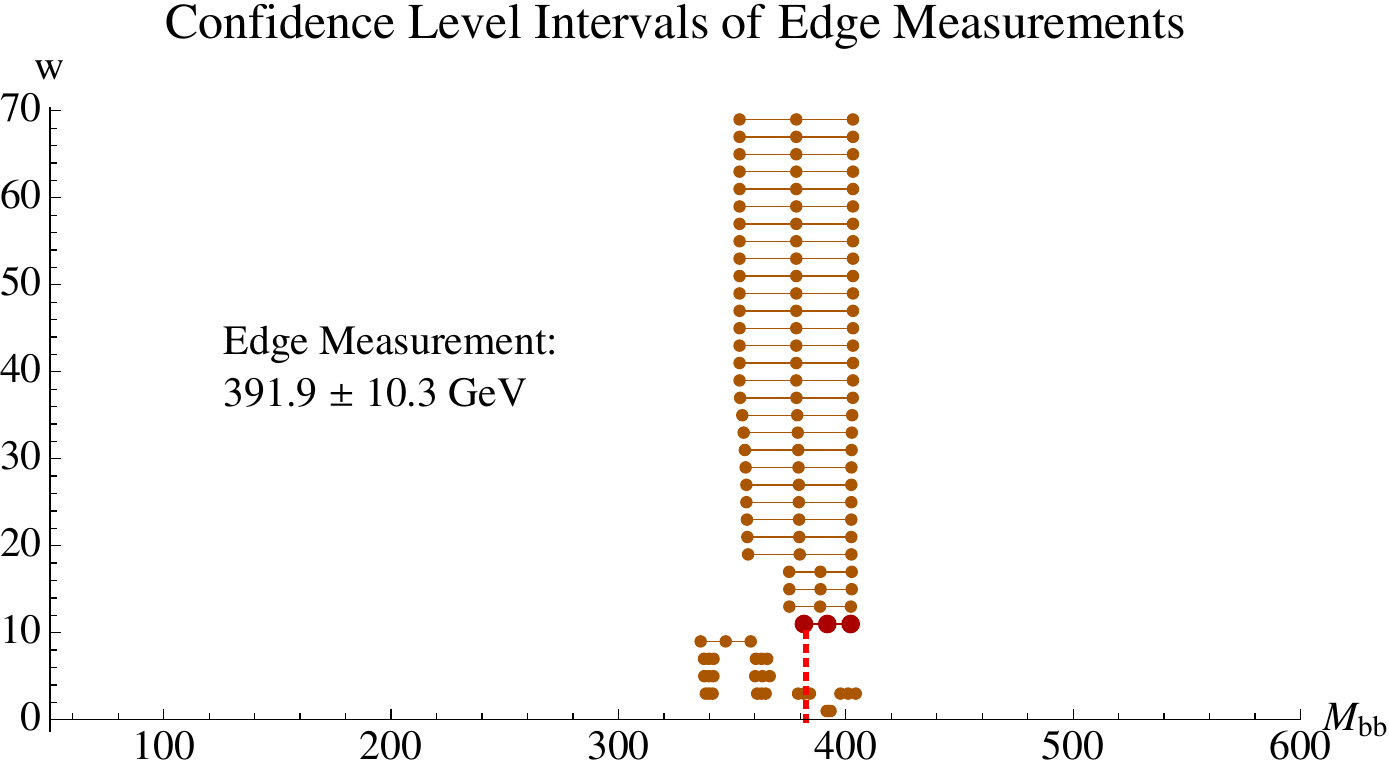}
\end{tabular}
\caption{Applying the Edge-to-Bump method to measure the kinematic edge in the cleaned-up $M_{bb}$ distribution of 18770 points from our first Monte Carlo study. Note how the physical edge reliably shows up as a growing upside-down cone in the peak range plot and is distinguished from noise at small $w$.  The dotted line indicates the expected edge position, and the peak range and confidence interval used for the final edge measurement are marked in bold red.}
\label{f.edgefinderexampleMbb}
\end{center}
\end{figure}

\begin{figure}
\begin{center}
\begin{tabular}{cc}
\includegraphics[width=\temp]{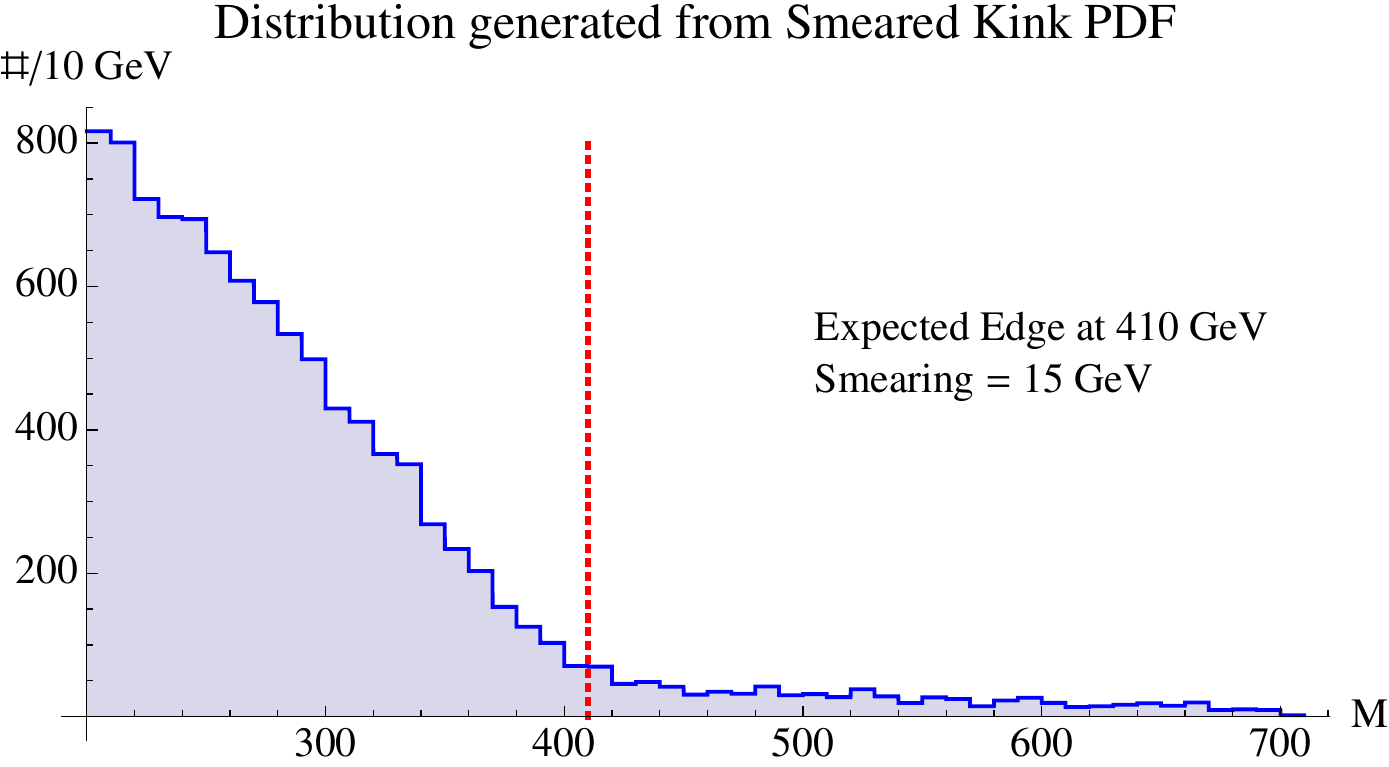}
&
\includegraphics[width=\temp]{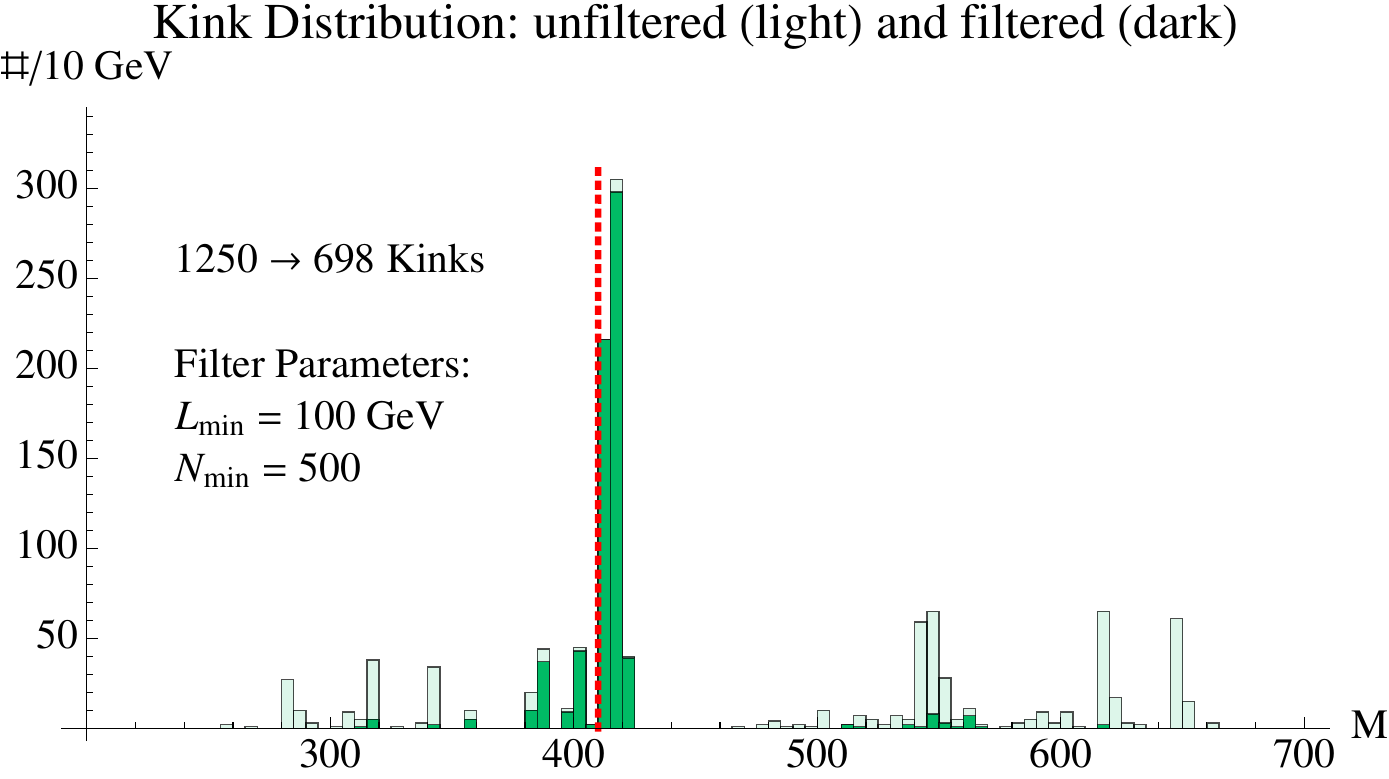}\\
\includegraphics[width=\temp]{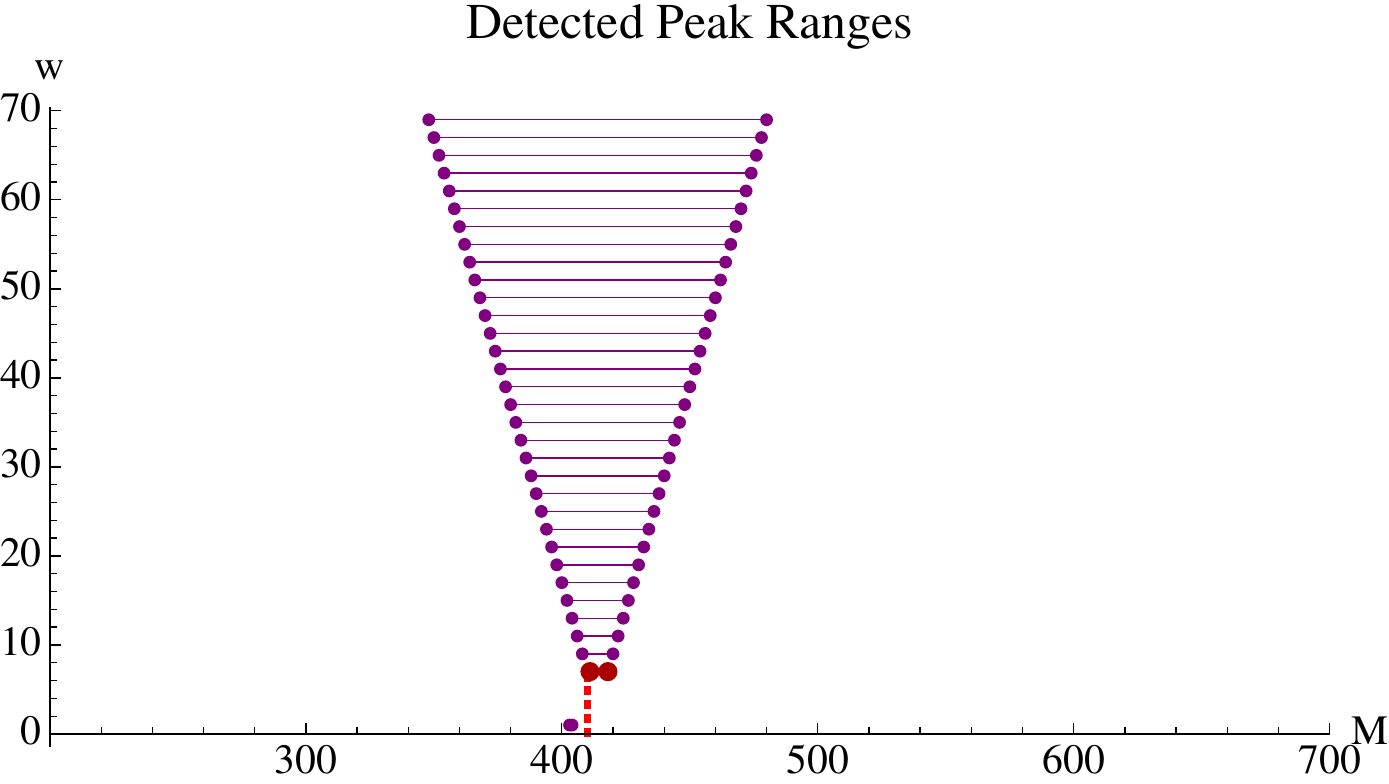}
&
\includegraphics[width=\temp]{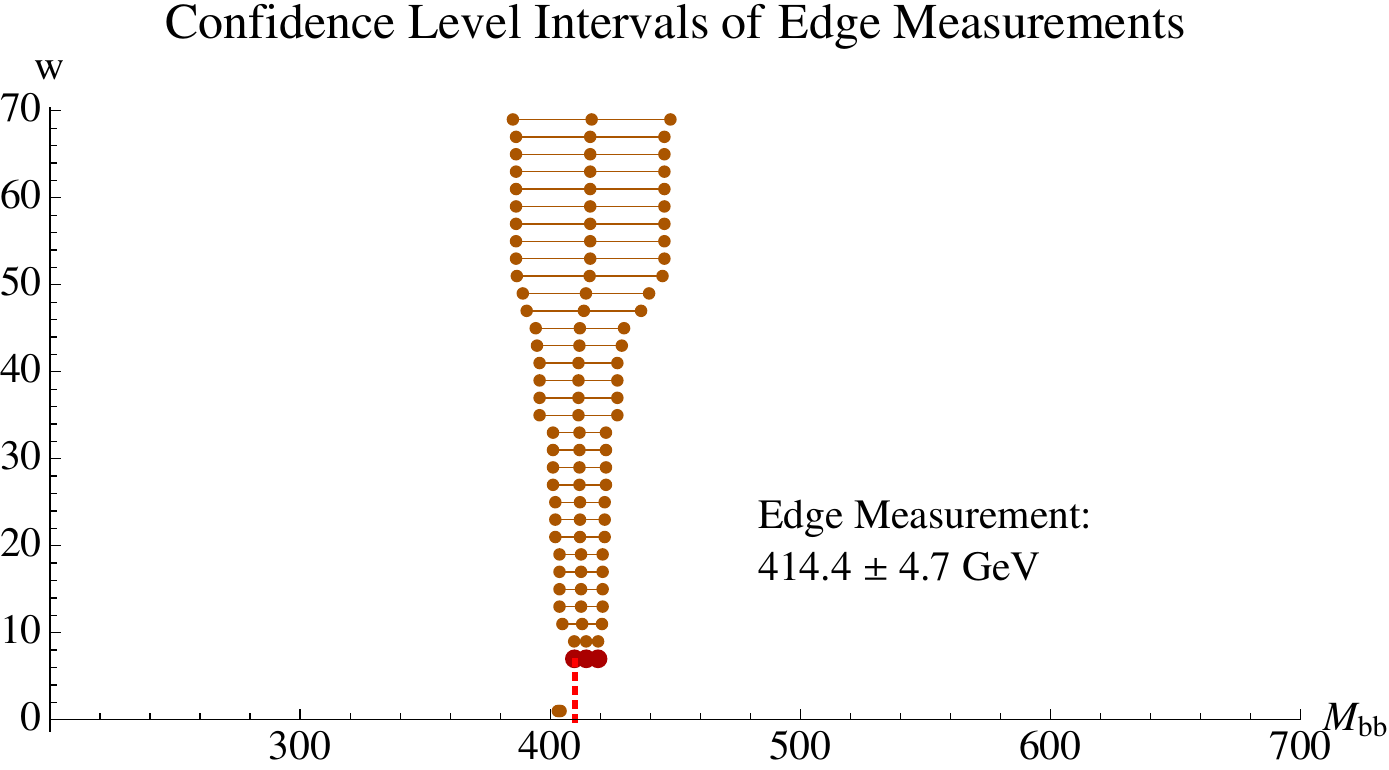}
\end{tabular}
\caption{Applying the Edge-to-Bump method to a distribution of 10000 points generated from a smeared kink PDF. The dotted line indicates the expected edge position, and the peak range and confidence interval used for the final edge measurement are marked in bold red.}
\label{f.edgefinderexamplegenerated}
\end{center}
\end{figure}

Consider a distribution for the invariant mass of two $b$-jets in the process $\tilde g \tilde g \rightarrow 2 \tilde b + 2 b \rightarrow 4 b + 2 \chi^0_1$ from our first Monte Carlo study conducted in \sref{casestudy1}. The first plot in  \fref{edgefinderexampleMbb} was obtained after making some cuts to reduce combinatorics background. Applying each of the five steps of our method produces the remaining plots of \fref{edgefinderexampleMbb} and yields an edge measurement of $M_{bb}^{max} = 391.1 \pm 10.3 \gev$, which is in good agreement with the expected value of 382.3 GeV. 

The second example in \fref{edgefinderexamplegenerated} uses data generated from a smeared kink PDF, i.e. the function shown in \fref{sampleedge}(b) convoluted with a gaussian whose variance acts as a smearing parameter. Again the measurement agrees very well with the expected edge. Note that this measurement was performed using only about 1000 kink fits, compared to 8000 for the kinematic edge of our first example. This is a general property of our method, that edges of higher quality or less smearing can be measured using fewer kink fits, and is entirely expected since a broad peak needs more data points to be reliably sampled. At any rate, performing more kink fits is merely a computational task in no way limited by the data, so as a general rule  more is better, though of course at some point increasing the number of kink fits will not increase the edge measurement precision. 

We have applied this method to many different distributions generated with the smeared kink PDF, and in all cases the edge was accurately determined. With increasing numbers of kink fits the measurement error will usually approach the smearing parameter or plateau at a somewhat smaller but similar value, which is pleasingly in line with expectations.

\subsection{\texttt{EdgeFinder} Mathematica Code}
We have implemented the Edge-to-Bump Method in Mathematica and make our code publicly available as the \texttt{EdgeFinder} package at the website: \vspace*{-1mm}
\begin{center}
\href{http://insti.physics.sunysb.edu/~curtin/edgefinder/}{
\texttt{http://insti.physics.sunysb.edu/\string~curtin/edgefinder/}
}
\end{center} \vspace*{-1mm}
\texttt{EdgeFinder} is very simple to use and can analyze any binned or unbinned distribution and find edges of both the start-point and end-point type. As mentioned above, performing the (usually thousands of) kink fits needed to analyze a typical distribution takes a few hours on a 2 GHz CPU core.

\section{Determining Decay Chain Assignment Event-by-Event}
\label{s.kinedge} \setcounter{equation}{0} \setcounter{footnote}{0}

Consider a symmetric decay chain arising from, for example, pair production of gluinos: $\tilde g \tilde g \rightarrow 4 j + 2 \chi^0_1 = 4 j + \mathrm{MET}$. On can construct the invariant masses of two jets from the same decay chain $M_{j_1 j_2}, M_{j_3 j_4}$ to measure $M_{jj}^{max}$, but for each event there are three possible ways of constructing this invariant mass pair. The two wrong-sign combinations make the measurement of the kinematic edge more difficult, and of course this combinatorial ambiguity applies to any other kinematic variable we might want to form for this decay.  There is  currently, as far as we are aware, no certain way to determine the correct decay-chain assignment of the daughter particles event-by-event (even at parton-level),  tough a number of possible approaches towards this problem exist in the literature.\vspace{-2mm}
\begin{itemize} \itemsep=0mm
\item The mixed event technique \cite{MixedEventTechnique} applied to the $M_{jj}$ invariant mass distribution creates artificial `pure' wrong-sign combinatorics background by mixing particles from different events in the construction of a kinematic variable. With the shape of the background known one can subtract it (after normalization) from the real distribution (wrong + correct combinations) to obtain a purified distribution from which the kinematic edge can be more easily measured. This works well for invariant mass endpoint measurements, but it is not clear whether this method is suitable for more complicated kinematic variables like $M_{T2}$, where the combinatorics background itself can have nontrivial structure with its own set of edges and features that can occur close to the physical edge of the correct combinations. Also note that this method does not give any even-by-event combinatorics information.

\item One method to reduce combinatorics ambiguity event-by-event is the hemisphere method (used for example in \cite{MT2analytical, MT2hemispheremethod}), which provides an approximate way to decide decay chain assignment if the parent particles are highly boosted. This basic idea was developed further in \cite{combinatoricsRY}, where a cut in the  $M_{jj}$-$p^T$ plane was used to select a purified sample of events with known decay chain assignments, with efficiencies of $\mathcal{O}(3\%)$ and purities of $\sim 90\%$ for the cases studied. (This method assumes that the kinematic edge $M_{jj}^{max}$ is known.)  Using cuts in the $M_{T2}-p^T$ plane can increase the efficiency by an $\mathcal{O}(1)$ factor \cite{combinatoricsBKMN}. Other methods using $M_{T2}$ as a selection variable can be found in \cite{combinatoricsCGP}.

\item While we focus on model-independent techniques,  a matrix-element method can be helpful in dealing with the combinatorial ambiguities if details of the underlying physics are known \cite{combinatoricsmatrixelement}. 
\end{itemize}

Note that the measurement of $M_{jj}^{max}$ itself is generally not extremely difficult. The distribution of the wrong invariant mass combinations is fairly flat, and the edge due to the correct distributions tends to stand out quite clearly. For the invariant mass distribution the mixed-event method can be used very effectively, and any number of selections or cuts can reduce the impact of the combinatorics background (as we will show in our collider studies). Our real motivation for resolving the combinatorial ambiguity event-by-event is for the application to more powerful but less robust variables like $M_{T2}$.

\begin{figure}
\begin{center}
\includegraphics[width=14cm]{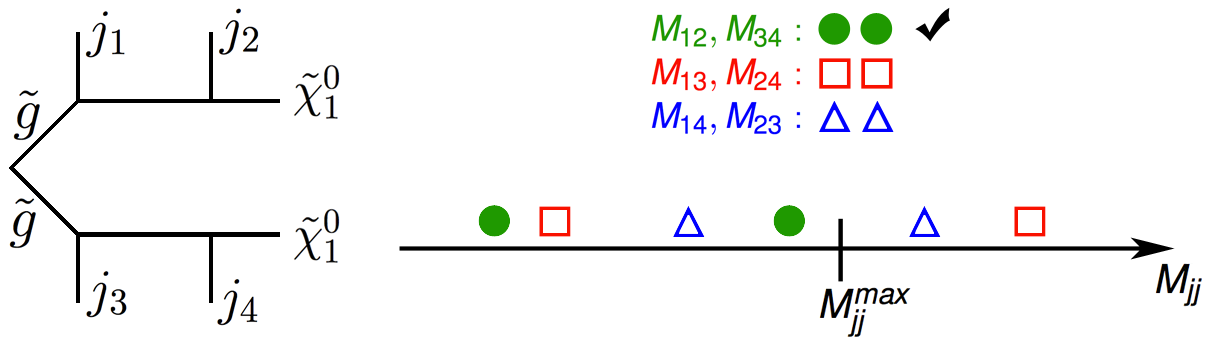}
\end{center}
\caption{This illustrates a special type of event for which we can identify the correct decay chain assignment. The correct decay chain assignment is labelled with a circle, while the incorrect ones are labelled by a square and triangle. If one or both of the invariant masses of each of the wrong pairings lie above  $M_{jj}^{max}$  and both invariant masses of the correct pairing lie below  $M_{jj}^{max}$ (which is guaranteed at parton level in the absence of measurement errors) then we can identify the correct pairing for this event. }
\label{f.MbbCBRexplanation}
\end{figure}

We propose an extremely simple method for determining the decay chain assignment of the four jets event-by-event, which we first used in \cite{susysumrule}. Like many of the above methods, we require that a measurement of the invariant mass edge $M_{jj}^{max}$ has already been made.

Consider any particular event where the $\tilde g \tilde g \rightarrow 4 j + 2 \chi^0_1$ decay takes place. Ignore showering/hadronization and detector effects, and assume a perfect measurement of $M_{jj}^{max}$. There are three possible ways to assign the four jets into two decay chains, each possibility yielding a pair of invariant masses, six in total. In \fref{MbbCBRexplanation} we labelled the three assignments (and the associated invariant mass pair) with the symbols `circle', `square' and `triangle'. Let `circle' be the correct assignment (though of course we don't know that yet). For some fraction of events we find that one or both of the invariant masses of the wrong pairings (square and triangle) lie above the measured kinematic edge $M_{jj}^{max}$, while both invariant masses of the other pairing (circle) lie below the edge. This allows us to identify the correct decay chain assignment (the only one with both invariant masses below the kinematic edge).

The appeal of this method lies both in its simplicity and its relatively high yield. In \cite{susysumrule} we examined the same MSSM parameter point we consider in our first Monte Carlo study (\sref{casestudy1}). At parton level with gaussian momentum smearing we found that about $30\%$ of events  \cite{susysumrule}  were of the type discussed above where identification of the correct decay chain assignment was possible. When including hadronization/showering and detector simulation, we found an efficiency of about $15\%$ for our two Monte Carlo studies. 

At parton level without measurement error, the purity of the obtained sub-sample with known decay chain assignments is trivially 100\%. This is affected by detector effects, showering/hadronization and the imperfect measurement of $M_{jj}^{max}$, though our two collider studies indicate that the method still works very well in the presence of those effects. A systematic study of the efficiency and purity obtainable with this method for a variety of spectra, at and beyond parton-level, is beyond the scope of this paper but should be conducted in the future.

There is an obvious elaboration on this basic idea. For a much larger fraction of events not two but only one decay-chain-assignment can be excluded because one of the corresponding invariant masses lies above $M_{jj}^{max}$. For these events we have also gained information, effectively halving the amount of combinatorics background. In our collider studies we use the information obtained for these events as well.

\section{$M_{T2}$ Measurements with Combinatorics Background}
\label{s.MT2} \setcounter{equation}{0} \setcounter{footnote}{0}

Much effort has gone into the formal and analytical definition, understanding and generalization of the $M_{T2}$-based family of kinematic variables \cite{MT2original, MT2subsystem, MT2analytical, MT2perp, MT2diffmothers, MT2diffdaughters, MT2other,MT2hemispheremethod,ZurekMT2perp}. However, their application in the presence of large combinatorial uncertainty has not been studied in detail and is not well understood. Since this represents an obvious hurdle to any realistic collider application, the development of reliable methods to conduct $M_{T2}$-based mass measurements in the presence of maximal combinatorial background is one of the key aims of this paper.

Considering a symmetric two-step decay chain like   $\tilde g \tilde g \rightarrow 2 \tilde b + 2 b \rightarrow 4 b + 2 \chi^0_1$ as an example, the problems posed by combinatorics background are qualitatively different for $M_{T2}$ variables compared to $M_{bb}$. Combinatorics background to invariant mass measurements merely serves to reduce the \emph{quality} of an edge measurement. By contrast, the shallower edges as well as the more complicated structure and larger amount of the corresponding combinatorics background make actual \emph{mismeasurement} of $M_{T2}^{max}$ the primary concern. This necessitates a very conservative $M_{T2}$ edge measurement approach with various cross-checks, and all the techniques introduced in the previous two sections come into play.

\subsection{Brief $M_{T2}$ Review}
\label{ss.MT2review}
The basic $M_{T2}$ variable \cite{MT2original} can be constructed for symmetric decay chains like the one shown in \fref{MT2chain}(a), where a pair of $X_1$-particles is produced by a hard process in a proton-(anti)proton collision, $X_0$ is an invisible decay product and $x_1$ is a visible SM daughter. One can think of $M_{T2}$ as a generalization of the simple transverse mass. Let us ignore the effect of ISR for now. The construction of the $M_{T2}$ variable can then be understood as follows:
\begin{enumerate} \itemsep=2mm
\item If we knew the transverse momenta $p_{X_0}^{T(1)}, p_{X_0}^{T(2)}$ of the invisible particles we could construct the transverse mass $M_T^{(i)}$ for each chain, which are lower bounds for $m_{X_1}$. Hence the best (highest) lower bound on $m_{X_1}$ is
\begin{equation}
\mathrm{max}[M_T^{(1)}, M_T^{(2)}] \leq m_{X_1}
\end{equation}

\item However, we only know the \emph{total} missing transverse momentum $\cancel{p}^T$. If we \emph{minimize} the above lower bound for all possible momentum splittings $\vec p_{X_0}^{T(1)} +  \vec p_{X_0}^{T(2)} = \cancel{\vec p}^T$, we will obtain the most \emph{conservative} (worst) but necessarily \emph{correct} lower bound on $m_X$:
\begin{equation}
{}^{\ \ \ \ \mbox{min}}_{\vec p_{X_0}^{T(1)} +  \vec p_{X_0}^{T(2)} = \cancel{\vec p}^T} \left\{\mathrm{max}[M_T^{(1)}, M_T^{(2)}] \right\} \leq m_{X_1}
\end{equation}

\item The calculation of the transverse mass has to make an assumption about the mass of $X_0$. Not knowing what that mass is, we have to use a testmass $\tilde M_{X_0}$. This leads to the  definition for $M_{T2}$:
\begin{equation}
\label{e.basicMT2}
M_{T2}^2(\vec p_{x_1}^{T(1)}, \vec p_{x_1}^{T(2)}, \tilde M_{X_0}) = {}^{\ \ \ \ \mbox{min}}_{\vec p_{X_0}^{T(1)} +  \vec p_{X_0}^{T(2)} = \cancel{\vec p}^T} \left\{\mathrm{max}[M_T^{(1)}(\vec p_{x_1}^{T(1)}, \vec p_{X_0}^{T(1)} , \tilde M_{X_0}), M_T^{(2)}(\vec p_{x_1}^{T(2)}, \vec p_{X_0}^{T(2)} , \tilde M_{X_0})] \right\} 
\end{equation}
The $M_{T2}$ distribution has an endpoint which satisfies  $M_{T2}^{max} = m_{X_1}$ if $\tilde M_{X_0} = M_{X_0}$. 
\end{enumerate}
In general, the endpoint depends on both the test mass $\chi$ and the total transverse momentum $p_{ISR}^T$ carried away by ISR. There are analytical expressions \cite{MT2subsystem, MT2analytical} for $M_{T2}^{max}(\tilde M_{X_0}, p_{ISR}^T)$, the simplest case being
\begin{equation}
M_{T2}^{max}(\tilde M_{X_0} = 0, p_{ISR}^T = 0) = \frac{m_{X_1}^2 - m_{X_0}^2}{m_{X_1}},
\end{equation}
so effectively a single $M_{T2}^{max}$ measurement can give us one unknown mass as a function of the other.  To calculate the $M_{T2}$ for a given event (and a given testmass $\tilde M_{X_0}$) analytical expressions exist only for $p^T_{ISR} = 0$. For realistic cases, a numerical minimization must be performed for each event (and each choice of testmass). 

\begin{figure}
\begin{center}
\begin{tabular}{cc}
\includegraphics[height=6cm]{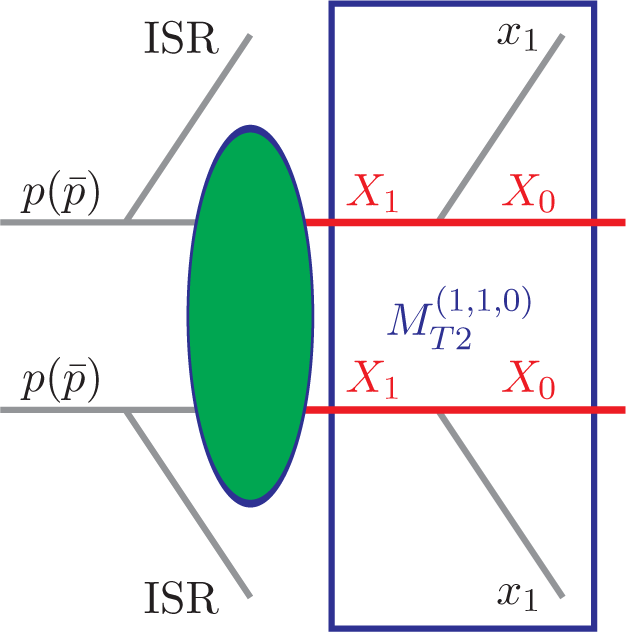}
&
\includegraphics[height=6cm]{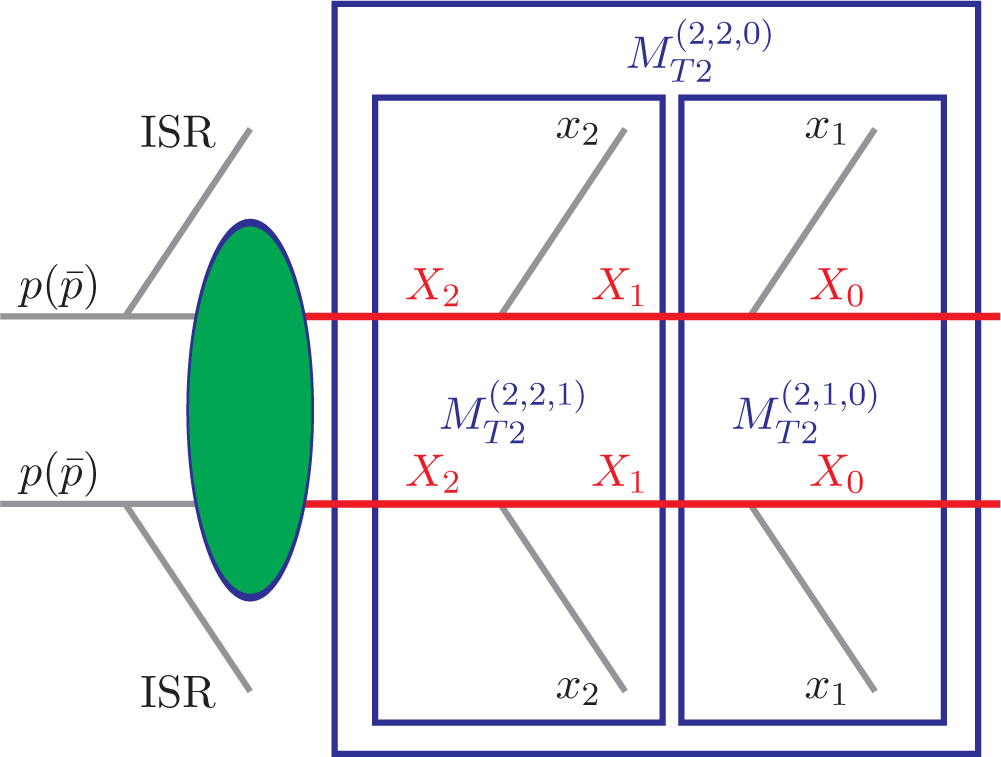}\\
(a)&(b)
\end{tabular}
\end{center}
\caption{(a) A simple 1-step symmetric decay chain. A hard process produces two particles $X_1$, which decay to invisible particles $X_0$ and SM daughters $x_1$. The blue box represents the particles entering the construction of the simple $M_{T2}$ variable. (b) A 2-step symmetric decay chain with the particles entering the respective  $M_{T2}$-subsystem  variables indicated. In each case we use $(1)$ and $(2)$ superscripts to distinguish transverse momenta from the two decay chains. Figure reproduced from \cite{MT2subsystem} with permission of the authors.}
\label{f.MT2chain}
\end{figure}

Ignoring ISR (more on that later), full mass determination is not possible for a 1-step decay chain. However, the $M_{T2}$ variable can be generalized to longer decay chains by considering only a part of the chain and forming $M_{T2}$ \emph{subsystem variables} \cite{MT2subsystem}. This proceeds in analogy to the steps described above for basic $M_{T2}$, and the three subsystem variables one can construct for a 2-step decay chain (the case we will be considering in our collider studies) are shown in \fref{MT2chain}(b). To calculate each of these for each event (and for each different choice of testmass) a numerical minimization must be performed, but analytical expressions for the endpoints of each subsystem variable as a function of the masses (with $p_{ISR}^T$ dependence) are available in \cite{MT2subsystem}. Interestingly, even in the absence of ISR the endpoint of each subsystem variable has a different functional form in terms of the underlying masses depending on whether the testmass is above or below the mass of the last $X_i$ particle in the subchain. This means that for each $M_{T2}$ subsystem variable we in effect get \emph{two} independent kinematic endpoints which each reveal unique information about the underlying particle masses, one for zero testmass and one for an extremely high testmass (e.g. take $\tilde M_{X_i} = E_b$ = beam energy). This means that a 2-step decay chain yields \emph{six} $M_{T2}$-subsystem endpoints, making complete mass determination (i.e. measurements of $M_{X_2}, M_{X_1}$ and $M_{X_0}$) possible.


Finally, let us discuss the impact of Initial State Radiation, which enters event-by-event via the momentum-conservation imposed in the sum in \eref{basicMT2}. One can imagine the dependence of an $M_{T2}$ endpoint on ISR by putting all events into very narrow $p^T_{ISR}$-bins; for each testmass $\tilde M_{X_0}$, the endpoint of the events in each bin give $M_{T2}^{max}(\tilde M_{X_0}, p^T_{ISR})$. Since ISR provides a transverse boost to the hard-scattering process, it is not surprising that increasing $p^T_{ISR}$ increases $M_{T2}^{max}$. In fact, this dependence on ISR can itself reveal additional information about the underlying masses and in principle allow for complete mass determination in a single-step decay chain (for a modern application see \cite{MT2perp, ZurekMT2perp}). However, this is unlikely to work in the presence of combinatorics background, since the effect is very subtle and the precision of the measured edges is unlikely to be high enough. We shall therefore take the opposite approach and try to remove as much ISR-dependence as possible to reduce its smearing effect on the $M_{T2}$-edges. There are two ways to do this:
\begin{itemize}
\item One can use simple ISR binning and ignore the ISR variation within each bin, accepting therefore some intrinsic smearing of the edge and the associated systematic error, as well as some reduced statistics. 
\item The variable $M_{T2\perp}$ was proposed by Konar, Kong, Matchev and Park \cite{MT2perp}. It is a one-dimensional projection of $M_{T2}$ with all transverse momenta replaced by their 1D component transverse to both the beam axis and $\vec p^T_{ISR}$. Its endpoints and their testmass dependence are identical to regular $M_{T2}^{max}$ with $p^T_{ISR} = 0$. This is an especially appealing solution since it allows us to use all the events in a sample, but $M_{T2\perp}$ edges are somewhat shallower than the corresponding $M_{T2}$ edge, making their measurement in the high-background scenarios we are considering more difficult. (This is because the 1D projection of the momenta makes it even less likely that an event with the momentum configuration to maximize $M_{T2\perp}$ occurs.)
\end{itemize}
Since these two methods have complementary advantages and drawbacks it is best to simply use both (the only cost is CPU time) and see which one works best for each variable.

\subsection{The Combinatorics Problem for $M_{T2}$}
Fundamentally, there are two types of combinatorics problems with kinematic variables like $M_{T2}$. Firstly, one must obviously distinguish between ISR and hard process jets. A number of techniques have been proposed to deal with this issue (for example \cite{ISRCOMBnojiri, ISRtagging,ISRCOMBalwall,ISRCOMBgripaios}). The second problem arises when some or all of the hard process final states are indistinguishable. We will focus most of our discussion on this latter difficulty. 

For 2-step or longer decay chains, the $M_{T2}$ subsystem variables are potentially very powerful tools for conducting mass measurements. However, compared to kinematic edges they are much more affected by combinatorial ambiguity. This is due to the shallower nature of $M_{T2}$ edges (which makes them generally more vulnerable to smearing), but also the sheer amount of combinatorial background as well as the background's intrinsic structure, which can create fake edges in the distribution that are very difficult to filter out reliably.

\begin{figure}
\begin{center}
\includegraphics[height=9cm]{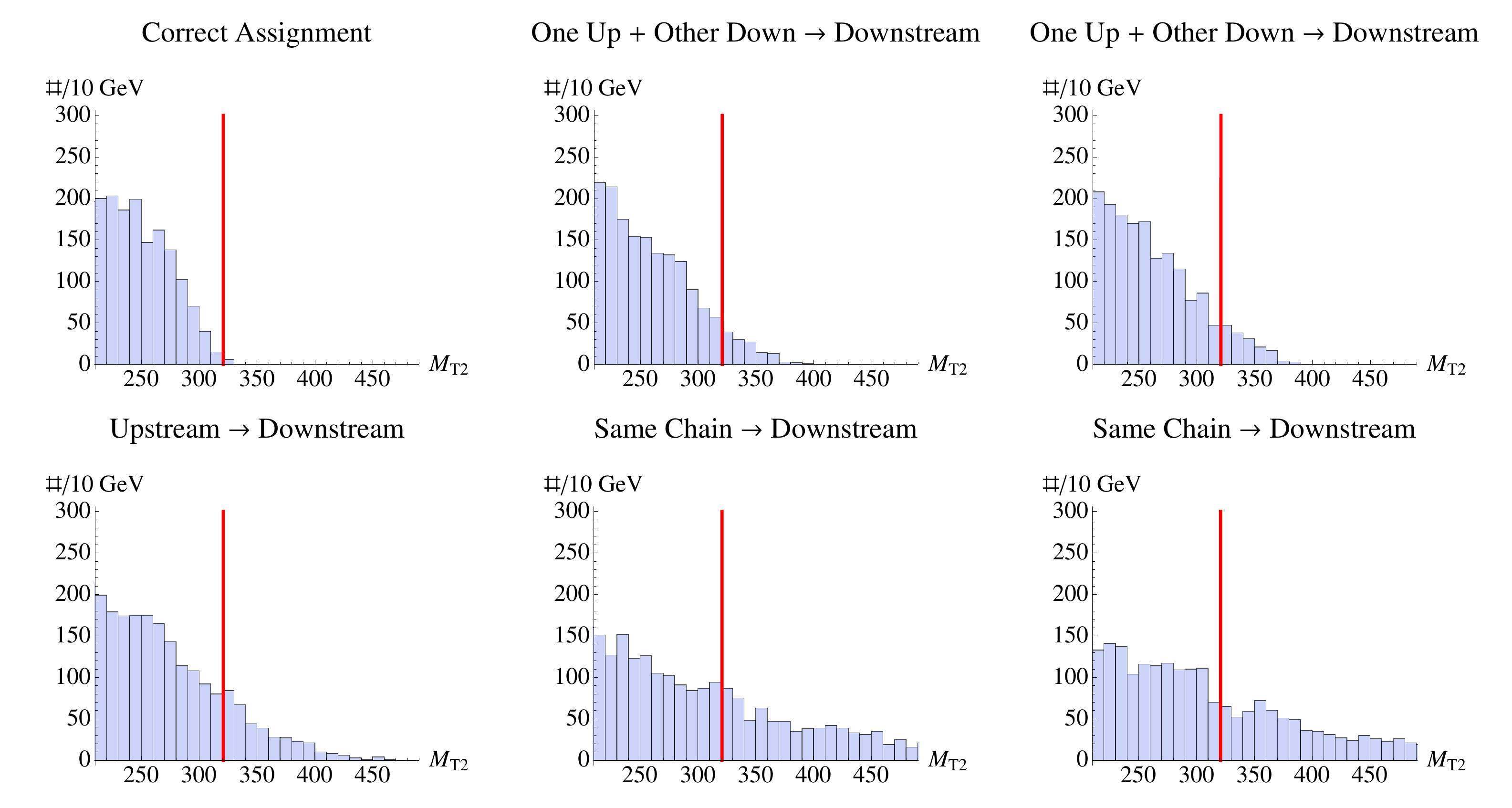}
\caption{The $M_{T2}^{210}(0)$ distributions for the six possible ways to assign two of the four $b$-jets in the process $pp \rightarrow \tilde g \tilde g \rightarrow 2 \tilde b + 2 b \rightarrow 4 b + 2 \chi^0_1$ as downstream. The testmass is zero and the MSSM parameters for the parton-level simulation were the same as for our first Monte Carlo study in \sref{casestudy1}. The red line indicates the expected position of $M_{T2}^{210}(0)^{max}$.}
\label{f.MT2allcomb}
\end{center}
\end{figure}

We can illustrate the problem by considering the $M_{T2}^{210}$ subsystem variable for the process $pp \rightarrow \tilde g \tilde g \rightarrow 2 \tilde b + 2 b \rightarrow 4 b + 2 \chi^0_1$, which is the subject of our two collider studies in Sections \ref{s.casestudy1} and \ref{s.casestudy2}. As illustrated in \fref{MT2chain}, this subsystem variable is constructed using the transverse momenta of the two downstream SM daughters. Using $b$-tags to distinguish the hard process jets from ISR this leaves six possibilities for assigning two $b$'s as downstream. The $M_{T2}^{210}(0)$ distributions for all six possible assignments are shown in \fref{MT2allcomb}, where we used the same MSSM parameter point as the first Monte Carlo study but used parton-level events to emphasize the problems arising from pure combinatorics. Apart from the fact that there is 5 times as much combinatorics background as signal, the wrong-sign combinations also feature their own edges/endpoints that can be very close to the real one! This will pollute the total sample and not only make accurate determination of the real edge extremely difficult but also introduce the danger of mistakenly measuring one of these fake background edges, giving not only an $M_{T2}^{max}$ measurement of poor quality but one that is just plain \emph{wrong}, which is much worse. As we will find, guarding against these fake edges is the main challenge arising in these mass measurements.




\subsection{Reducing $M_{T2}$ Combinatorics Background}
\label{ss.MT2CBR}
We will use two methods of reducing combinatorics background in $M_{T2}$-subsystem distribution for the process $pp \rightarrow \tilde g \tilde g \rightarrow 2 \tilde b + 2 b \rightarrow 4 b + 2 \chi^0_1$.

\subsubsection*{KE (Kinematic-Edge) Method}
We can use the method outlined in \sref{kinedge} to determine the decay chain assignments of the four $b$'s for a subset of the events. This obviously reduces the combinatorial ambiguity for the construction of $M_{T2}$-subsystem variables as well, though milage varies depending on the variable. 

$M_{T2}^{220}$ is a special case, since for its construction we need to assign the four $b$'s to decay chains but needn't specify their ordering. It therefore has the same combinatorial structure as the invariant mass kinematic edge (three possible ways of constructing $M_{T2}^{220}$ for each event) and this method is expected to be quite effective.  To use the maximal amount of information and make use of the identical combinatorial structure we use a weighing procedure. For each event there are three decay chain assignments, and as explained in \sref{kinedge}, one can exclude a  decay chain assignment if one or both of the corresponding invariant masses lie above the measured $M_{bb}^{max}$ edge. If all three decay chain assignments are excluded this way we discard the event, since the measured momenta are unlikely to be trustworthy. For all other events we can discard 0, 1 or 2 decay chain assignments (and hence 0, 1 or 2 of the 3 possibilities for $M_{T2}^{220}$). Each event is given a total weight of 1, which is evenly split according to the remaining possible $M_{T2}^{220}$'s. This can work very well, as shown in the example of \fref{CBRexample} (top) where the physical edge seems to be unambiguously revealed. 

$M_{T2}^{210}$ and $M_{T2}^{221}$ require the separation of the four $b$'s into an upstream and a downstream pair, giving a total of 6 possibilities. We will only consider the subset of events where the decay chain assignment can be uniquely determined, which reduces the number of possibilities for constructing these variables to 4. This makes the physical edge visible some of the time.

\subsubsection*{DL (Drop-Largest) Method}
This method is much simpler. Since $M_{T2}$ by its very nature represents a \emph{lower} bound on some mass, if there are several possible ways of constructing an $M_{T2}$-subsystem variable for a given event, the largest possibilities are least likely to be correct. For $M_{T2}^{220}$, we merely discard the largest of the three possibilities for each event, while for $M_{T2}^{210}$ and $M_{T2}^{221}$ we discard the largest two of the six possibilities for each event. This trivial method can be surprisingly effective, as \fref{CBRexample} (bottom) demonstrates.

\begin{figure}
\begin{center}
\hspace*{-9mm}
\begin{tabular}{m{17mm}m{7.5cm} m{7.5cm}}
$M_{T2}^{220}(E_b)$: & 
\includegraphics[width=7.5cm]{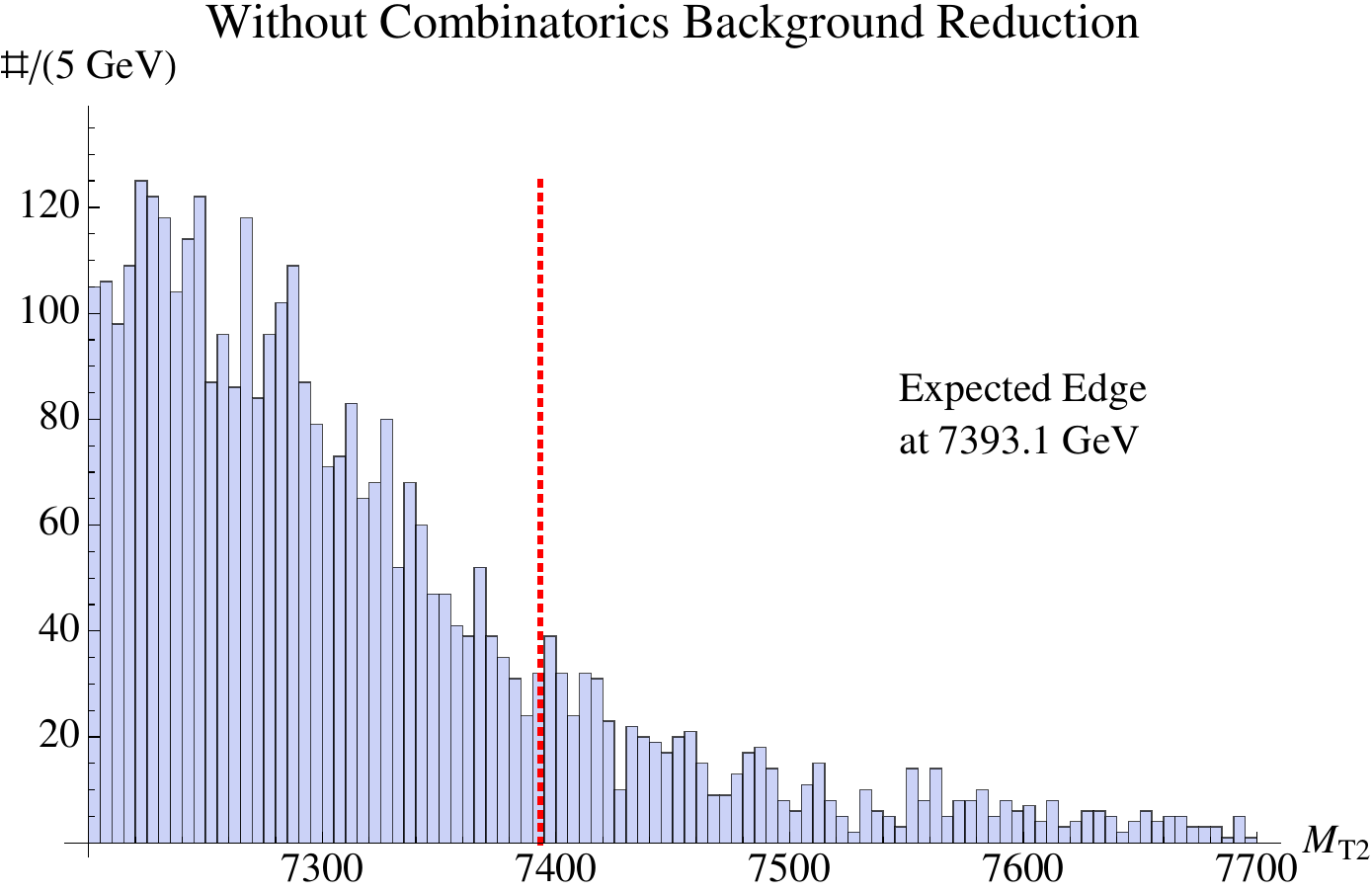}
&
\includegraphics[width=7.5cm]{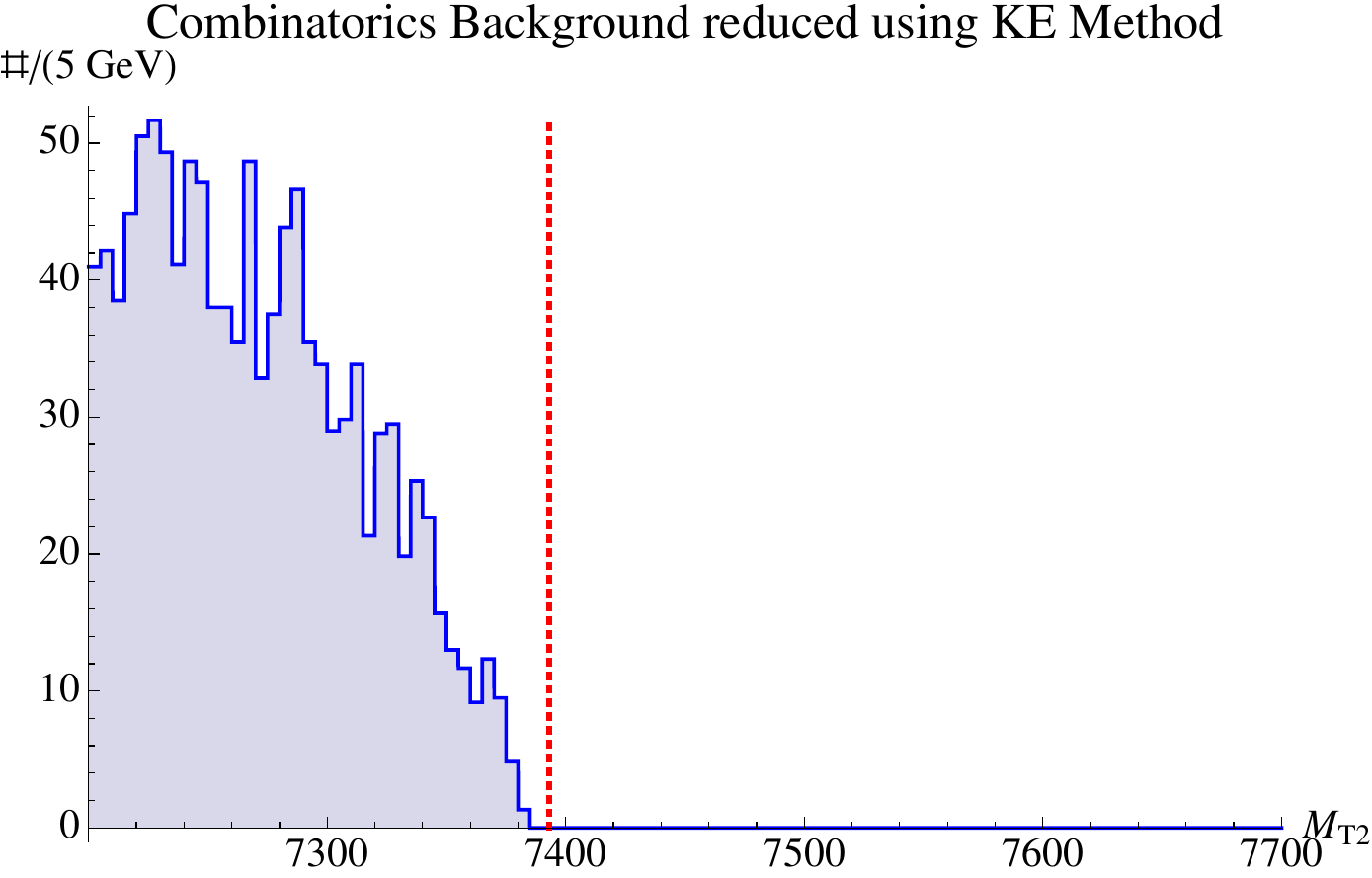}
\\ \\
$M_{T2}^{221}(0)$: & \includegraphics[width=7.5cm]{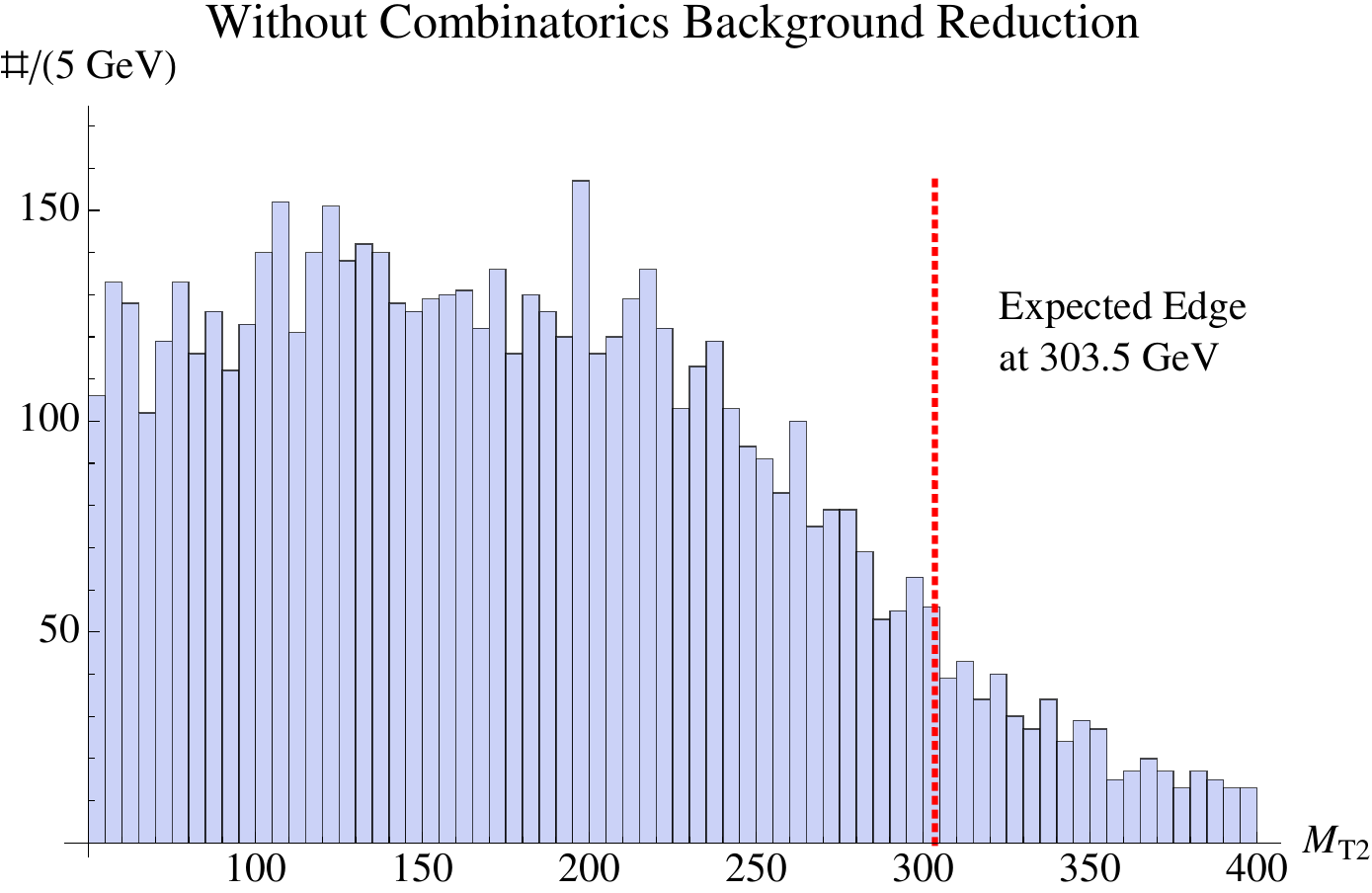}
&
\includegraphics[width=7.5 cm]{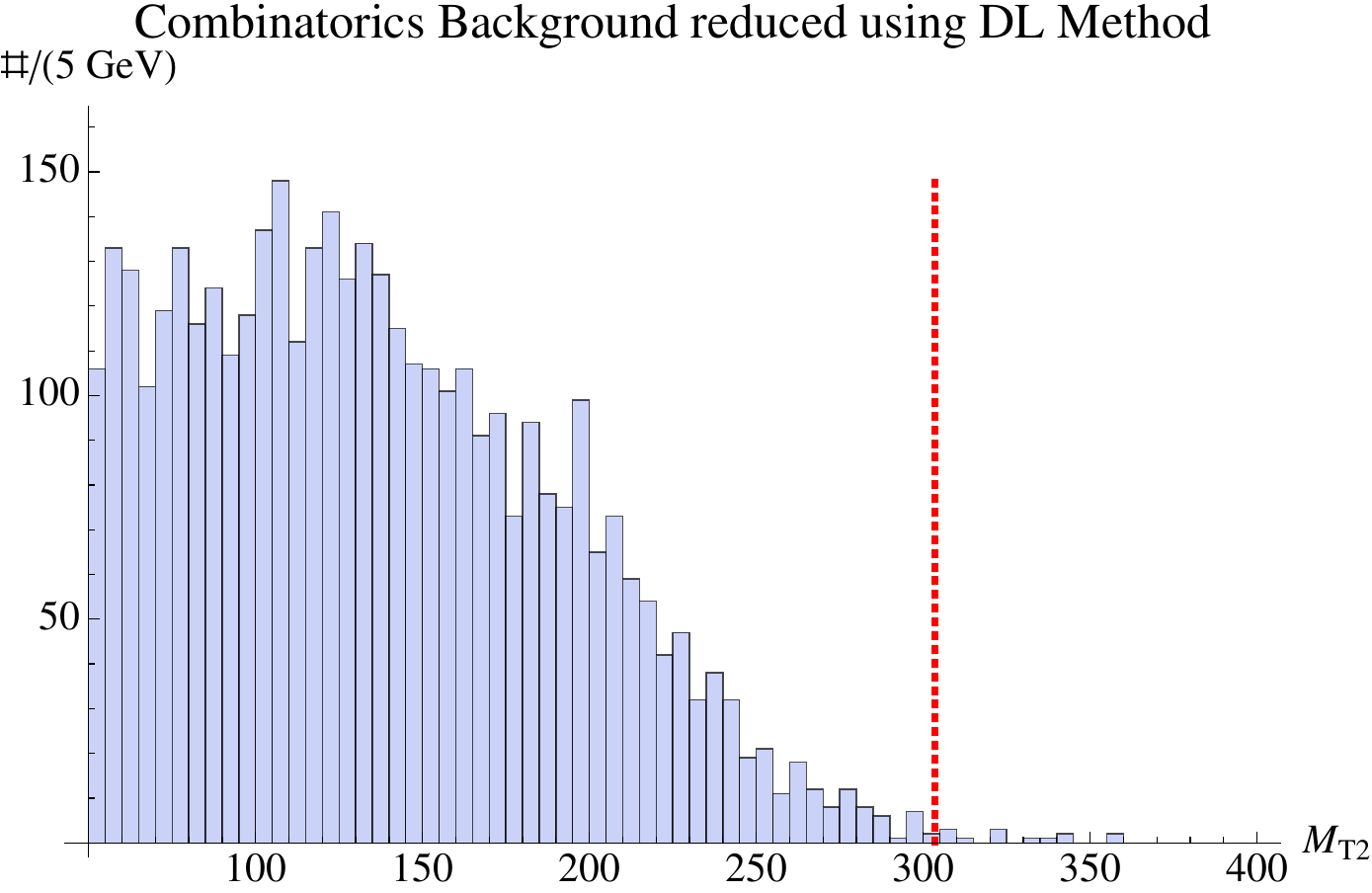}
\end{tabular}
\end{center}
\caption{Examples of reduced combinatorics background: KE method applied to the $M_{T2}^{220}(E_b)$ distribution (top), and DL method applied to the $M_{T2}^{221}(0)$ distribution (bottom) from our first Monte Carlo study in \sref{casestudy1}. ($p_T^{ISR}$ cutoff used to control ISR smearing. All units in GeV. Includes detector effects and hadronization/showering.)}
\label{f.CBRexample}
\end{figure}

\subsubsection*{Performance}

\begin{table}
\hspace*{-7mm}
\footnotesize
\begin{tabular}{|l|m{14mm}|m{56mm}|m{73mm}|}
\hline variable & real edge & KE method & DL method\\ 
 
\hline \hline 
$M_{T2\perp}^{221}(0)$ & 303.5 & overestimates edge by $\sim 50 \gev$ & two fake edges which underestimate real edge by $\sim 50 \gev$ and $\sim 130 \gev$ \\
$M_{T2}^{221}(0)$ & & extremely smeared but consistent with real edge& works well\\

 \hline 
 $M_{T2\perp}^{221}(E_b)$ & 7153.4 & overestimates edge by $\sim 50 \gev$ & smeared edge but works well\\
$M_{T2}^{221}(E_b)$ &  & overestimates edge by $\sim 50 \gev$& works \emph{extremely} well\\

 \hline 
 $M_{T2\perp}^{210}(0)$ & 320.9 & works well & smeared edge but works well\\
$M_{T2}^{210}(0)$ &  & works \emph{extremely} well & works well\\
 
 \hline 
 $M_{T2\perp}^{210}(E_b)$ & 7239.8 & runs out of points: underestimates edge by $\sim  70 \gev$ & runs out of points: underestimates edge by $\sim 110 \gev$\\
$M_{T2}^{210}(E_b)$ &  & runs out of points: underestimates edge by $\sim 40 \gev$ & runs out of points: underestimates edge by $\sim 90 \gev$\\

 \hline 
 $M_{T2\perp}^{220}(0)$ & 506.7 & extremely smeared, underestimates edge by $\sim 100 \gev$ & extremely smeared, underestimated edge by $\sim 80 \gev$\\
$M_{T2}^{220}(0)$ &  & works well & works well\\
 
 \hline 
 $M_{T2\perp}^{220}(E_b)$ & 7393.1 & works \emph{extremely} well& very smeared, multiple edges, overestimate real edge by $\sim 100$ and $\sim 200 \gev$\\
$M_{T2}^{220}(E_b)$ &  & works \emph{extremely} well & extremely smeared, overestimates edge by $\sim 100 \gev$\\

 \hline  $M_{T2\perp\mathrm{all}}^{210}(0)$ & 312.8 & works well & underestimates edge by $\sim 50 \gev$\\
 $M_{T2\perp\mathrm{all}}^{210}(E_b)$ & 7158.2 & works well & underestimates edge by $\sim 50 \gev$\\

 \hline 
\end{tabular}
\caption{Performance of the KE and DL method of reducing combinatorics background when applied to the $M_{T2}$-subsystem variables in the first Monte Carlo study. A method was evaluated to work well when it revealed the \emph{correct} edge instead of an artifact. Note that the edges of $M_{T2\perp}$ and $M_{T2}$ with ISR binning reveal the same information, as do $M_{T2\perp\mathrm{all}}^{210}$ with different test masses.  (All units in GeV. $E_b = 7000 \gev$.)
}
\label{t.CBRoverviewBP5}
\end{table}


\tref{CBRoverviewBP5} gives a rough overview of the KL and DL method's effectiveness in the case of our first Monte Carlo study. This demonstrates that the situation is quite complicated: for some $M_{T2}$-subsystem variables, both methods work quite well; sometimes one or both methods fail. This failure can manifest itself by measuring a fake edge (significantly over- or under-estimated position) or multiple edges, one or none of which may be correct. In any case, no one method can be trusted all of the time for all variables, and from looking at a cleaned up distribution it is hard or impossible to tell whether the method was successful or not. 

While it is possible that for each $M_{T2}$-subsystem variable there exists a different specific method of reducing combinatorics background that is reliable regardless of the mass spectrum, identifying such methods would require a very large-scale study that is far beyond the scope of this paper. Moreover, even such highly optimized methods would risk failing simply by running out of statistics to the left of an edge and hence measuring a fake endpoint that underestimates the true edge position in an undetectable way. At any rate, comparing the KL and DL effectiveness for  $M_{T2\perp}$ and ISR-binned $M_{T2}$, which contain the same information and should be amenable to the same methods of reducing combinatorics background, reveals no obvious pattern of which of our two methods work for which variable.

\subsection{Performing Reliable $M_{T2}$ Edge Measurements}
\label{ss.MT2edgetobump}

While it seems that for some subsystem variables an $M_{T2}^{max}$ measurement is possible using our (or any other) methods of reducing combinatorics background, \emph{it is clear that the most important challenge is  identifying the cases where these methods fail}, so that we may either ignore the corresponding $M_{T2}$ variable \emph{or} (equivalently) get an edge measurement with large error bars that reflect the unreliable nature of the measurement. Since we only need to measure two independent $M_{T2}$ endpoints (in addition to the $M_{bb}$ kinematic edge, which is easy to measure) to determine all the masses in a two-step decay chain, we can afford to impose very stringent quality requirements on an edge measurement.

\subsubsection*{Golden Rule for $M_{T2}$ Measurements}
The only potentially reliable approach to measuring $M_{T2}$-subsystem edges is the simultaneous use of \emph{at least} two different methods of reducing combinatorics background. For each distribution the two methods act as cross-checks on each other, and an edge measurement is only accepted if \emph{both measurements yield the same clear edge.} Our collider studies demonstrate the validity of this approach.

While we study the specific 2-step decay chain $pp \rightarrow \tilde g \tilde g \rightarrow 2 \tilde b + 2 b \rightarrow 4 b + 2 \chi^0_1$, the above principle should apply to any multi-step decay chain with combinatorics background. It is also not unique to our KE and DL methods, and they can be substituted for two or more different procedures for cleaning up $M_{T2}$-subsystem distributions (though of course results may vary depending on the methods' performance). \emph{The important principle is that no sole method of reducing $M_{T2}$ combinatorics background is trustworthy by itself. }

\subsubsection*{Implementing the Golden Rule: Extending Edge-to-Bump to $M_{T2}$ Edges}
How do we implement this general idea? Consider the distribution of a particular $M_{T2}$-subsystem variable, e.g. first row of \fref{MT2example}. Applying our two methods of reducing combinatorics background yields two `cleaned up' distributions, call them the KE- and DL-distributions (second row of \fref{MT2example}). We perform steps 1 - 5 of the Edge-To-Bump method, obtaining an edge distribution (third row), detected peak ranges (fourth row) and an edge measurement vs peakwidth plot (fifth row) for each of the two cleaned up distributions. 

The next step is to somehow combine the two sets of edge measurements. The error bars of the combined measurement should reflect (a) the quality of the individual edges in the KE and DL distributions (i.e. the amount of smearing); (b) the degree of (dis)agreement between the edges of the two distributions; (c) the overall quality of the data, in the sense that we should put more faith into a measurement where both distributions only have one clear edge each than if both distributions have many edges (where the chance of random coincidence between two edge measurement is higher). 

We need to satisfy the above criteria while also minimizing error bars where reliably possible and extracting as much information as we can, even from very unclear $M_{T2}$ distributions. Therefore, we define four different procedures for extracting a combined edge measurement, depending on the quality of the DL and KL distributions for each $M_{T2}$-subsystem variable.
\begin{itemize}
\item[\texttt{Case A}] The best case scenario is if the individual KE and DL contributions only have \emph{one clear edge each}. (Recall how a clear edge will show up as a characteristic `broadening river' shape in the measurement plot, see Figures \ref{f.peakintervals}(c), \ref{f.edgefinderexampleMbb}, \ref{f.edgefinderexamplegenerated}). 

In that case we simply merge the two plots of edge measurement vs peakwidth $w$. The procedure for this is very simple: imagine overlaying the two plots, deleting any 1-$\sigma$ confidence level interval (1$\sigma$CLI) that does not overlap with an interval in the other plot, and then merging the ones that do overlap (this merging reflects the increased uncertainty due to any disagreement between the overlapping edges). The result is a single \emph{overlapping edge measurement plot} which we interpret as if it came from just one distribution, as explained in \ssref{edgefitterprocedure}. Note that this could give a null result (if the edges do not overlap), in which event we move on to Case B. For an example from the first Monte Carlo study see \fref{MT2example} (left). 

\item[\texttt{Case B}] This applies if there are more than one clear edges in either the KL or DL distributions, or if there is one edge each but the merged measurement plot does not show a clear edge candidate.  In this case we do \emph{not} use the overlapping edge measurement plot. Instead, we determine all the individual edges in the DL and KE distributions independently. This will yield a set of 1$\sigma$CLI's. The 1$\sigma$CLI of the final $M_{T2}$ edge measurement is taken to be the smallest interval that contains all these intervals. See \fref{MT2example} (right) for an example from the first Monte Carlo study.

At first glance this procedure might appear overly conservative. After all, if there is one clear edge that shows up in both distributions as well as other edges that do not,  one might think that the two overlapping edges are likely to be physical. Unfortunately, the same would be the case if the KE and DL method \emph{both} failed to remove one (or more) combinatorial artifact. Furthermore, if both distributions have many edges the chance of random agreement between two of them is high. (Of course the above arguments could also apply to Case A, but it is less likely and has not occurred in our two collier study.)

\item[\texttt{Case C}] If there are no clear edges in either the KL or DL distributions we can still learn something about the general \emph{scale} of $M_{T2}^{max}$ by taking the corresponding 1$\sigma$CLI to be the smallest interval that contains \emph{all} the edge measurements in both distributions.

\item[\texttt{Case D}] No Measurement. This only applies if all the edges found for one or both distribution are very obvious Filter Artifacts or red herrings very close to the origin of the distribution. This indicates a complete failure of our combinatorics background reduction methods, and the measurement should not be kept. This only occurs once in our analyses. 

\end{itemize}
Note that we ignore filter artifact edges in all of the above, as explained in \ssref{edgefitterprocedure}. For illustrations of this process for \emph{all} the $M_{T2}$ edge measurements in both collider studies see the Appendix. 

\section{First Monte Carlo Study}
\label{s.casestudy1} \setcounter{equation}{0} \setcounter{footnote}{0}

We now show how all these techniques can be put together to determine all the masses in the decay chain $pp \rightarrow \tilde g \tilde g \rightarrow 2 \tilde b + 2 b \rightarrow 4 b + 2 \chi^0_1$ at the LHC with center-of-mass energy of $14 \tev$. This Monte Carlo study was used as a benchmark to develop the analysis tools introduced in this paper and included showering/hadronization and detector effects. In \sref{casestudy2} we discuss a blind study that verifies our methods.

\subsection{MSSM Parameters}

In \cite{susysumrule} we measured all the masses in $pp \rightarrow \tilde g \tilde g \rightarrow 2 \tilde b + 2 b \rightarrow 4 b + 2 \chi^0_1$ using some very prototypical versions of the ideas presented in this paper, and claimed the measurement could be performed in a more realistic setting as well. To verify that claim and develop our measurement techniques further,  we decided to use the same MSSM benchmark point for our first Monte Carlo study. It is defined by the following weak-scale inputs (all masses in GeV unless otherwise noted):
\begin{center}
\begin{tabular}{|l|l|l|l|l|l|l|l|l|l|}
\hline
$\tan \beta$ & $M_1$ & $M_2$ & $M_3$& $\mu$ & $M_A$ &
$M_{Q3L}$ & $M_{tR}$ & $M_{bR}$ &  $A_t$\\
\hline
 10 & 100 & 450 & 450 & 400 & 600 & 310.6 & 778.1 & 1000 &  392.6\\
\hline
\end{tabular}
\end{center}
with all other $A$-terms zero and all other sfermion soft masses set at $1 \tev$. The relevant spectrum (calculated with {\texttt SuSpect} \cite{SuSpect}) is the following:
\begin{center}
\begin{tabular}{|l|l|l|l|l|l|l|l|}
\hline
$m_{\tilde t_1}$ & $m_{\tilde t_2}$ & $\sin \theta_{\tilde t}$ & $m_{\tilde b_1}$& $m_{\tilde b_2}$ & $\sin \theta_{\tilde b}$ &
$m_{\tilde g}$ & $m_{\tilde \chi^0_1}$\\
\hline
 371 & 800 & -0.095 & 341 & 1000 & -0.011& 525 & 98
 \\
\hline
\end{tabular}
\end{center}
This benchmark point was originally chosen for its absence of any SUSY background to our process of interest. Its spectrum has already been excluded by LHC searches \cite{LHCsearches}, but since we end up performing our analysis with pure signal and the main challenges are combinatorics, it still serves well to develop and demonstrate our statistical analysis techniques.

\subsection{Generating Event Sample for the  Analysis}

\subsubsection{Signal}

MadGraph 5 \cite{MGME} was used to simulate  the process $pp \rightarrow \tilde g \tilde g\rightarrow 2 \tilde b + 2 b \rightarrow 4 b + 2 \chi^0_1$ at lowest order, with Pythia 6.4 \cite{pythia} for showering/hadronization and PGS with the standard CMS card for detector effects.  We use the CTEQ6l1~\cite{cteq} parton distribution functions throughout, with the { MGME} default ($p_T$-dependent) factorization/renormalization scale choice. The gluino pair production cross section for our benchmark point is 11.6 pb at a center-of-mass energy of 14 TeV, and we ran the study with 50 fb$^{-1}$ of integrated luminosity, giving a total of $5.8 \times 10^{5}$ signal events.

\subsubsection{Selection Rules}
To keep an event for our analysis we require four $b$-tags and $\mathrm{MET} > 150 \gev$, as well as some standard jet-acceptance cuts: $|\eta| < 2.5, p_T > 20 \gev$. The four $b$-tags have an efficiency of 4.0\% ,  with the addition kinematic cuts bringing an additional 40\% penalty, giving a total signal efficiency of 1.6\%. The number of surviving signal events with four identified $b$-jets + MET + ISR jets is 9385. Note that actual $b$-tag rate at LHC14 is likely to be significantly higher than what PGS assumed ($\sim 45\%$ per tag), so our signal efficiencies are quite pessimistic.

\subsubsection{Backgrounds}
The main Standard Model backgrounds for our signal process are  $Z + 4j$ BG (simulated using ALPGEN \cite{ALPGEN}); $\mathrm{Diboson} + 4j + \mathrm{escaped\ lepton}$ (smaller than $Z + 4j$ \cite{CMSTDR}); fully leptonic $t \bar t$ with mistagged $\tau$'s or escaped light leptons (simulated in MGME); and QCD background. The QCD background is effectively eliminated by the four $b$-tags and MET cut\footnote{We thank Julia Thom-Levy (CMS) for clarifying this for us.}, while the remaining backgrounds end up contributing only about $\sim 10\%$ as many events as the signal after cuts. In light of the two-orders-of-magnitude-larger combinatorics background within the signal itself, and since the SM backgrounds are highly unlikely to be similarly malicious in polluting our $M_{T2}$-distributions with fake edges and artifacts, \emph{ we ignore all SM background completely and perform the following analyses with the 9385 pure signal events.}

\subsection{Kinematic Variables}
There are a total of 9 kinematic edges we can attempt to measure for this decay chain, which all depend on the underlying masses in a different way: \vspace{-2mm}
\begin{itemize} \itemsep=0mm
\item The endpoint of $M_{bb}$, the invariant mass of two $b$'s from the same decay chain.
\item We can construct three $M_{T2}$-subsystem variables \cite{MT2subsystem} as shown in \fref{MT2chain}(b). Setting the testmass to zero and the beam energy ($E_b = 7000 \gev$) gives six independent kinematic edges. As explained in \ssref{MT2review}, we use two methods to eliminate the $p^T_{ISR}$ dependence:  constructing $M_{T2\perp}$-subsystem variables \cite{MT2perp} and using ISR binning. We attempt to measure the edge for each subsystem variable using both methods, keeping the measurement with the smallest error bar. 
\item We can construct $M_{T2\perp\mathrm{all}}^{210}$, which we define to be the 1-dimensional projection of $M_{T2}^{210}$ following \cite{MT2perp}, but treating the two upstream momenta as `ISR' as well. Since the endpoint dependence on the masses is that of classical $M_{T2}$, measuring the endpoint for different test masses does not, in principle, provide additional information. However, since the effect of testmass on combinatorics background is not understood, we choose to measure this endpoint with both a testmass of zero and the beam energy, keeping the measurement that gives the smallest uncertainty on the final mass determination. 
\end{itemize}
In principle, only three edge measurement are  required to determine the gluino, sbottom and neutralino mass. However, since some of the measurements will have large error bars we want to measure as many as we can.

Our method of ISR binning is very simple. We include $p^T_{ISR} \neq 0$ effects in calculating $M_{T2}$ event-by-event, but for each subsystem variable we choose some $p^T_{max}$, and we only include a events with $p^T_{ISR} < p^T_{max}$ in the distribution for that variable. We then measure the edge and interpret it as a $p^T_{ISR} = 0$ edge measurement. The non-zero $p^T_{ISR}$ of the events will smear the edge and cause some positive systematic error, but that smearing will be included in the error bars when using the Edge-to-Bump method to measure the edge position. Therefore, the only complication is how to choose $p^T_{max}$ low enough to minimize smearing but high enough to give sufficient statistics. Our choices were motivated by the different $p^T_{ISR}$-dependencies of the $M_{T2}$-subsystem variable endpoints, and are as follows (all in GeV):
\begin{center}
\begin{tabular}{c|cccccc}
& $M_{T2}^{221}(0)$ & $M_{T2}^{221}(E_b)$ & $M_{T2}^{210}(0)$ & $M_{T2}^{210}(E_b)$ & $M_{T2}^{220}(0)$ & $M_{T2}^{220}(E_b)$\\ \hline
$p^T_{max}$ & 30 & 45 & 100 & 50 & 50 & 40
\end{tabular}

\end{center}
This guarantees an edge smearing of less than 10 GeV for the large majority of the allowed $(m_{\tilde g}, m_{\tilde b_1}, m_{\chi^0_1})$ mass space, including our particular spectrum. (If we were unlucky enough to have a spectrum for which the edge smearing due to ISR is extremely large, we might have to attempt more sophisticated binning methods.)



\subsection{Measuring the Invariant Mass Edge}
\label{ss.mbbmeasurement}
For each event there are 3 possible pairs of $M_{bb}$, six in total. Two of those pairs are combinatorics background. Plotting the total distribution of $M_{bb}$ with full combinatorics background still shows a clear edge at about 400 GeV. One can then try out a large variety of cuts for reducing the combinatorics background.  (i) For each event, drop the $M_{bb}$ pair that includes the invariant mass formed by combining the jet pair with the largest $\Delta R$ separation. (ii) For each event, only include an $M_{bb}$ pair if all of the corresponding jet pairs have $\Delta R < 1.5$. (iii) For each invariant mass pair in an event define $M_{bb}^{larger}$, the larger of the two $M_{bb}$'s. Only keep the invariant mass pair with the smallest $M_{bb}^{larger}$. One can also try combinations of the above. 
All these cuts yield distributions with the feature at 400 GeV significantly enhanced, which gave us confidence that this is the feature we need to determine. Cut (iii) seemed to work best, and was used to conduct the final edge measurement. 

The cleaned up $M_{bb}$ distribution, as well as the edge distribution, the peak width plot and the final measurement plot from the application of the Edge-to-Bump method were shown in \fref{edgefinderexampleMbb}. The final endpoint measurement is $M_{bb}^{max} = 391.9 \pm 10.3 \gev$, which agrees well with the expected value of $382.3 \gev$.

We can then use this measurement to determine the decay chain assignment uniquely for 1570 (16.7\%) of the original 9385 Events. One of the three possibilities can be excluded for 2304 events (24.5\%), while no information is gained for 5300 events (56.5\%). For 211 events (2.2\%) all three possible assignments are excluded, indicating badly measured momenta.

\subsection{Measuring $M_{T2}$ Edges}
\label{ss.mt2measurement}
For each $M_{T2}$-subsystem variable we use the KE and DL methods (\ssref{MT2CBR}) to obtain two distributions with reduced combinatorics background. We then apply the Edge-to-Bump method (extended for $M_{T2}$ edges) as explained in \ssref{MT2edgetobump} to obtain an edge measurement. \fref{MT2example} shows the complete measurement procedure for two examples. For details on the remaining measurements see the Appendix. All the edge measurements are summarized in \tref{edgemeasurementsfirst}.

None of the edge determinations deviate significantly from the prediction, meaning we were successful in avoiding false measurements. Many of the error bars are fairly large, but for the most part this truthfully reflects the obfuscating effect of combinatorics background, as well as the poor quality of the edge itself (recall that this measurement was performed using jets only).

\begin{table}
\begin{center}
\begin{tabular}{|llllc|}
\hline
Variable & Prediction & Measurement & Deviation/$\sigma$ & Quality  \\ \hline \hline
$M_{bb}$ & $382.3$ & $391.8 \pm 10.3$ & $+0.93$ &  --- \\ \hline
$M_{T2\perp}^{221}(0)$ & $303.5$ & $240 \pm 140$ &  $-0.45$ & C \\
$M_{T2}^{221}(0)$    &    & $301 \pm 47$  & $-0.05$  & A \\ \hline
$M_{T2\perp}^{221}(E_b)$ & $7153.4$ & $7154 \pm 42$ & $+0.01$ & A   \\
$M_{T2}^{221}(E_b)$ & & $7171 \pm 42$ & $+0.42$ & A  \\ \hline
$M_{T2\perp}^{210}(0)$ & $320.9$ & $283 \pm 44$ & $-0.86$ & A  \\
$M_{T2}^{210}(0)$ & & $327.2 \pm 8.7$ &  $+0.72$ & A \\ \hline 
$M_{T2\perp}^{210}(E_b)$  & $7239.8$ & $7141 \pm 54$ & $-1.84$ & A  \\
$M_{T2}^{210}(E_b)$ & & $7176 \pm 37$ & $-1.75$  & A \\ \hline 
$M_{T2\perp}^{220}(0)$ &  $506.7$ & $509 \pm 211$ & $+0.01$ & C  \\
$M_{T2}^{220}(0)$ & & $528 \pm 56$  & $+0.38$ & B \\ \hline 
$M_{T2\perp}^{220}(E_b)$ & $7393.1$ & $7484 \pm 106$ & $+0.86$ & B \\
$M_{T2}^{220}(E_b)$ & & $7456 \pm 70$ & $+0.90$  & B  \\ \hline 
$M_{T2\perp\mathrm{all}}^{210}(0)$ & $312.8$ & $249 \pm 52$ & $-1.23$ & B  \\ 
$M_{T2\perp\mathrm{all}}^{210}(E_b)$ &$7158.2$  & $7129 \pm 40$ & $-0.73$  & A  \\ \hline
\end{tabular}
\end{center}
\caption{Edge Measurements for the first Monte Carlo study. $E_b = 7000 \gev$. The measurements are obtained from the 1$\sigma$ confidence level intervals. The Quality column specifies which method was used to merge the two sets of edge measurements, as explained in \ssref{MT2edgetobump}.}
\label{t.edgemeasurementsfirst}
\end{table}

\begin{figure}
\begin{center}
\vspace*{-10mm}
\hspace*{-20mm}
\begin{tabular}{|c|c|}
\hline 
\hline & \\
 $M_{T2}^{210}(0)$ with $p^T_{ISR} < 100 \gev$ & $M_{T2\perp\mathrm{all}}^{210}(0)$

\\
&\\ \hline\hline

\renewcommand{\tempstringtwo}{3cm}
\renewcommand{\tempstringone}{BP5_MT2subsystem210ListZero}
\begin{tabular}{rr}
\multicolumn{2}{c}{\includegraphics[height=\tempstringtwo]{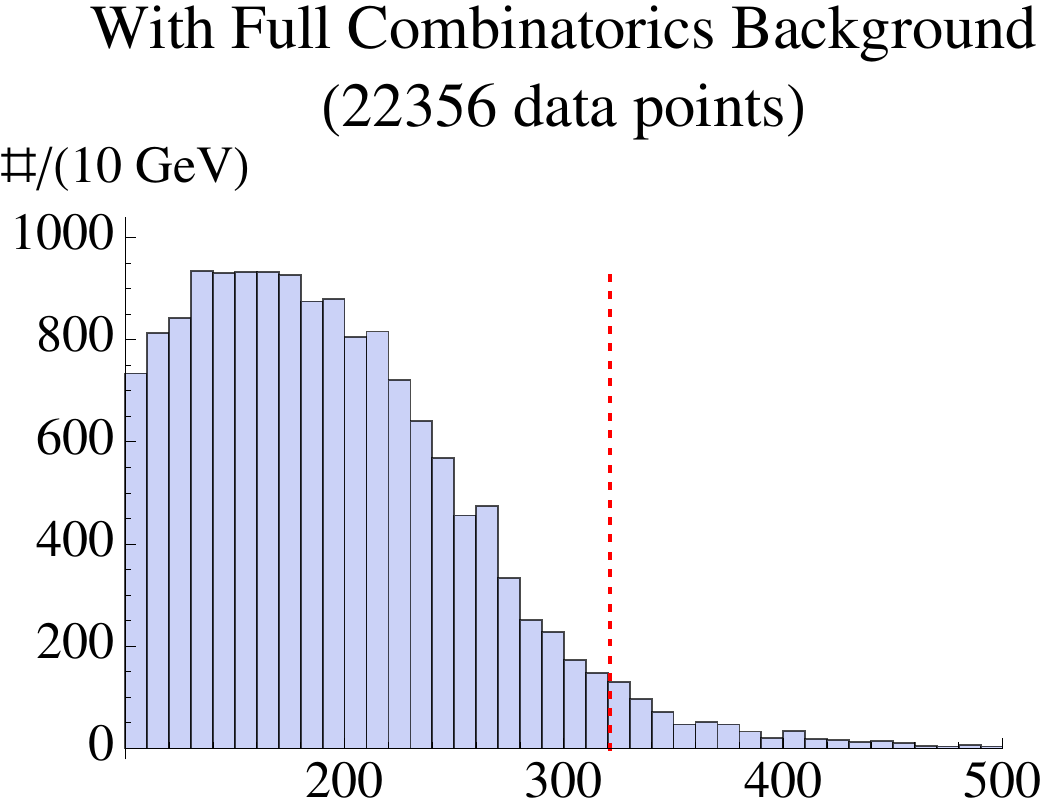}}
\\ \includegraphics[height=\tempstringtwo]{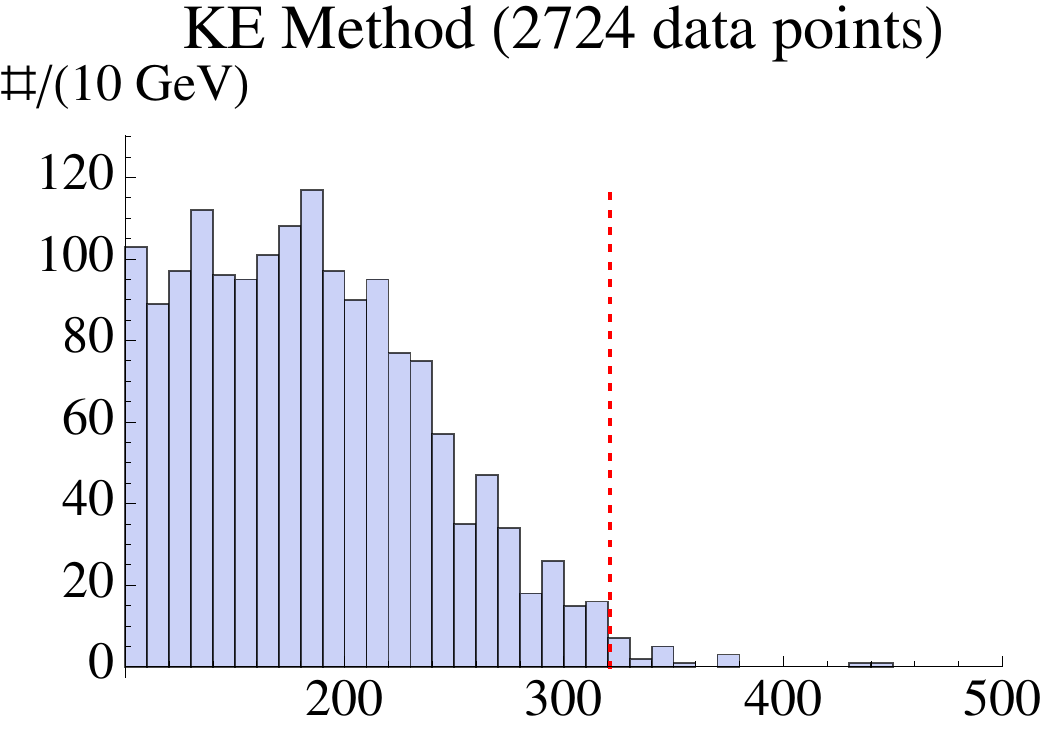}
& \includegraphics[height=\tempstringtwo]{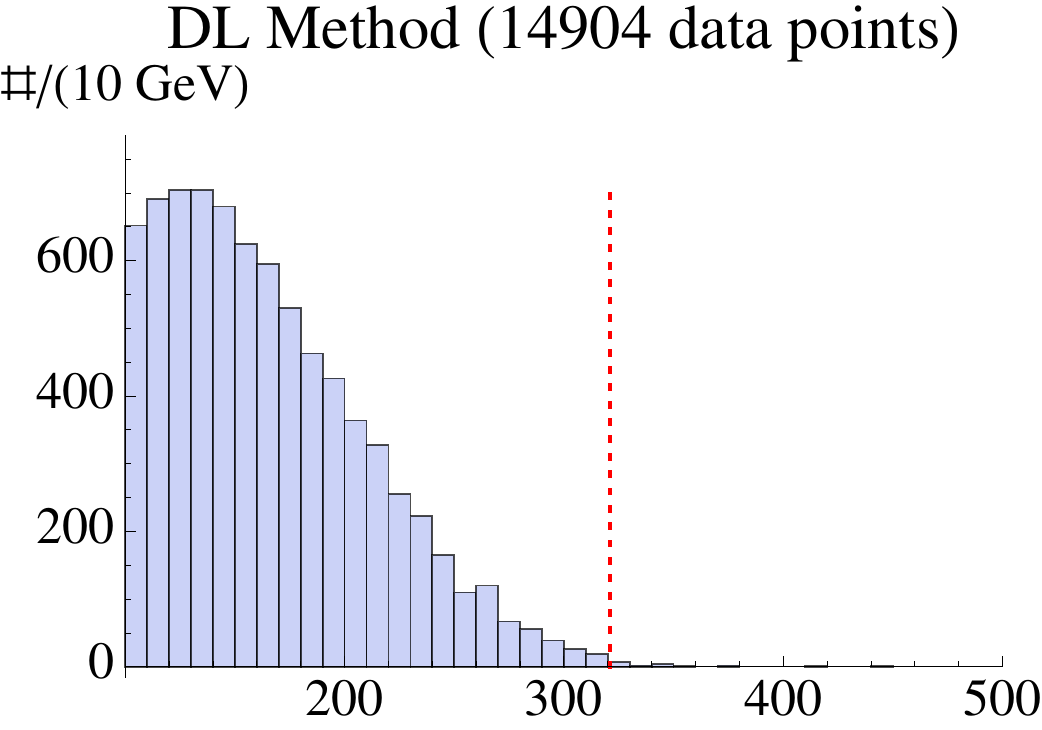}
\\ \includegraphics[height=\tempstringtwo]{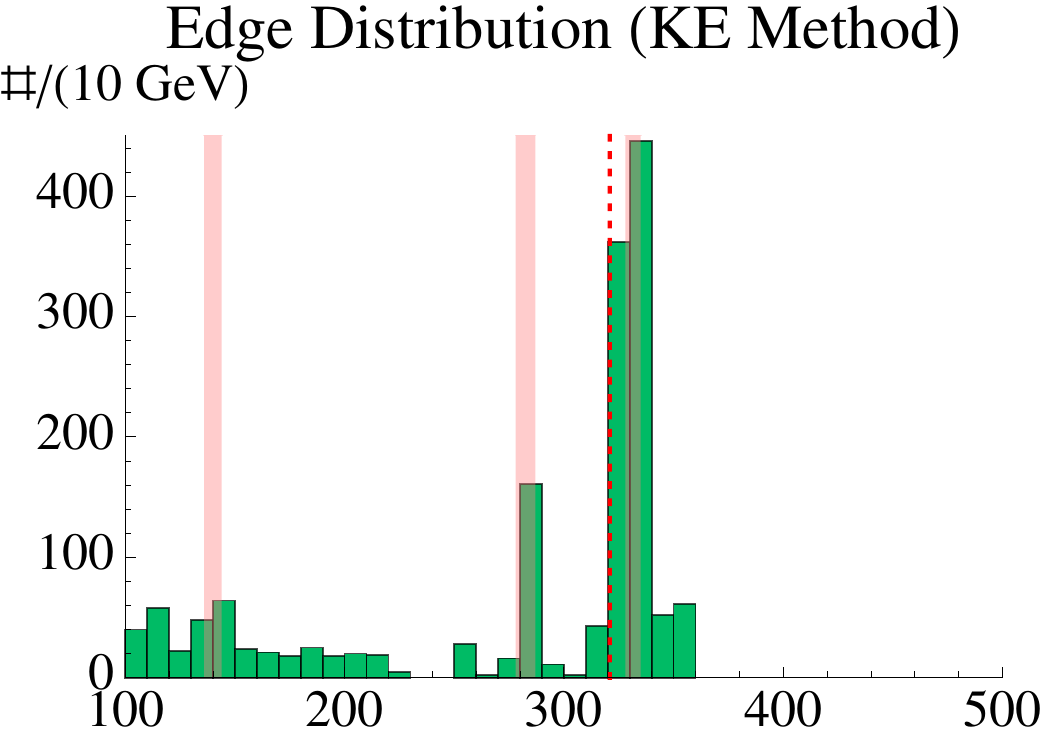}
& \includegraphics[height=\tempstringtwo]{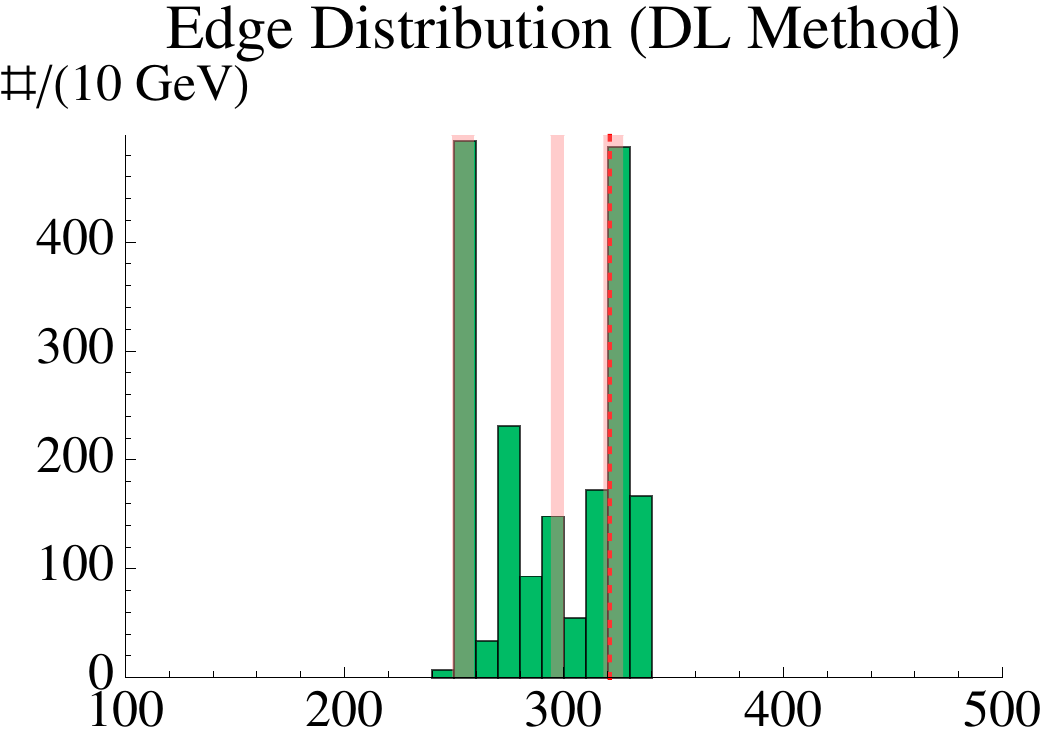}
\\ \includegraphics[height=\tempstringtwo]{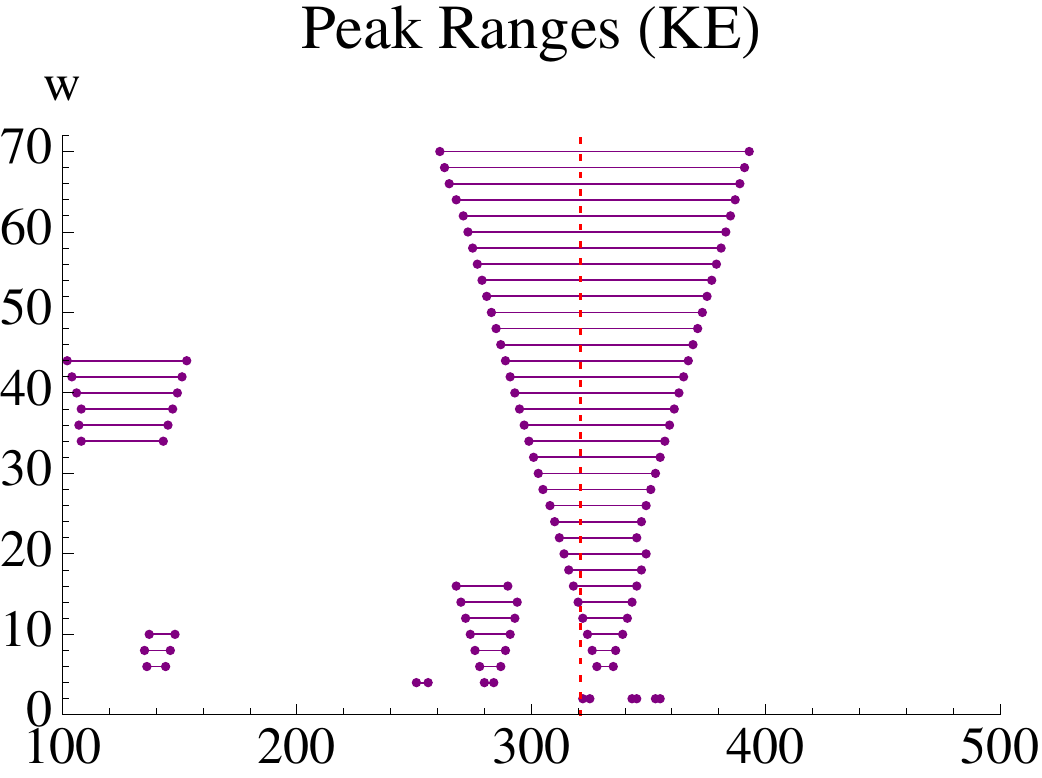}
& \includegraphics[height=\tempstringtwo]{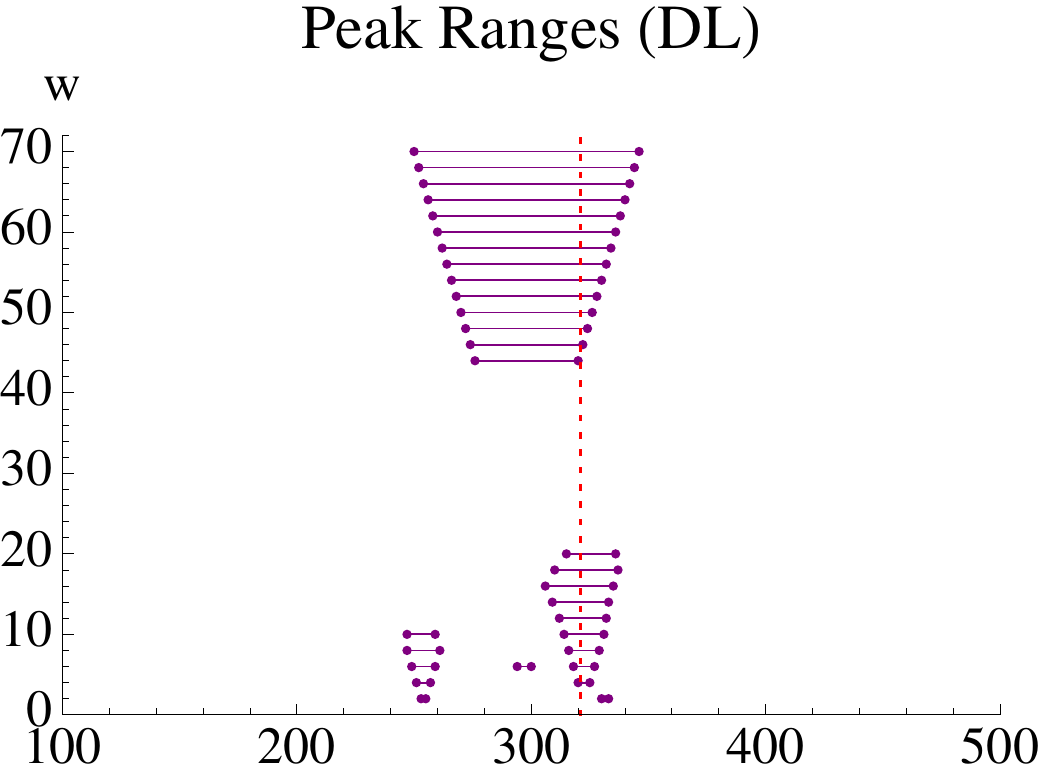}
\\ \includegraphics[height=\tempstringtwo]{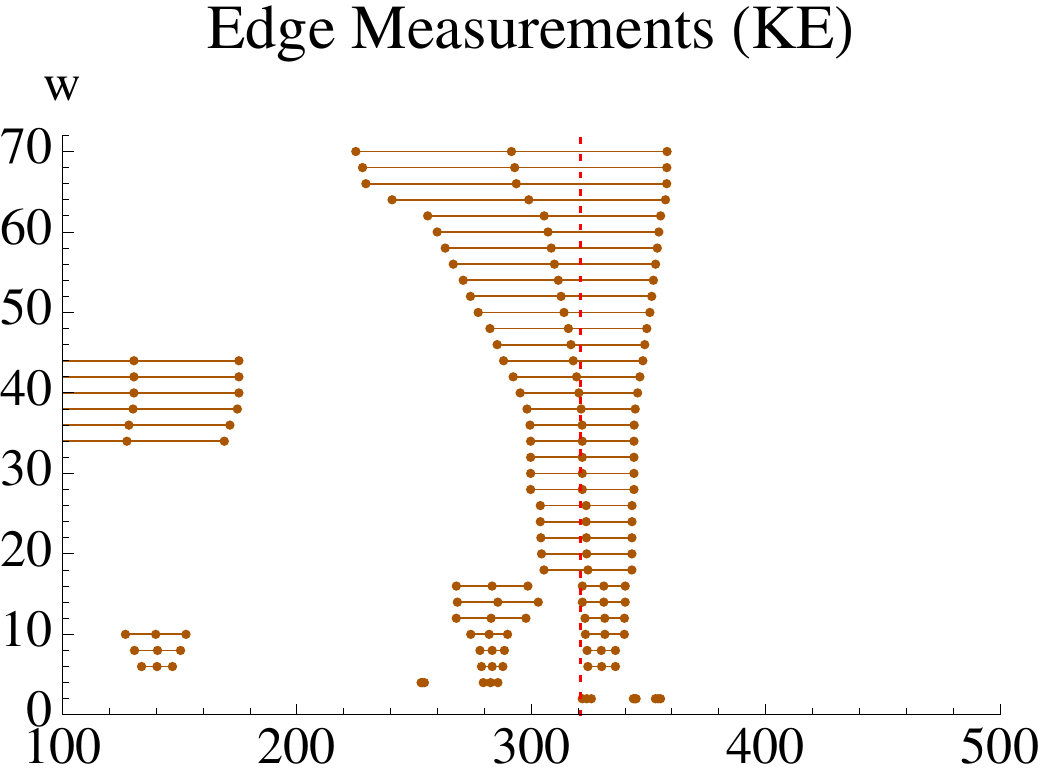}
& \includegraphics[height=\tempstringtwo]{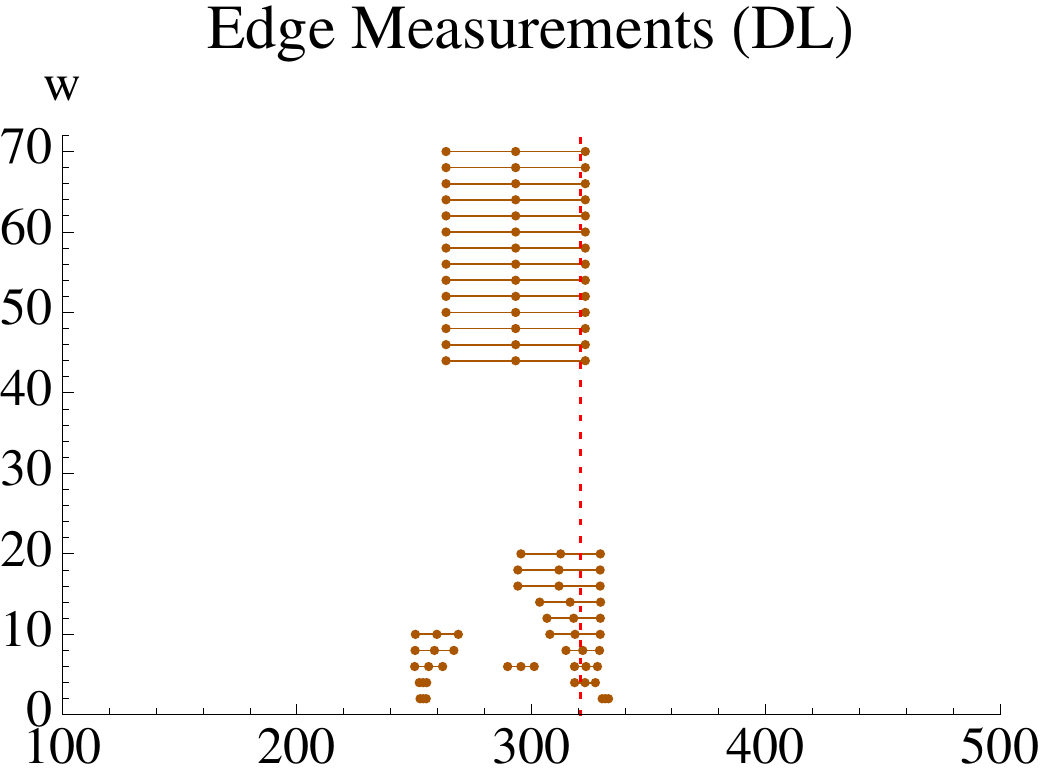}
\\ \multicolumn{2}{c}{\includegraphics[height=\tempstringtwo]{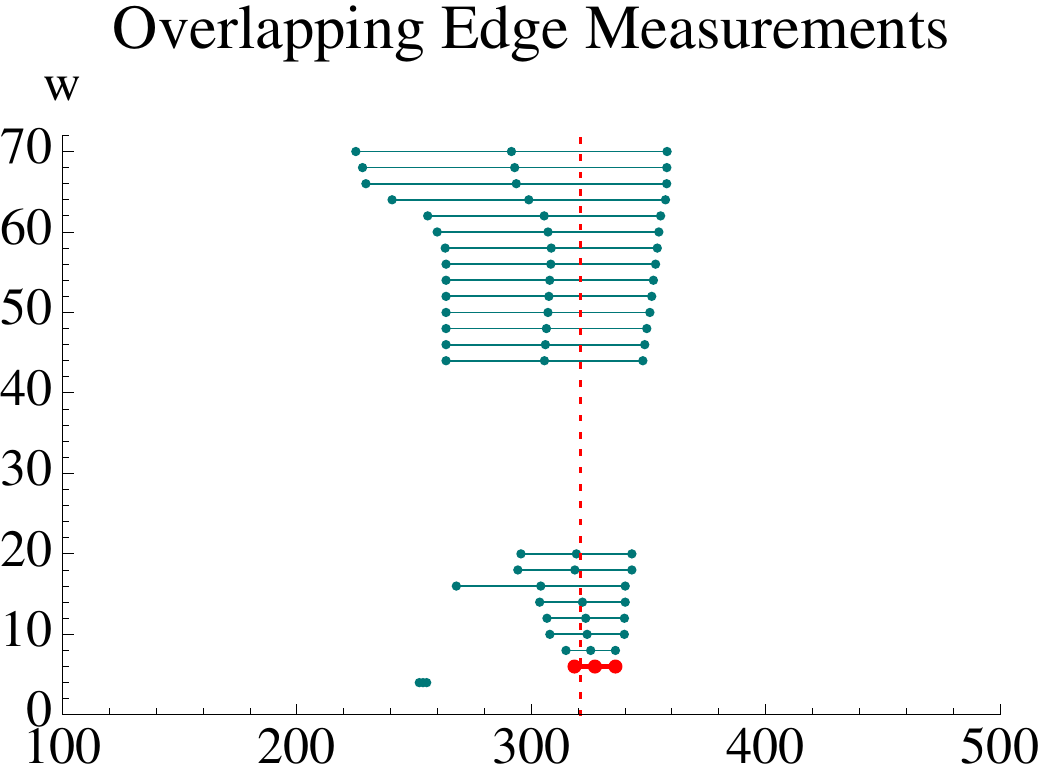}}
\end{tabular}

&

\renewcommand{\tempstringtwo}{3cm}
\renewcommand{\tempstringone}{BP5_MT2subsystem210perpallListZero}
\begin{tabular}{rr}
\multicolumn{2}{c}{\includegraphics[height=\tempstringtwo]{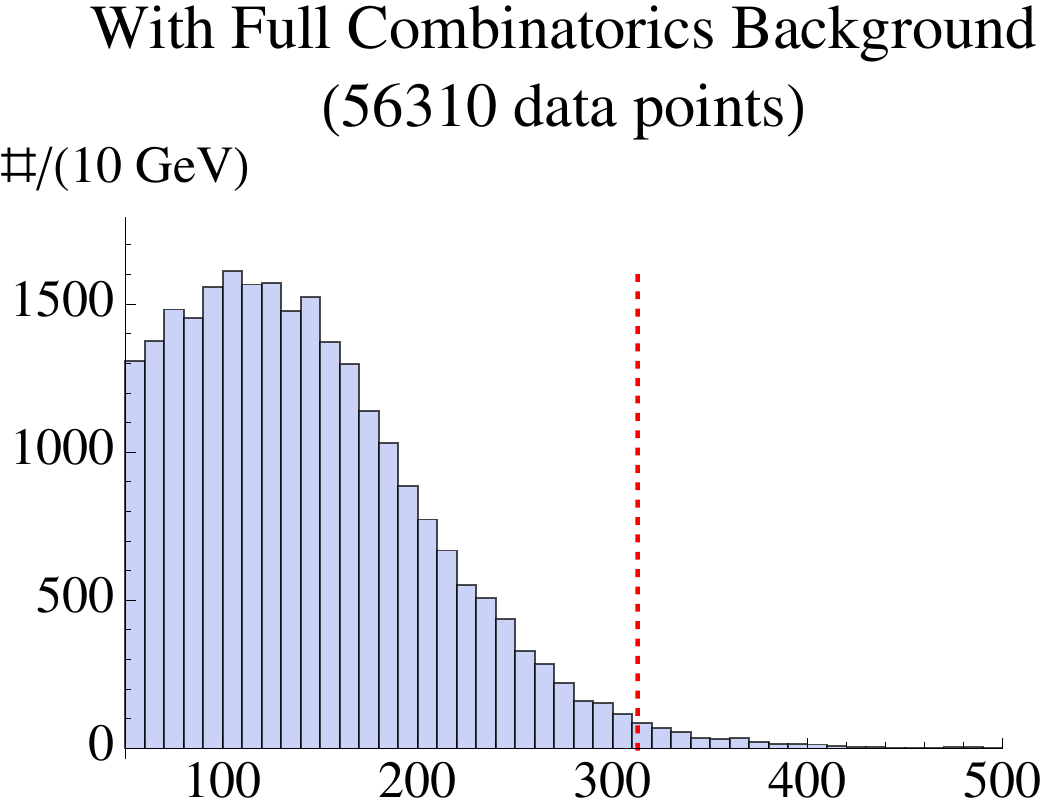}}
\\ \includegraphics[height=\tempstringtwo]{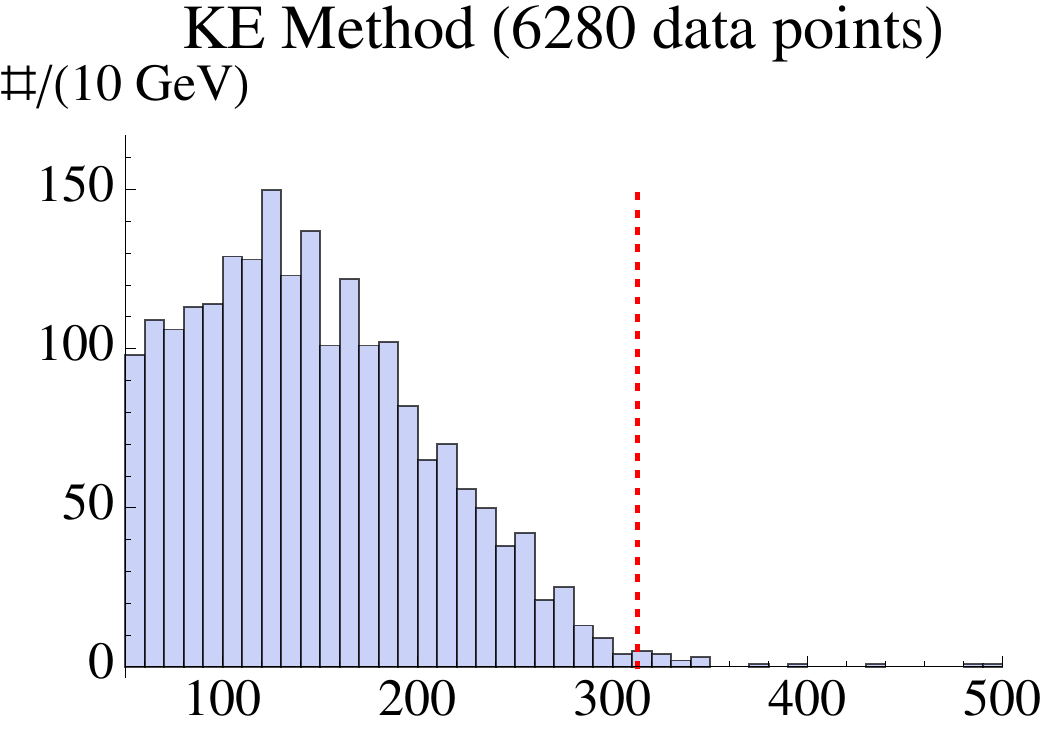}
& \includegraphics[height=\tempstringtwo]{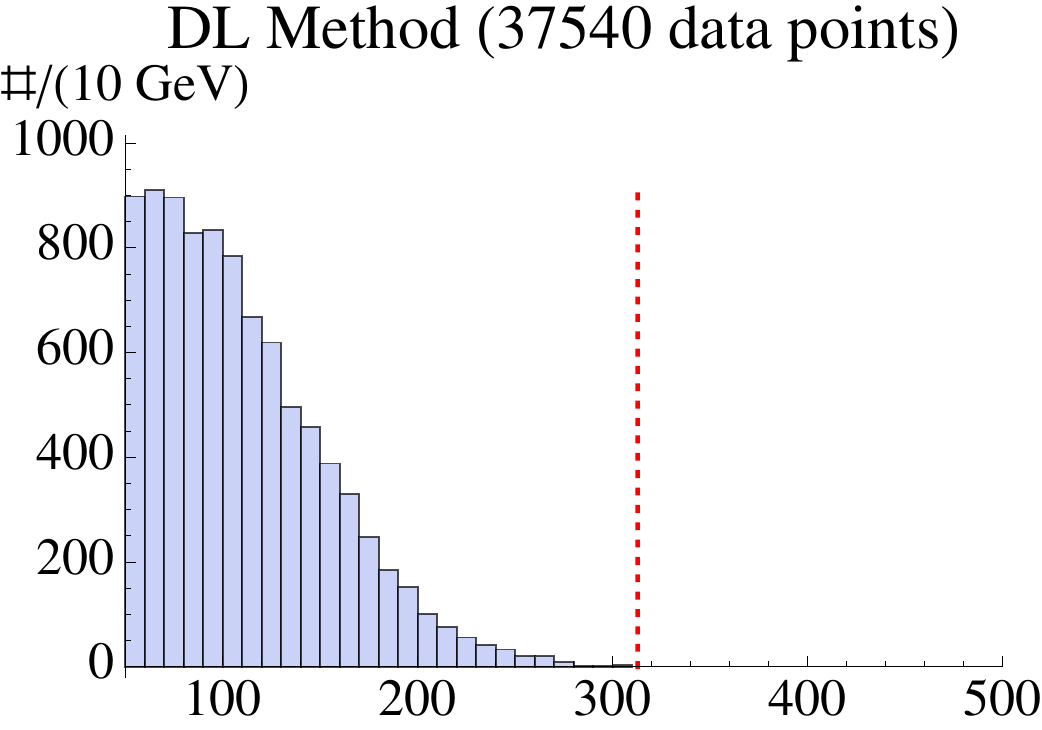}
\\ \includegraphics[height=\tempstringtwo]{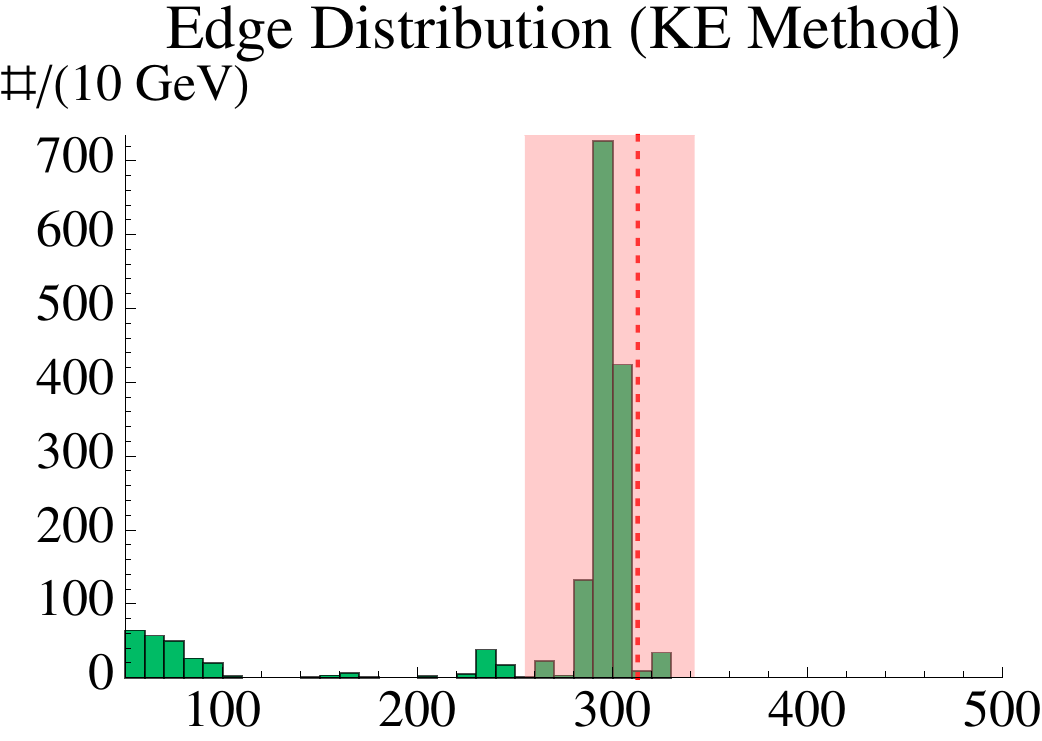}
& \includegraphics[height=\tempstringtwo]{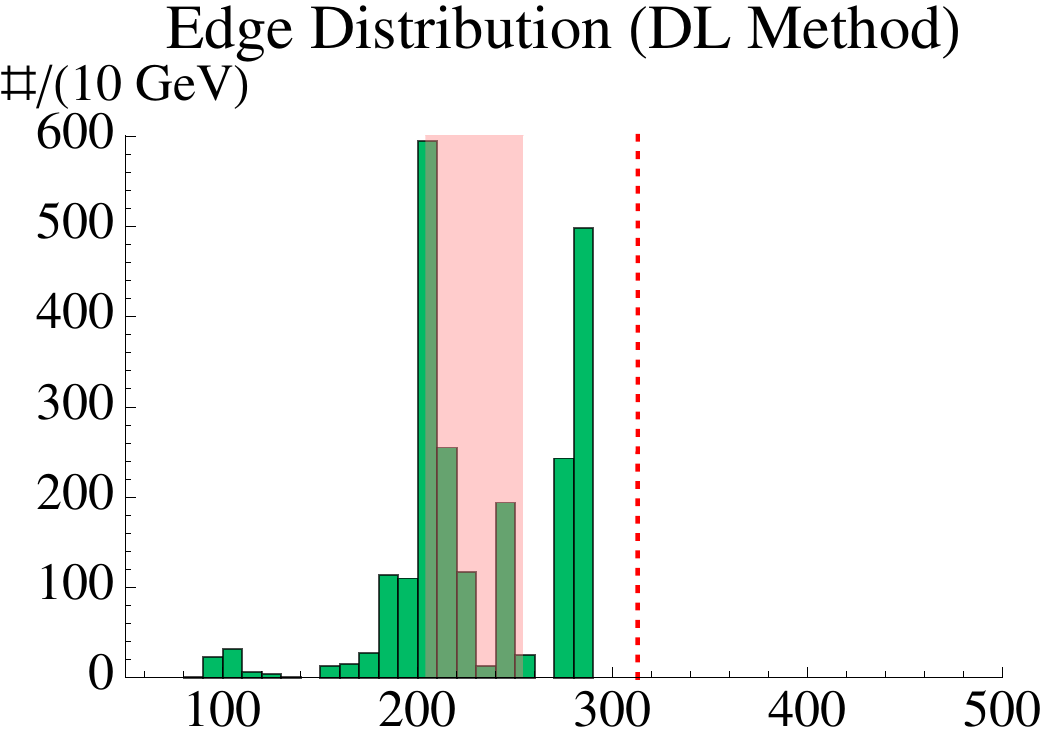}
\\ \includegraphics[height=\tempstringtwo]{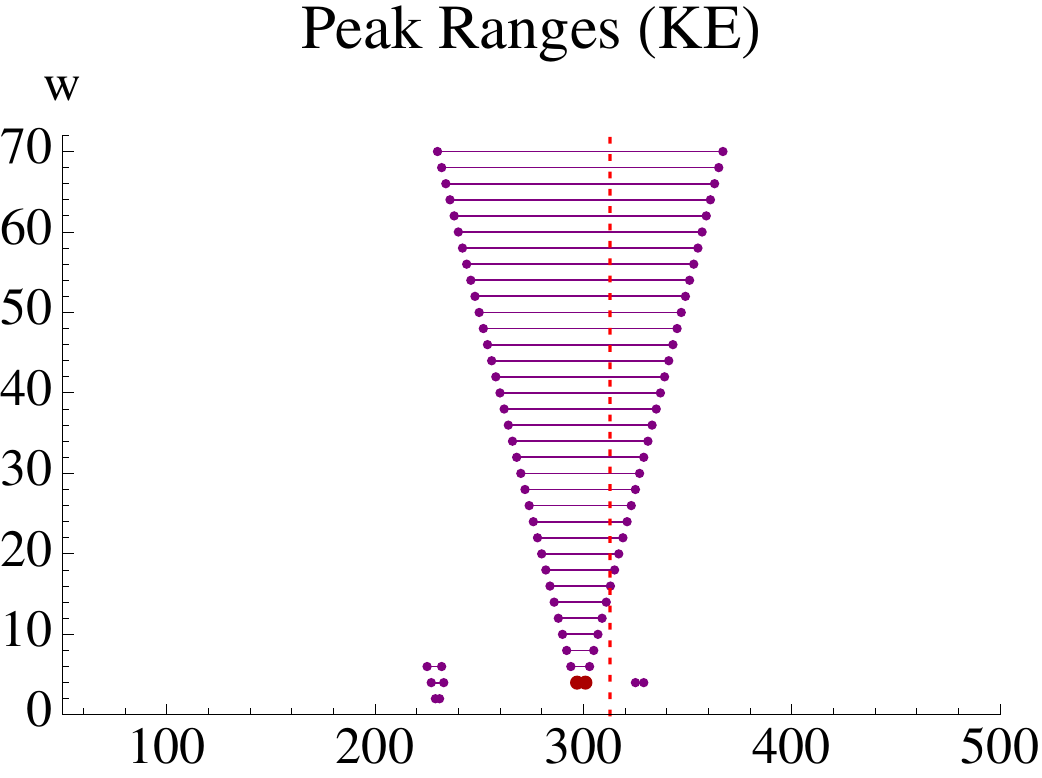}
& \includegraphics[height=\tempstringtwo]{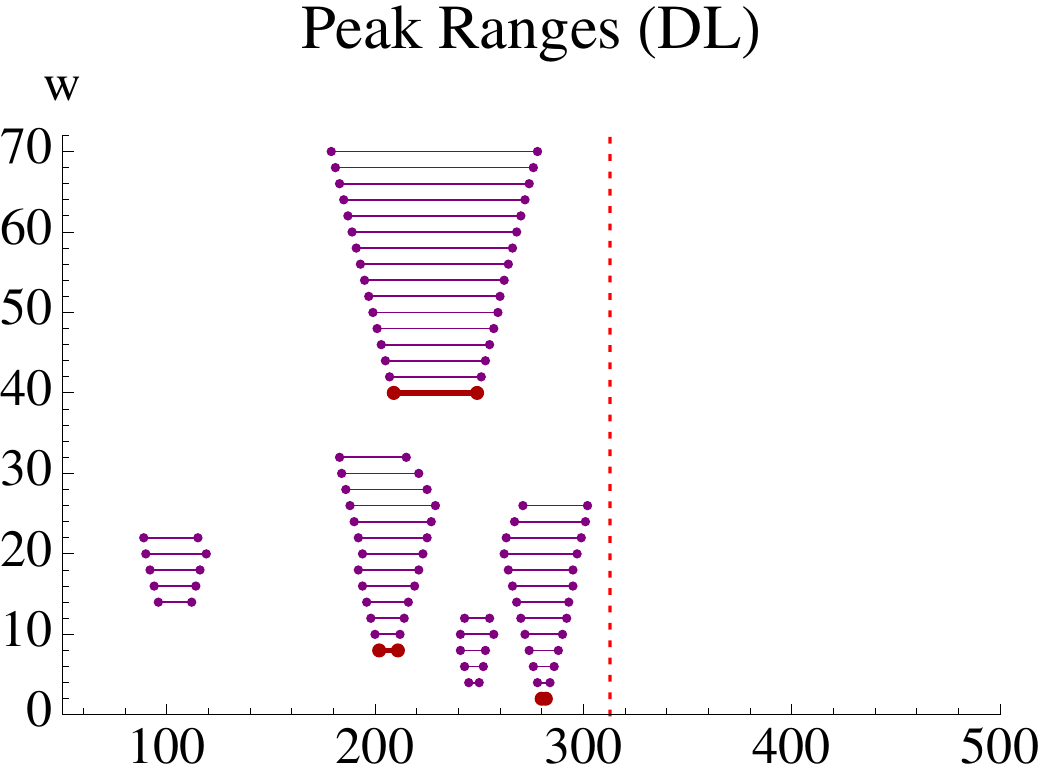}
\\ \includegraphics[height=\tempstringtwo]{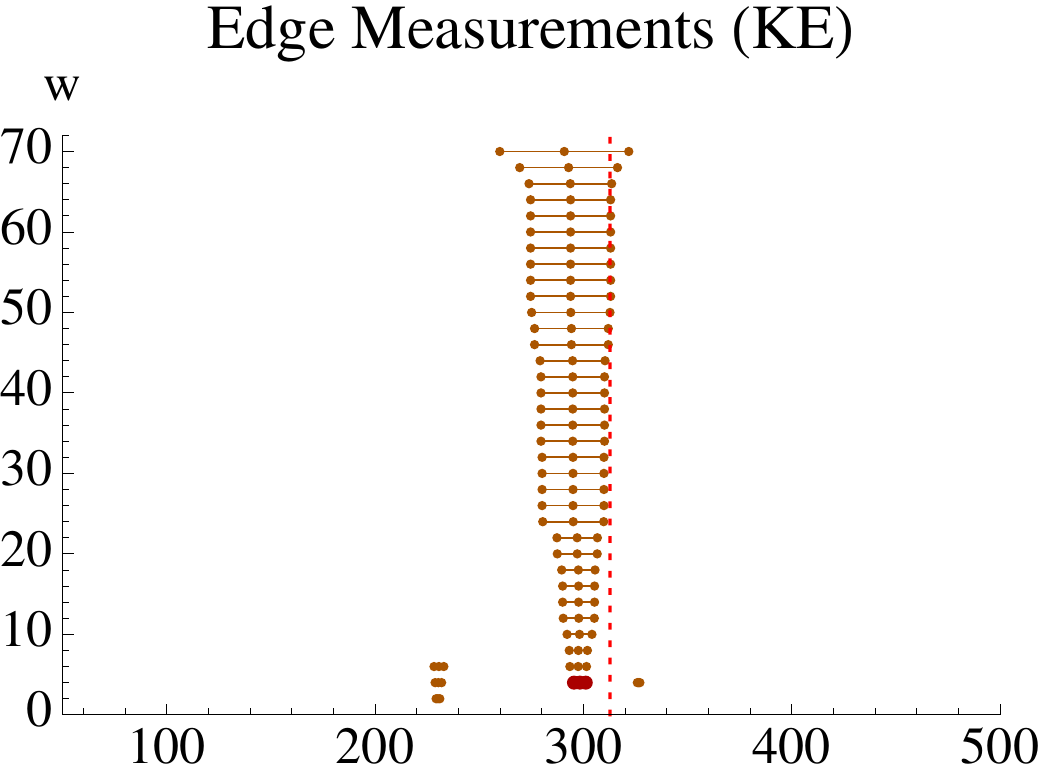}
& \includegraphics[height=\tempstringtwo]{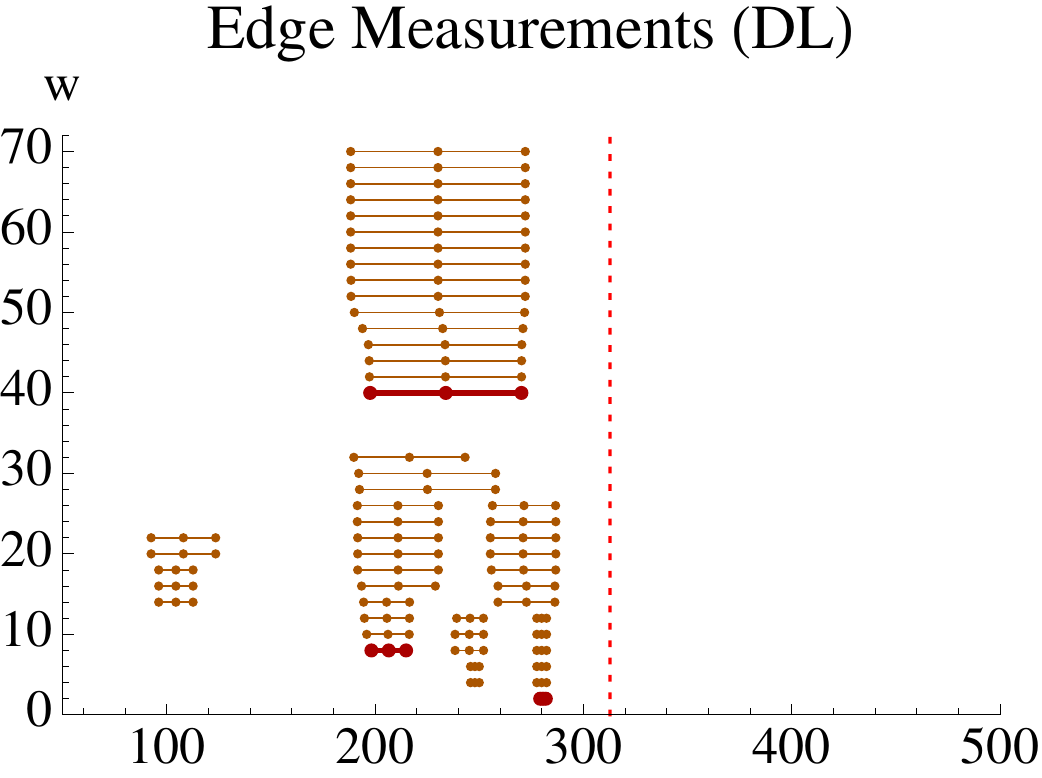}
\\ \multicolumn{2}{c}{\includegraphics[height=\tempstringtwo]{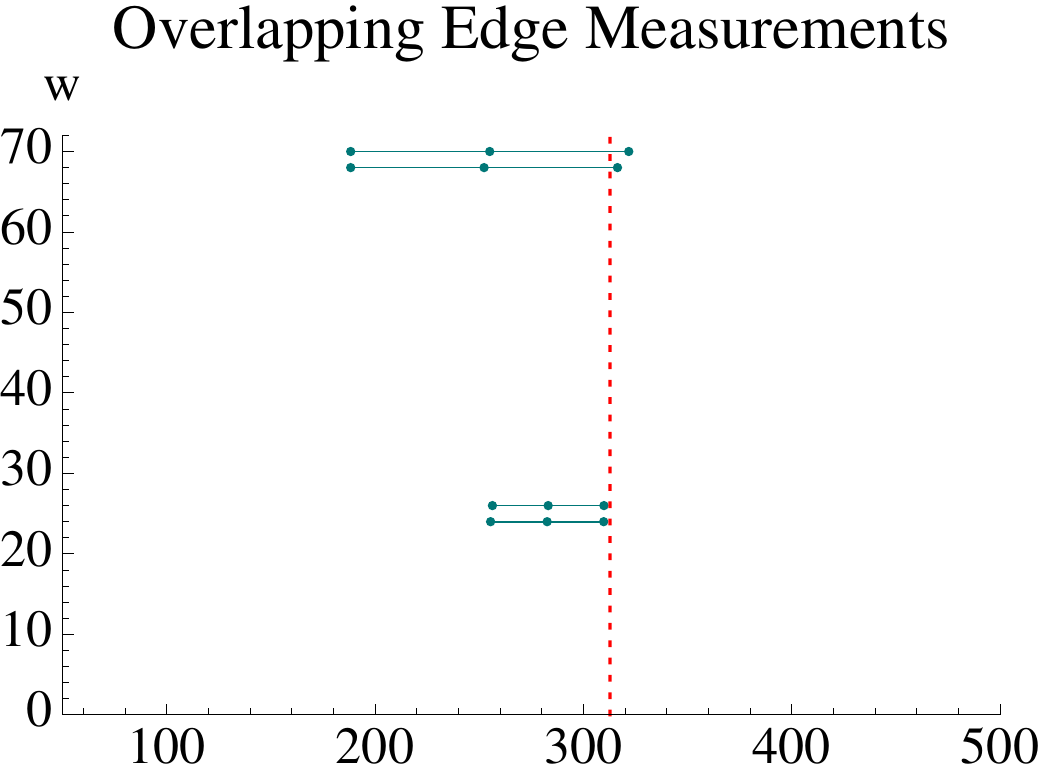}}
\end{tabular}

\\
\hline 
Edge Measurement: $327.2 \pm 8.7$ GeV  [320.9]  (Case A)
& 
Edge Measurement: $249 \pm 52$ GeV [312.8]  (Case B)
\\
\hline
\end{tabular} \vspace{-6mm}
\end{center}
\caption{The complete edge measurements for two of the 14 examined $M_{T2}$ distributions in the first Monte Carlo study. [Expected endpoint locations in square brackets.] See the Appendix the other measurements. 
}
\label{f.MT2example}
\end{figure}

\subsection{Determining Masses from Edge Measurements}
\label{ss.massmeasurement}

The space of possible masses for this decay is the quarter-cube of $m_{\tilde g}, m_{\tilde b_1}, m_{\tilde \chi^0_1}$ masses with the constraint $E_b = 7 \tev > m_{\tilde g} > m_{\tilde b_1} > m_{\tilde \chi^0_1}$. (For simplicity express all masses in GeV and regard them as dimensionless numbers in this section.)

Now imagine measuring, say,  $M_{T2}^{210}(0)^{max}$ and knowing its value to be exactly $M_{T2}^{210}(0)^{max}_{meas}$. The known analytical dependence  of that endpoint on the three masses \cite{MT2subsystem} defines a surface in mass-space: $M_{T2}^{210}(0)^{max}\{m_{\tilde g},m_{\tilde b_1},m_{\tilde \chi^0_1}\} = M_{T2}^{210}(0)^{max}_{meas}$ (where curly brackets indicate we are treating the endpoint as a function of the three masses). If we knew the endpoint exactly we would know that the point in mass-space corresponding to the correct spectrum must lie somewhere on that surface.

In reality our endpoint measurement has some error: $M_{T2}^{210}(0)^{max} = M_{T2}^{210}(0)^{max}_{meas} \pm \delta M_{T2}^{210}(0)^{max}_{meas}$.  Interpreting this uncertainty as a gaussian 1-$\sigma$ error, the clearly defined surface in mass space now becomes some \emph{gaussian density} 
\begin{equation}
\mathcal{D}\{m_{\tilde g}, m_{\tilde b_1}, m_{\tilde \chi^0_1}\} = \frac{1}{\sqrt{2 \pi}} \exp\left[\frac{1}{2} \left(\frac{M_{T2}^{210}(0)^{max}\{m_{\tilde g},m_{\tilde b_1},m_{\tilde \chi^0_1}\} - M_{T2}^{210}(0)^{max}_{meas}}{ \delta M_{T2}^{210}(0)^{max}_{meas}}\right)^2\right]
\end{equation}
that is a function of the three masses and peaked at the surface  $M_{T2}^{210}(0)^{max}\{m_{\tilde g},m_{\tilde b_1},m_{\tilde \chi^0_1}\} = M_{T2}^{210}(0)^{max}_{meas}$. We can then define a \emph{1-$\sigma$ Confidence Level Volume} for the possible values of the masses by the constraint
\begin{equation}
\label{e.CLV}
\left| \mathcal{D}\{m_{\tilde g}, m_{\tilde b_1}, m_{\tilde \chi^0_1}\} \right| > \mathcal{D}_{min},
\end{equation}
where $\mathcal{D}_{min}$ is chosen such that the \emph{total integrated weight} enclosed in this volume is $\mathrm{Erf}(1/2) \approx 0.68$. 

This is easily extended to a set of endpoint measurements $M_i^{max} = {M_i}^{max}_{meas} \pm \delta {M_i}^{max}_{meas}$ (with known analytical dependence on the masses). The gaussian density is simply
\begin{equation}
\tilde {\mathcal{D}}\{m_{\tilde g}, m_{\tilde b_1}, m_{\tilde \chi^0_1}\} = {\prod_i} \frac{1}{\sqrt{2 \pi}} \exp\left[\frac{1}{2} \left(\frac{M_i\{m_{\tilde g},m_{\tilde b_1},m_{\tilde \chi^0_1}\} - {M_i}^{max}_{meas}}{ \delta {M_i}^{max}_{meas}}\right)^2\right].
\end{equation}
We renormalize this by defining
\begin{equation}
 {\mathcal{D}}\{m_{\tilde g}, m_{\tilde b_1}, m_{\tilde \chi^0_1}\} = \frac{\tilde {\mathcal{D}}\{m_{\tilde g}, m_{\tilde b_1}, m_{\tilde \chi^0_1}\}}{\mathcal{D}_{tot}},
 \end{equation}
 where 
 \begin{equation}
 \ \mathcal{D}_{tot} = \int_{m_{\tilde g}^{min}}^{E_b} d m_{\tilde g}  \int_{m_{\tilde b_1}^{min}}^{m_{\tilde g}} d m_{\tilde b_1} \int_{m_{\tilde \chi^0_1}^{min}}^{m_{\tilde b_1}} d m_{\tilde \chi^0_1}\ \ {\tilde {\mathcal{D}}\{m_{\tilde g}, m_{\tilde b_1}, m_{\tilde \chi^0_1}\}}
\end{equation}
so that \eref{CLV} again defines the 1-$\sigma$ Confidence Level Volume. 

It is illustrative to obtain uncorrelated 1-$\sigma$ Confidence Level \emph{Intervals} for the individual masses. We define the \emph{gaussian density projections}
\begin{eqnarray}
\mathcal{D}_{\tilde g}\{m_{\tilde g}\} & = &  \int_{m_{\tilde b_1}^{min}}^{m_{\tilde g}} d m_{\tilde b_1} \int_{m_{\tilde \chi^0_1}^{min}}^{m_{\tilde b_1}} d m_{\tilde \chi^0_1}\ \ { {\mathcal{D}}\{m_{\tilde g}, m_{\tilde b_1}, m_{\tilde \chi^0_1}\}},
\\
\mathcal{D}_{\tilde b_1}\{m_{\tilde b_1}\} & = & \int_{m_{\tilde g}^{min}}^{E_b} d m_{\tilde g}  \int_{m_{\tilde \chi^0_1}^{min}}^{m_{\tilde b_1}} d m_{\tilde \chi^0_1}\ \ { {\mathcal{D}}\{m_{\tilde g}, m_{\tilde b_1}, m_{\tilde \chi^0_1}\}},
\\
\mathcal{D}_{\tilde \chi^0_1}\{m_{\tilde \chi^0_1}\} & = & \int_{m_{\tilde g}^{min}}^{E_b} d m_{\tilde g}  \int_{m_{\tilde b_1}^{min}}^{m_{\tilde g}} d m_{\tilde b_1} \ \ { {\mathcal{D}}\{m_{\tilde g}, m_{\tilde b_1}, m_{\tilde \chi^0_1}\}}.
\end{eqnarray}
\eref{CLV} then defines the 1-$\sigma$ Confidence Level Intervals for each of the masses.

\begin{figure}
\begin{center}
\hspace*{-14mm}
\begin{tabular}{ccc}
\includegraphics[width=6cm]{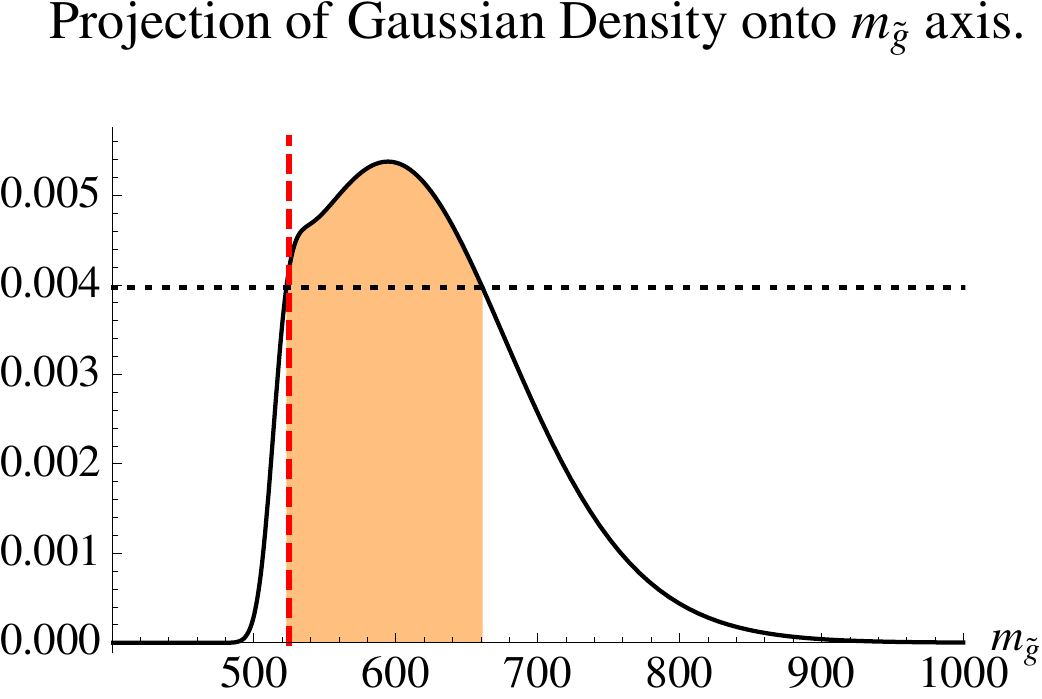}&
\includegraphics[width=6cm]{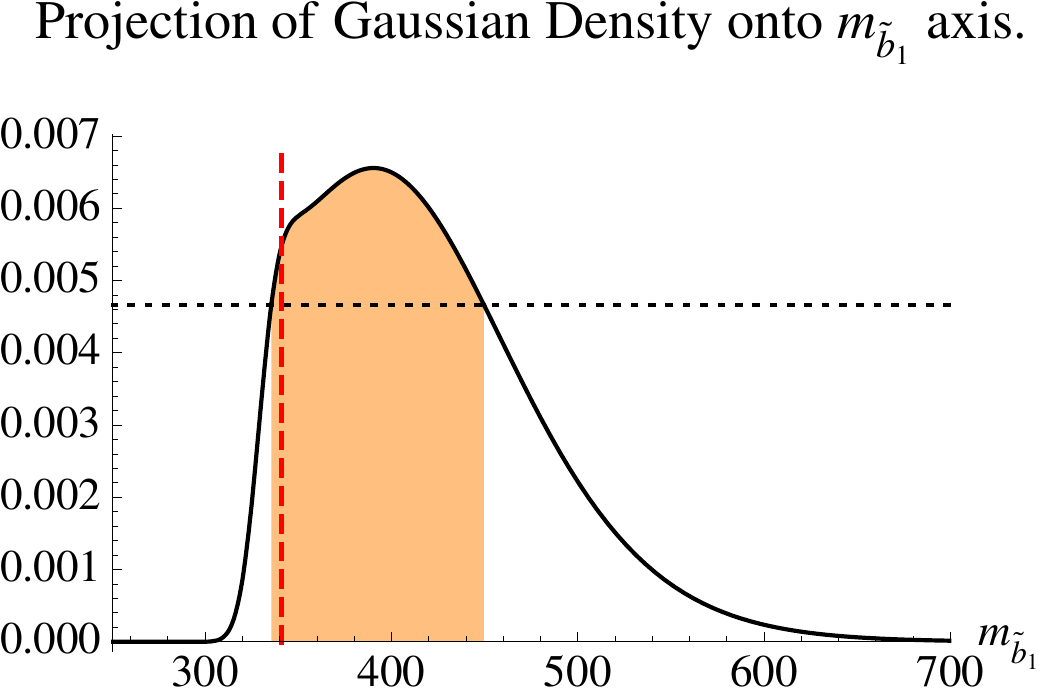}&
\includegraphics[width=6cm]{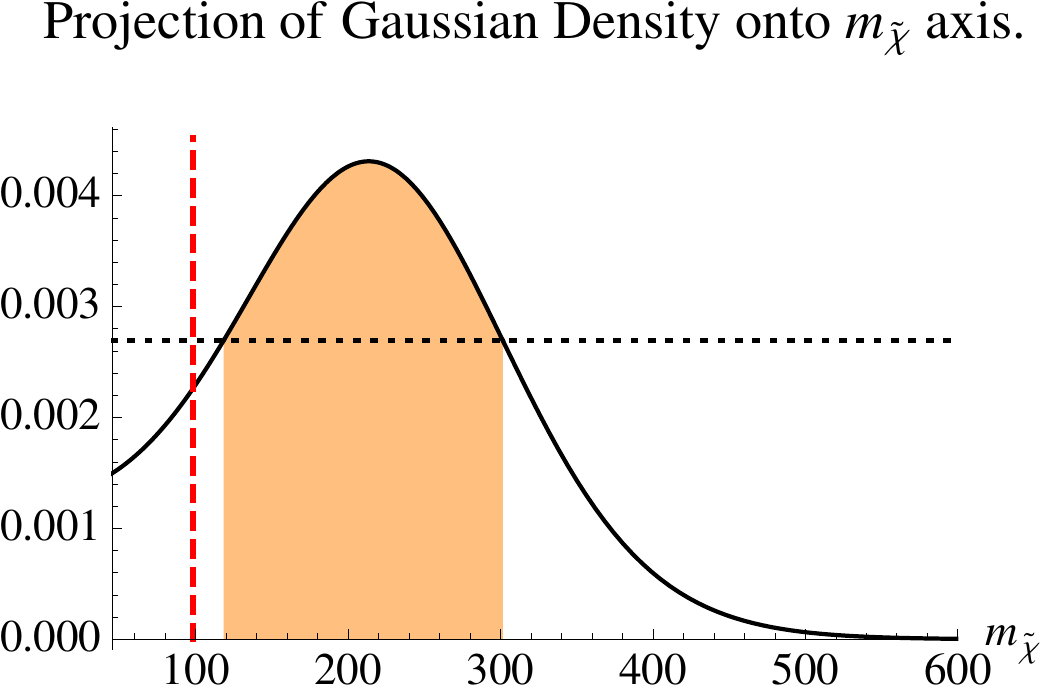}
\\ \\
$m_{\tilde g}^{meas} = 592 \pm 69 \ \  (525)$
& 
$m_{\tilde b_1}^{meas} = 393 \pm 57   \ \ (341)$
&
$m_{\tilde \chi^0_1}^{meas} =  210 \pm 92 \ \   (98)$
\end{tabular}\vspace*{-5mm}
\end{center}
\caption{Mass measurements for the first Monte Carlo study in GeV (actual masses in brackets). The plots show the gaussian density projections for the three masses. The 1-$\sigma$ confidence level interval is shaded, and the true mass value is indicated with the vertical dashed line. The dotted line indicates the value of $\mathcal{D}_{min}$ which defines the confidence interval.
}
\label{f.massmeasurementsfirst}
\end{figure}

\subsection{Results}

We are now ready to extract the mass measurements for our first Monte Carlo study from the edge measurements in \tref{edgemeasurementsfirst}. Since the endpoints of $M_{T2}$-subsystem variables formulated using the $\perp$-projection or with ISR-binning contain the same mass information, for each such variable we discard the edge measurement with larger error bars. We also used the $M_{T2\perp\mathrm{all}}^{210}(E_b)$ edge instead of $M_{T2\perp\mathrm{all}}^{210}(0)$ since that gave smaller error bars on the masses. In defining the gaussian density projections, a priori the values of $m^{min}$ for the three masses should be zero, but we set $m_{\chi^0_1}^{min} = 45 \gev$ to satisfy the LEP invisible $Z$ decay width measurement~\cite{LEPbound}. (The other minimum values do not matter since the gaussian density vanishes for small sbottom and gluino masses.) \fref{massmeasurementsfirst} shows the gaussian density projections for each of the three masses and the extracted mass measurement with 1-$\sigma$ error bars.

The precision of the $\tilde \chi^0_1$ mass measurement is very poor, we do not learn much more than the assumption  $m_{\chi^0_1} < m_{\tilde b_1}$. However, the gluino and sbottom masses are determined with an error of about $10\%$, which seems quite satisfactory considering the difficulty of this fully hadronic measurement.

\section{Blind Verification Study}
\label{s.casestudy2} \setcounter{equation}{0} \setcounter{footnote}{0}

The second Monte Carlo study was meant as a blind trial of our measurement methods. Maxim Perelstein prepared a MadGraph \texttt{param\_card.dat} MSSM model file which we used to generate events. We emphasize that the mass measurements in this study were undertaken \emph{without} prior knowledge of the actual spectrum. 

 The weak-scale inputs for the blind benchmark point are
 \begin{center}
\begin{tabular}{|l|l|l|l|l|l|l|l|l|l|}
\hline
$\tan \beta$ & $M_1$ & $M_2$ & $M_3$& $\mu$ & $M_A$ &
$M_{Q3L}$ & $M_{tR}$ & $M_{bR}$ &  $A_t$\\
\hline
 10 & 85 & 1000 & 630 & 500 & 1000 & 1000 & 1000 & 380 &  392.6\\
\hline
\end{tabular}
\end{center}
with all other $A$-terms zero and all other sfermion soft masses set at $1 \tev$. The relevant spectrum (calculated with {\texttt SuSpect} \cite{SuSpect}) is the following:
\begin{center}
\begin{tabular}{|l|l|l|l|l|l|l|l|}
\hline
$m_{\tilde t_1}$ & $m_{\tilde t_2}$ & $\sin \theta_{\tilde t}$ & $m_{\tilde b_1}$& $m_{\tilde b_2}$ & $\sin \theta_{\tilde b}$ &
$m_{\tilde g}$ & $m_{\tilde \chi^0_1}$\\
\hline
 1016 & 1029 & 0.76 & 404 & 1012 & 1& 703 & 84
 \\
\hline
\end{tabular}
\end{center}
This spectrum, with a gaugino pair production cross section of 1.61 pb at the LHC14, has not yet been excluded \cite{LHCsearches}.
To best verify the statistical methods used in the previous study we emulate it as closely as possible. We first generated $5.8 \times 10^{5}$ events (with hadronization/showering and detector effects), then applied the same $b$-tag and kinematic cuts with efficiencies of $4.4 \%$ and $48\%$ respectively. This left us with 12427 events, somewhat more than we had for our first study since the jets were harder. To reproduce the conditions of the first study in all ways except underlying spectrum, we discarded the excess events and only used 9385. This corresponds to using $\sim 270 \mathrm{fb}^{-1}$ of integrated luminosity at the LHC14 (though given our pessimistic $b$-tag efficiencies it could easily only be $100 \mathrm{fb}^{-1}$).

\begin{table}
\begin{center}
\begin{tabular}{lcccc}
Variable & Prediction & Measurement & Deviation/$\sigma$ & Quality  \\ \hline \hline
$M_{bb}$ & $563.4$ & $556.5 \pm 14.9$ &  $-0.46$ &  --- \\ \hline
$M_{T2\perp}^{221}(0)$ & $472.0$ & $340 \pm 148$ & $-0.89$ & B \\
$M_{T2}^{221}(0)$  &  & $426 \pm 83$   & $-0.55$   & B  \\ \hline
$M_{T2\perp}^{221}(E_b)$ & $7239.5$ & $7218 \pm 67$ & $-0.33$ & A \\
$M_{T2}^{221}(E_b)$ & & $7239 \pm 48$ & $-0.01$ & A \\ \hline
$M_{T2\perp}^{210}(0)$ & $391.3$ & $343 \pm 83$ & $-0.58$ & B \\
$M_{T2}^{210}(0)$ & & $406.8 \pm 10.8$ & $+1.43$ & A \\ \hline 
$M_{T2\perp}^{210}(E_b)$  & $7333.1$ & $7215 \pm 71$ & $-1.67$ & A \\
$M_{T2}^{210}(E_b)$ & & N/A & &  D\\ \hline 
$M_{T2\perp}^{220}(0)$ & $693.0$ & $598 \pm 165$ & $-0.57$ & C \\
$M_{T2}^{220}(0)$ & & $681 \pm 64$ & $-0.19$  & B\\ \hline 
$M_{T2\perp}^{220}(E_b)$  & $7572.9$ & $7663 \pm 125$ & $+0.73$ & B \\
$M_{T2}^{220}(E_b)$ & & $7642 \pm 93$ & $+0.74$   &  B \\ \hline 
$M_{T2\perp\mathrm{all}}^{210}(0)$ & $385.5$ & $327 \pm 128$ & $-0.45$ & C \\ 
$M_{T2\perp\mathrm{all}}^{210}(E_b)$ & $7195.4$ & $7184 \pm 47$ & $-0.24$ & A \\ \hline
\end{tabular}
\end{center}
\caption{Edge Measurements for the second Monte Carlo study. $E_b = 7000 \gev$. The measurements are obtained from the 1$\sigma$ confidence level intervals. The Quality column specifies which method was used to merge the two sets of edge measurements, as explained in \ssref{MT2edgetobump}.}
\label{t.edgemeasurementssecond}
\end{table}

In keeping with the first study we ignored SM backgrounds, but in this case their contributions seem comparable to the SUSY signal. We avoided changing the cuts to reproduce the kinematic conditions of the first study, but one could certainly sharpen them to dramatically reduce SM backgrounds with relatively minor signal cost. Even if there is a sizable fraction of SM events in the distributions, they are unlikely to pollute the kinematic edges in the same fashion as the combinatorics background.

We performed the $M_{bb}$ and $M_{T2}$ endpoint measurements in exactly the same way as described in Sections \ref{ss.mbbmeasurement} and \ref{ss.mt2measurement}. It is interesting to point out that the efficiencies associated with the KE method of reducing combinatorics background (the fraction of events for which one or both of the decay chain assignments could be excluded) are practically identical to the first study. The harder jet spectrum in the blind study reduced the efficiency of the $p^T_{max}$ cut for the ISR-binned $M_{T2}$ edge measurements by an $\mathcal{O}(1)$ factor. To improve our measurement we increased the $p^T_{max}$ values for $M_{T2}^{221}$. This is not inconsistent -- a higher choice of $p^T_{max}$ gives more statistics at the expense of more intrinsic smearing in the edge, which will be automatically incorporated into the error bars of the edge measurement. 
\begin{center}
\begin{tabular}{c|cccccc}
& $M_{T2}^{221}(0)$ & $M_{T2}^{221}(E_b)$ & $M_{T2}^{210}(0)$ & $M_{T2}^{210}(E_b)$ & $M_{T2}^{220}(0)$ & $M_{T2}^{220}(E_b)$\\ \hline
$p^T_{max}$ & 60 & 80 & 100 & 50 & 50 & 40
\end{tabular}
\end{center}
The endpoint measurements are summarized in  \tref{edgemeasurementssecond}. Overall the edges seemed more shallow, but the methods performed well, again avoiding all mismeasurements. See the Appendix for more plots. 

Proceeding identically to \ssref{massmeasurement}, and again using $M_{T2\perp\mathrm{all}}^{221}(E_b)$ instead of $M_{T2\perp\mathrm{all}}^{221}(0)$, we obtained the mass measurements shown in \fref{massmeasurementssecond}. The mass measurements actually seem better than in the first study, with 1-$\sigma$ agreement across the board and somewhat smaller errors. This shows that our methods are applicable beyond our particularly chosen first benchmark point.

\begin{figure}
\begin{center}
\hspace*{-14mm}
\begin{tabular}{ccc}
\includegraphics[width=6cm]{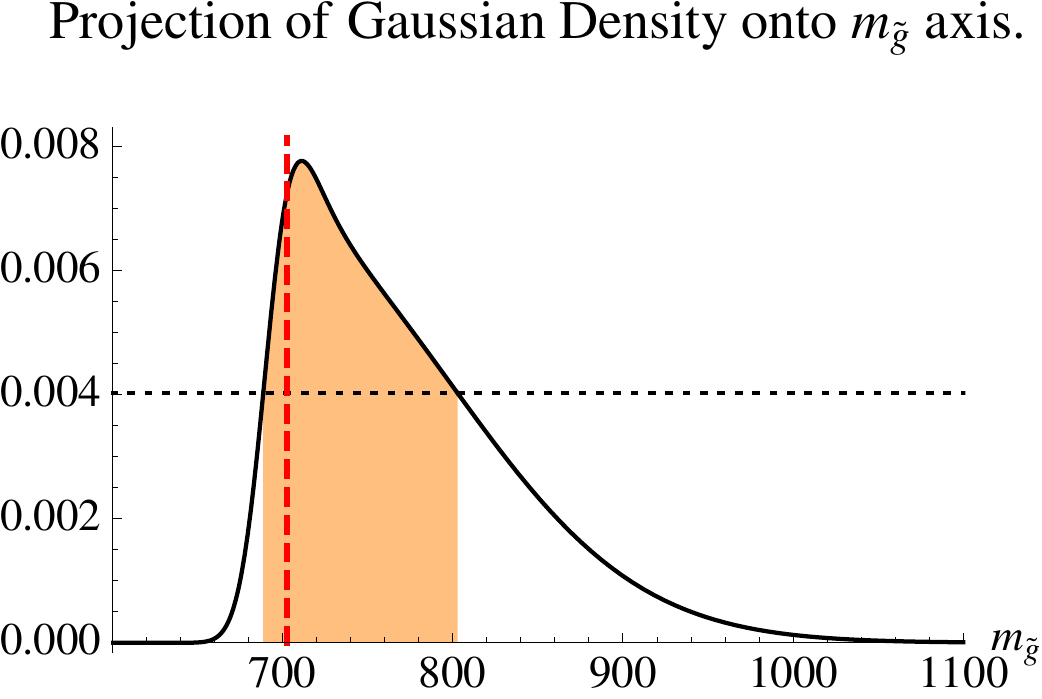}&
\includegraphics[width=6cm]{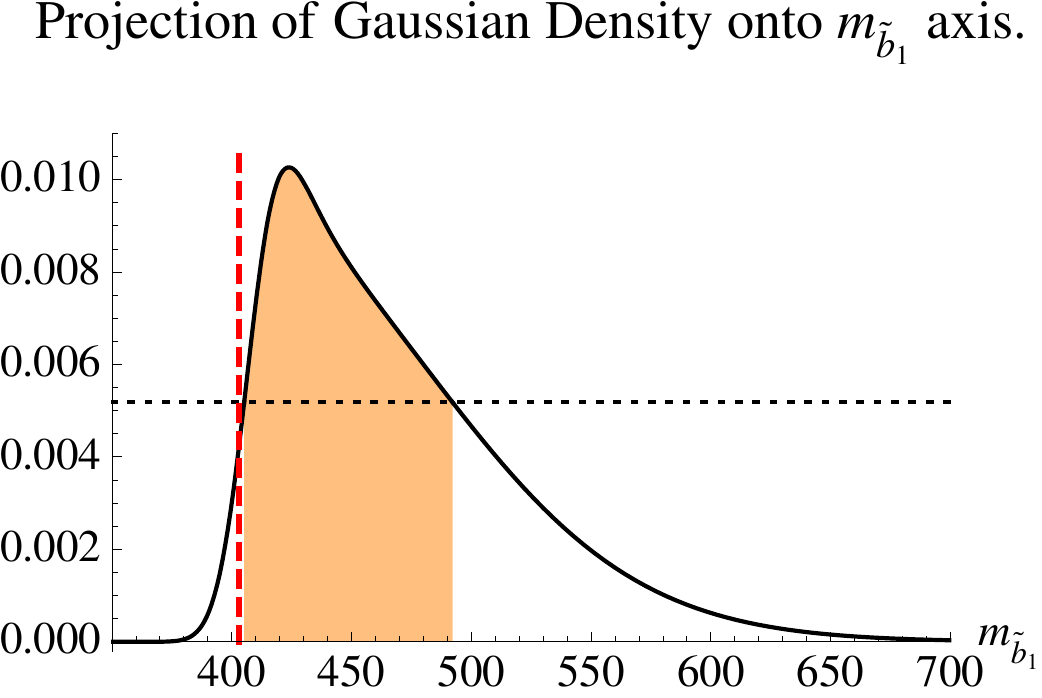}&
\includegraphics[width=6cm]{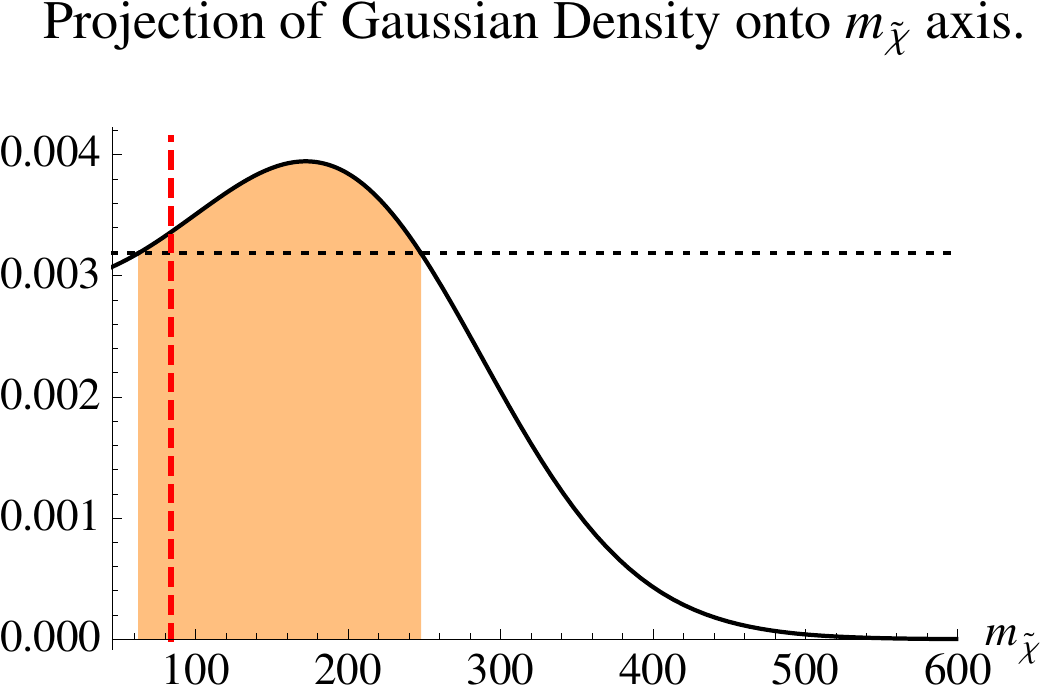}
\\ \\
$m_{\tilde g}^{meas} = 746 \pm 57 \ \  (703)$
& 
$m_{\tilde b_1}^{meas} = 449 \pm 44   \ \ (403)$
&
$m_{\tilde \chi^0_1}^{meas} =  155 \pm 92 \ \   (84)$
\end{tabular}\vspace*{-5mm}
\end{center}
\caption{Mass measurements for the second Monte Carlo study in GeV (actual masses in brackets). The plots show the gaussian density projections for the three masses. The 1-$\sigma$ confidence level interval is shaded, and the true mass value is indicated with the vertical dashed line. The dotted line indicates the value of $\mathcal{D}_{min}$ which defines the confidence interval.
}
\label{f.massmeasurementssecond}
\end{figure}

\section{Conclusion}
\label{s.conclusion} \setcounter{equation}{0} \setcounter{footnote}{0}

We introduced three new measurement techniques that address many of the realistic problems encountered at hadron colliders in applying $M_{T2}$ based variables. They make it possible to obtain mass measurements of all the particles in a \emph{fully hadronic} two-step decay chain with \emph{maximal combinatorial uncertainty} in the hard process. ISR is identified via $b$-tags, but issues of ISR-combinatorics could in general be addressed using the methods of \cite{ISRCOMBnojiri, ISRtagging, ISRCOMBalwall, ISRCOMBgripaios}.

These techniques are individually or together applicable beyond the example we studied, and we hope they will be helpful in determining the details of new physics found at the LHC. Given our example of a close-to worst-case scenario, we expect that dealing with less severe situations (e.g. only some combinatorics background with some leptons in the final state) would represent much less of a challenge by comparison. 

The Edge-to-Bump method represents a new approach to extracting interesting features from a distribution, and the basic idea should be adaptable to many applications. Focusing the analysis on a distribution-of-fits rather than a single fit on the original distribution fully or partially addresses issues of selection bias, choice of fit function and systematic error by sheer redundancy, and the possibilities for application as well as extensions and optimizations of the method are far from exhausted. 

Our  method of determining decay-chain assignments using a measured invariant-mass-edge 
is extremely simple and has a high yield of $\sim \mathcal{O}(10\%)$. Detailed exploration of this method should be the subject of a dedicated future study. 

Finally, we showed that $M_{T2}$ remains a viable variable in close to worst-case realistic scenarios (fully hadronic, little or no combinatorics information). No single method of reducing combinatorics background can be trusted for these powerful but fragile variables, but  application of our two methods as mutual cross-checks allows us to recover enough edge measurements to make a mass determination. The crucial issue of rejecting fake edges and supplying error bars that are not unrealistically small (without arbitrary and unmotivated error inflation) has been addressed by our extension of the Edge-to-Bump method to include the Golden Rule for $M_{T2}$ edge measurements.

The measured masses from both collider studies agree with the actual values in all cases, with precisions of $\sim 10\%$ for the sbottom and gluino mass at the LHC14 with $\mathcal{O}(100 \mathrm{fb}^{-1})$ of integrated luminosity.

Interestingly, in both studies there appears to be some systematic overestimation in the mass determination by about 1 $\sigma$. Looking at the first study one could think that this is due to overestimating the kinematic edges themselves (ISR effects \& smearing), but in the second Monte Carlo study most of the edges are in fact underestimated (except, notably, for the most precise measurement $M_{T2}^{210}(0)$). It would be helpful to understand this effect more completely. One could also try and determine how much data these methods require to complete a successful mass determination, and how the measurements scale with statistics. 

Our analysis used pure signal, so conducting this study with SM background and no (or fewer) $b$-tags would represent the true `worst-case' scenario. The only other assumption was that of a symmetric two-step decay chain. Generalization of these techniques to asymmetric chains \cite{MT2diffmothers, MT2diffdaughters} would be very interesting, as would be their possible combination with methods of detecting the decay chain topology in the first place. We leave such questions for future investigations.

\vspace*{-4mm}

\subsection*{Acknowledgements}
First and foremost I would like to thank Maxim Perelstein for many hours of helpful conversation, his comments on the manuscript and his preparation of the Madgraph model file used in the blind Monte Carlo study. I am also very grateful to Patrick Meade, Konstantin Matchev and Mihoko Nojiri for helpful discussions and their comments on the manuscript,  Julia Thom-Levy (of the CMS collaboration) for her explanations regarding QCD background and David Krohn for conversations about ISR-combinatorics. This work was supported in part by the National Science Foundation, under grants PHY-0757868 at Cornell and PHY-0969739 at Stony Brook.

\appendix

\section{Additional Plots for the Monte Carlo Studies}
\label{a.appendix}
Space constraints prevented us from including all the plots from both Monte Carlo studies in this paper. The interested reader can access them in a supplementary document online at \href{http://insti.physics.sunysb.edu/~curtin/edgefinder/}{\texttt{http://insti.physics.sunysb.edu/\string~curtin/edgefinder/}},  along with the \texttt{EdgeFinder} code.

\end{document}